\documentclass[fleqn,usenatbib]{mnras}

\usepackage{newtxtext,newtxmath}
\usepackage[T1]{fontenc}
\usepackage{ae,aecompl}

\usepackage{graphicx}	% Including figure files
\usepackage{amsmath}	% Advanced maths commands

\usepackage{amssymb}	% Extra maths symbols
\usepackage{multirow}
\usepackage{booktabs}
\newcommand{\lbol}{$L_{\mathrm{bol}}$}
\newcommand{\ledd}{$L_{\mathrm{Edd}}$}
\newcommand{\lx}{$L_{\mathrm{X}}$}
\newcommand{\mbh}{$M_\mathrm{BH}$}
\newcommand{\mcol}[1]{\multicolumn{1}{c}{#1}}
\newcommand{\msun}{M$_{\sun}$}
\newcommand{\mywidth}{0.67\columnwidth}
\newcommand{\mywidthtwo}{0.82\columnwidth}
\newcommand{\nbmc}{$N_\mathrm{BMC}$}
\newcommand{\nplbmc}{$N_\mathrm{PL}/N_\mathrm{BMC}$}
\newcommand{\nh}{$N_{\mathrm{H}}$}
\newcommand{\npl}{$N_{\mathrm{PL}}$}

\newcommand{\borus}{\texttt{Borus}}
\newcommand{\mytorus}{\texttt{MYTorus}}
\newcommand{\torus}{\texttt{Torus}}
\newcommand{\xspec}{\textsc{xspec}}

\newcommand{\chandra}{\textit{Chandra}}
\newcommand{\nustar}{\textit{NuSTAR}}
\newcommand{\suzaku}{\textit{Suzaku}}
\newcommand{\xmm}{\textit{XMM-Newton}}

\newcommand{\gro}{GRO J1655-40}
\newcommand{\gx}{GX 339-4}
\newcommand{\xte}{XTE J1550-564}

\title[Estimating black hole masses in obscured AGN]{Estimating black hole masses in obscured AGN using X-rays}

\author[Gliozzi, Williams, \& Michel]{
Mario Gliozzi$^{1}$\thanks{E-mail: mgliozzi@gmu.edu}
and James K. Williams$^{1}$
and Dina A. Michel$^{1}$
\\
$^{1}$ Department of Physics and Astronomy,
George Mason University, 4400 University Drive, Fairfax, VA 22030\\
}

\date{Accepted XXX. Received YYY; in original form ZZZ}

\pubyear{2015}

\begin{document}
\label{firstpage}
\pagerange{\pageref{firstpage}--\pageref{lastpage}}
\maketitle

\begin{abstract}
Determining the black hole masses in active galactic nuclei (AGN) is of crucial importance to constrain the basic characteristics of their central engines and shed light on their growth and co-evolution with their host galaxies. While the black hole mass (\mbh) can be robustly measured with dynamical methods in bright type 1 AGN, where the variable primary emission and the broad line region (BLR) are directly observed, a direct measurement is considerably more challenging if not impossible for the vast majority of heavily obscured type 2 AGN. In this work, we tested the validity of an X-ray-based scaling method to constrain the  \mbh\ in heavily absorbed AGN. To this end, we utilized a sample of type 2 AGN with good-quality hard X-ray data obtained by the \nustar\ satellite and with \mbh\ dynamically constrained from megamaser measurements. Our results indicate that, when the X-ray broadband spectra are fitted with physically motivated self-consistent models that properly account for absorption, scattering, and emission line contributions from the putative torus and constrain the primary X-ray emission, then the X-ray scaling method yields \mbh\ values that are consistent with those determined from megamaser measurements within their respective uncertainties. With this method we can therefore systematically determine the  \mbh\ in any type 2 AGN, provided that they possess good-quality X-ray data and accrete at a moderate to high rate.
\end{abstract}


%%%%%%%%%%%%%%%%%%%%%%%%%%%%%%%%%%%%%%%%%%%%%%%%%%

%%%%%%%%%%%%%%%%% BODY OF PAPER %%%%%%%%%%%%%%%%%%

\section{Introduction}
Historically, radio-quiet active galactic nuclei (AGN) have been divided into two main categories based on their optical spectroscopy: type 1 AGN, whose spectra are characterized by the presence of broad permitted lines (with full width at half maximum FWHM $> 2000\, \mathrm {km~s^{-1}}$) along with narrow forbidden lines, and type 2 AGN, where only narrow forbidden lines are detected \citep[e.g.,][]{khach74, ant83}. 

According to the basic AGN unification model, type 2 AGN can be considered as the obscured counterpart of type 1 AGN and  their main differences can be simply ascribed 
to different viewing angles, due to the presence of an obscuring toroidal structure made of gas and dust surrounding the AGN \citep[e.g.,][]{oster78,ant93,tad08,urry95}. However,
over the years, theoretical and observational studies have revealed that the simplest version of the unification model, based on a smooth donut-shaped torus, is unable to explain 
several observations, favoring instead a scenario where the torus is clumpy, with a covering factor depending on various AGN properties, and where the overall obscuration occurs on different scales 
with significant contribution from the galaxy itself. See \citet{netz15} and  \citet{ramo17} for recent comprehensive reviews on the unification model of AGN.  

Regardless of the nature of the 
obscuration, in type 2 AGN, the central engine -- an optical/UV emitting accretion disk, coupled with an X-ray emitting Comptonization corona -- and the broad line region (BLR) are not 
directly accessible to observations. This makes it more difficult to determine the properties of obscured AGN, which represent the majority of the AGN population and thus play a crucial 
role in our understanding of  the AGN activity, census, and cosmological evolution (see \citealt{hick18} for a recent review on obscured AGN).

%%%%%%%%%%%%%%%%%%%%%%%%%%%%%%%%%%%%%%%%%%%%%%%%%%%%%
%%%%%%%% TAB1: Sample properties  %%%%%%%%%%%%%%%%%%%%%%%%%%%%%%%%
\begin{table*}
	\caption{Properties of the sample}		
	\begin{center}
	\begin{tabular}{lrrrcc} 
			\toprule
			\toprule       
			\mcol{Source} & \mcol{Distance} &  \mcol{\mbh} & \mcol{$\lambda_{\mathrm{Edd}}$} & \mcol{\nustar} & \mcol{Exposure} \\
			\mcol{name} & \mcol{(Mpc)} & \mcol{($10^6$\ \msun)} & \mcol{(\lbol/\ledd)} & \mcol{observation ID} & \mcol{(ks)} \\
			\mcol{(1)} & \mcol{(2)} & \mcol{(3)} & \mcol{(4)} & \mcol{(5)} & \mcol{(6)} \\
			%			\noalign{\smallskip}
			\midrule
			NGC 1068 & $14.4^{\textrm{a}}$ & $8.0\pm0.3$ & $0.210\pm0.053$ & 60002033002 & 52.1 \\
			\noalign{\smallskip}
			NGC 1194 & $53.2^{\textrm{b}}$ & $65.0\pm3.0$ & $0.007\pm0.002$ & 60061035002 & 31.5 \\
			\noalign{\smallskip}
			NGC 2273 & $25.7^{\textrm{b}}$ & $7.5\pm0.4$ & $0.132\pm0.034$ & 60001064002 & 23.2 \\
			\noalign{\smallskip}
			NGC 3079 & $17.3^{\textrm{c}}$ & $2.4_{-1.2}^{+2.4}$ & $0.011\pm0.009$ & 60061097002 & 21.5 \\
			\noalign{\smallskip}
			NGC 3393 & $50.0^{\textrm{d}}$ & $31.0\pm2.0$ & $0.062\pm0.016$ & 60061205002 & 15.7 \\
			\noalign{\smallskip}
			NGC 4388 & $19.0^{\textrm{b}}$ & $8.5\pm0.2$ & $0.035\pm0.009$ & 60061228002 & 21.4 \\
			\noalign{\smallskip}
			NGC 4945 & $3.7^{\textrm{e}}$ & $1.4\pm0.7$ & $0.135\pm0.075$ & 60002051004 & 54.6 \\
			\noalign{\smallskip}
			IC 2560 & $26.0^{\textrm{f}}$ & $3.5\pm0.5$ & $0.175\pm0.050$ & 50001039004 & 49.6 \\
			\noalign{\smallskip}
			Circinus & $4.2^{\textrm{g}}$ & $1.7\pm0.3$ & $0.143\pm0.044$ & 60002039002 & 53.9 \\
			\bottomrule
		\end{tabular}	
	\end{center}
	\begin{flushleft}
		Columns: 1 = megamaser AGN name. 
		2 = distance used computing the \mbh\ from the maser measurements. References for the distances and black hole masses are (a) \citet{loda03}, (b) \citet{kuo11}, (c) \citet{kondra05}, (d) \citet{kondra08}, (e) \citet{green97}, (f) \citet{yama12}, and (g) \citet{green03}. 3 = black hole mass. 4 = Eddington ratio with Brightman's bolometric correction of $10\times$\ to \lx\ from \citet{bright16}. 5 = \nustar\ observation ID. 6 = exposure time.
	\end{flushleft}
	\label{table:tab1}
	%\footnotesize
\end{table*}
%%%%%%%%%%%%%%%%%%%%%%%%%%%%%%%%%%%%%%%%%%%%%%%%%%%%%
%%%%%%%%%%%%%%%%%%%%%%%%%%%%%%%%%%%%%%%%%%%%%%%%%%%%%

In order to shed light on the properties of the AGN central engine and its accretion state, we need to accurately determine the black hole mass (\mbh). In type 1 AGN, a reliable dynamical method frequently used is the so-called reverberation mapping method, where intrinsic changes in the continuum emission of the central engine, measured with some time delay in the line emission produced by the BLR, are used to constrain the \mbh, modulo a geometric factor \citep{bland82,pete04}. 
On the other hand, in  type 2 AGN, by definition the BLR is not visible and hence the reverberation mapping technique cannot be applied. Nevertheless, there is a small fraction of 
heavily obscured AGN for which it is still possible to measure the \mbh\ in a reliable way via a dynamical method. These are the sources that display water megamaser emission; if this emission is located in the accretion disk and is characterized by  the Keplerian motion, then the \mbh\ can be constrained with great accuracy \citep[e.g.][]{kuo11}. 
%%%% PROBLEM ADDING THE REFERENCE at the end of this sentence
%{\bf These are the sources that display water megamaser emission; if this emission is located in the outer part of the accretion disk and is characterized by the Keplerian motion, then the enclosed mass can be constrained with great accuracy \citep{kuo11}.}

In this work, we use a sample of heavily obscured type 2 AGN with \mbh\ constrained by megamaser measurements and with good-quality hard X-ray spectra obtained with the \textit{Nuclear Spectroscopic Telescope Array} (\nustar), a focusing hard X-ray telescope launched in 2012 with large effective area and excellent sensitivity in the energy range 3--78 keV, where the signatures of absorption and reflection are most prominent. Our main goal is to test whether an X-ray scaling method that yields \mbh\ values broadly consistent with those obtained from reverberation mapping in type 1 AGN can be extended to type 2 AGN.
 
The paper is structured as follows. In Section 2, we describe the sample properties and the X-ray data reduction. In Section 3, we report on the spectral analysis of \nustar\ data. The application of the X-ray scaling method and the comparison between the \mbh\ values derived with this method and those obtained from megamaser measurements are described in Section 4. We discuss the main results and draw our conclusions in Section 5.

\section{Sample Selection and Data Reduction}
We chose our sample of type 2 AGN based on the following two criteria:  these objects must have 1) the \mbh\ dynamically determined by megamaser disk measurements, and 2)  good-quality hard X-ray data. The former criterion is crucial to quantitatively test the validity of the X-ray scaling method applied to heavily obscured AGN, whereas the latter criterion is necessary to robustly constrain  the properties of the primary X-ray emission by accurately assessing the contributions of absorption and reflection caused by the putative torus. These criteria are fulfilled by the sample described by \citet{bright16}, which is  largely based on the sample of megamasers analyzed by \citet{mas16} and spans a range in X-ray luminosity between $10^{42}~ \mathrm{erg~s^{-1}}$ and a few units in $10^{43} ~\mathrm{erg~s^{-1}}$. The general properties of this sample, 
including the distance used to determine the \mbh\ from maser measurements, the \mbh\ itself, and the Eddington ratio $\lambda_{\mathrm{Edd}} = L_{\mathrm{bol}}/L_{\mathrm{Edd}}$, are reported in Table~\ref{table:tab1}.

The archival \nustar\ data of these nine objects were calibrated and screened using the \nustar\ data analysis pipeline \texttt{nupipeline} with standard filtering criteria and the calibration database \texttt{CALDB} version 20191219. From the calibrated and screened event files we extracted light curves and spectra, along with the RMF and ARF files necessary for the  spectral analysis, using the \texttt{nuproduct} script. The extraction regions used for both focal plane modules, FPMA and FPMB,  are circular regions of  radii  ranging from 40\arcsec\ to 100\arcsec\ depending on the brightness of the source, and centered on the brightest centroid. Background spectra and light curves were extracted by placing circles of the same size used for the source  in source-free regions of the same detector. No flares were found in the background light curves. All spectra were binned with a minimum of 20 counts per bin using the HEASoft task \texttt{grppha} 3.0.1 for the $\chi^2$ statistics to be valid.
%%%%%%%%%%%%%%%%%%%%%%%%%%%%%%%%%%%%%%%%%%%%%%%%%%%%%
%%%%%%%% FIG1: Spectral fits  %%%%%%%%%%%%%%%%%%%%%%%%%%%%%%%%%%%%
\begin{figure*}
	\includegraphics[width=\mywidth]{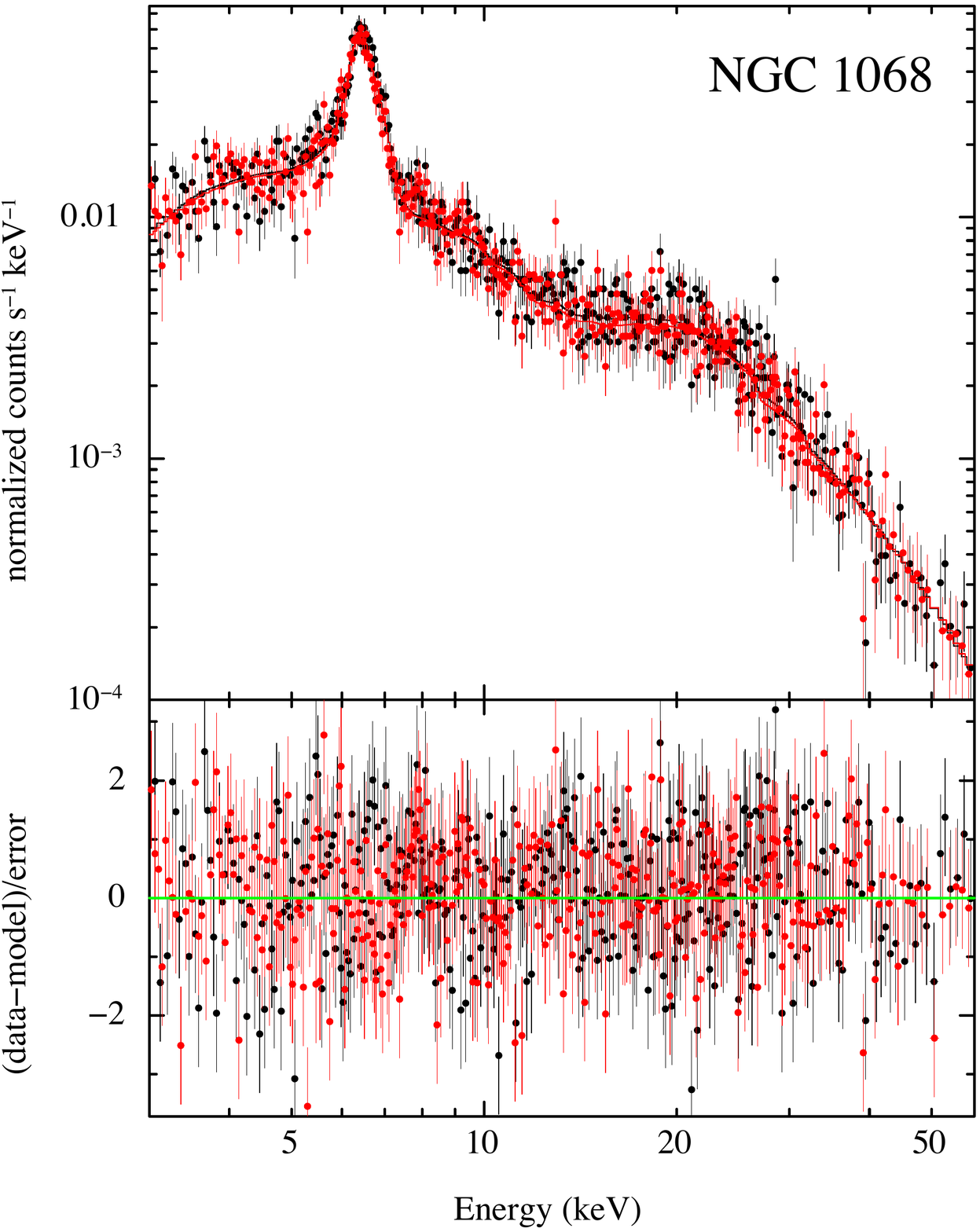}
	\includegraphics[width=\mywidth]{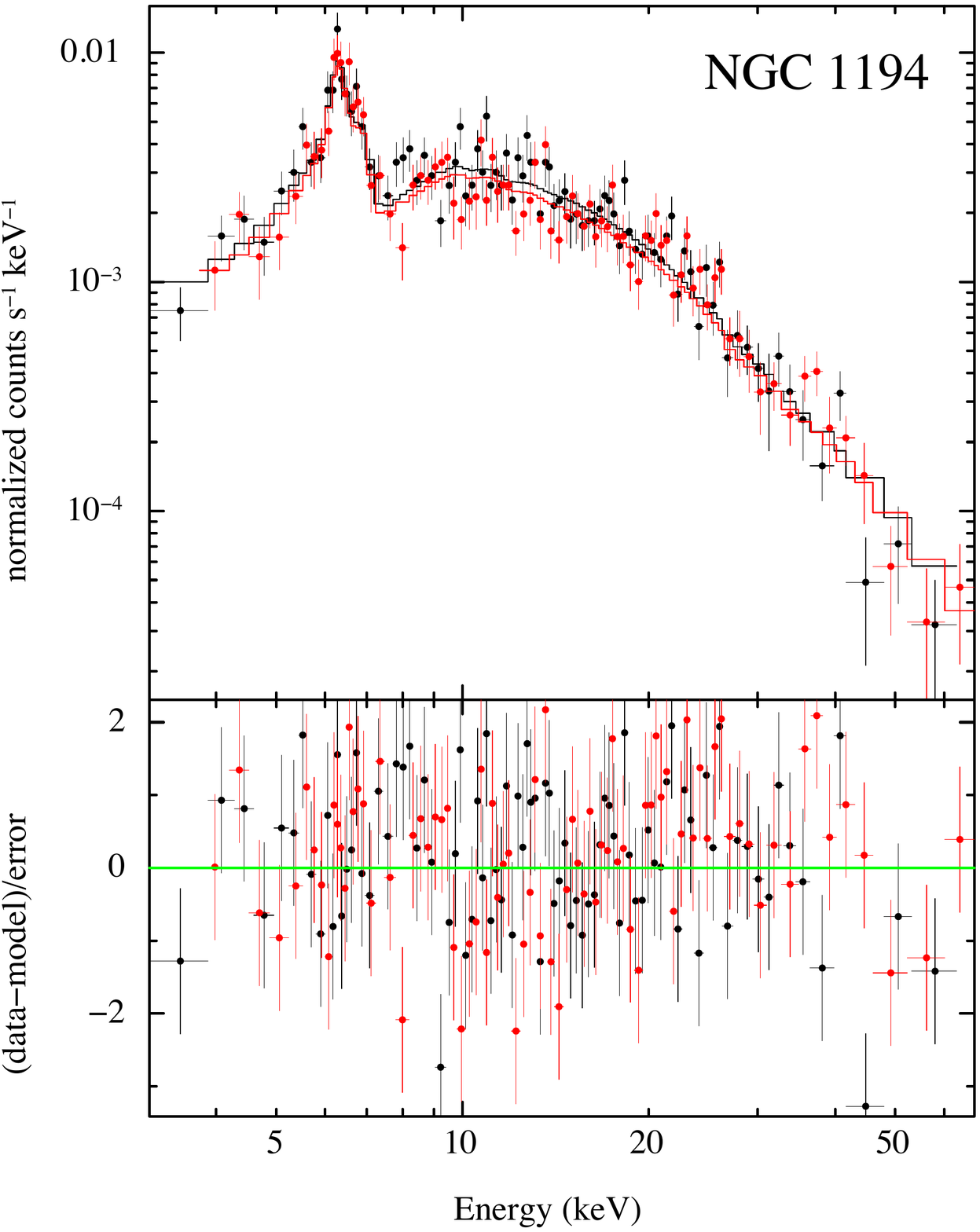}
	\includegraphics[width=\mywidth]{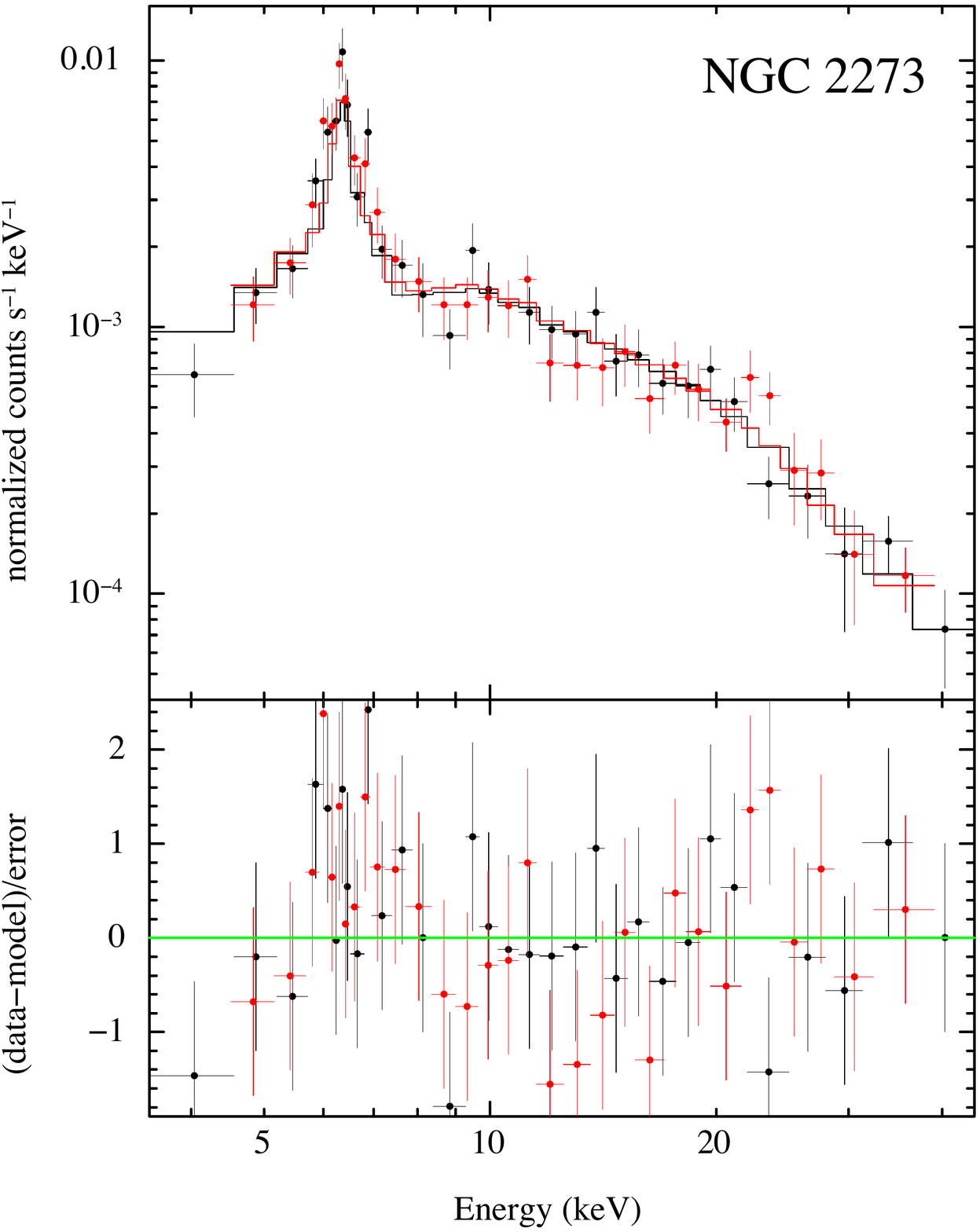}\\
	\includegraphics[width=\mywidth]{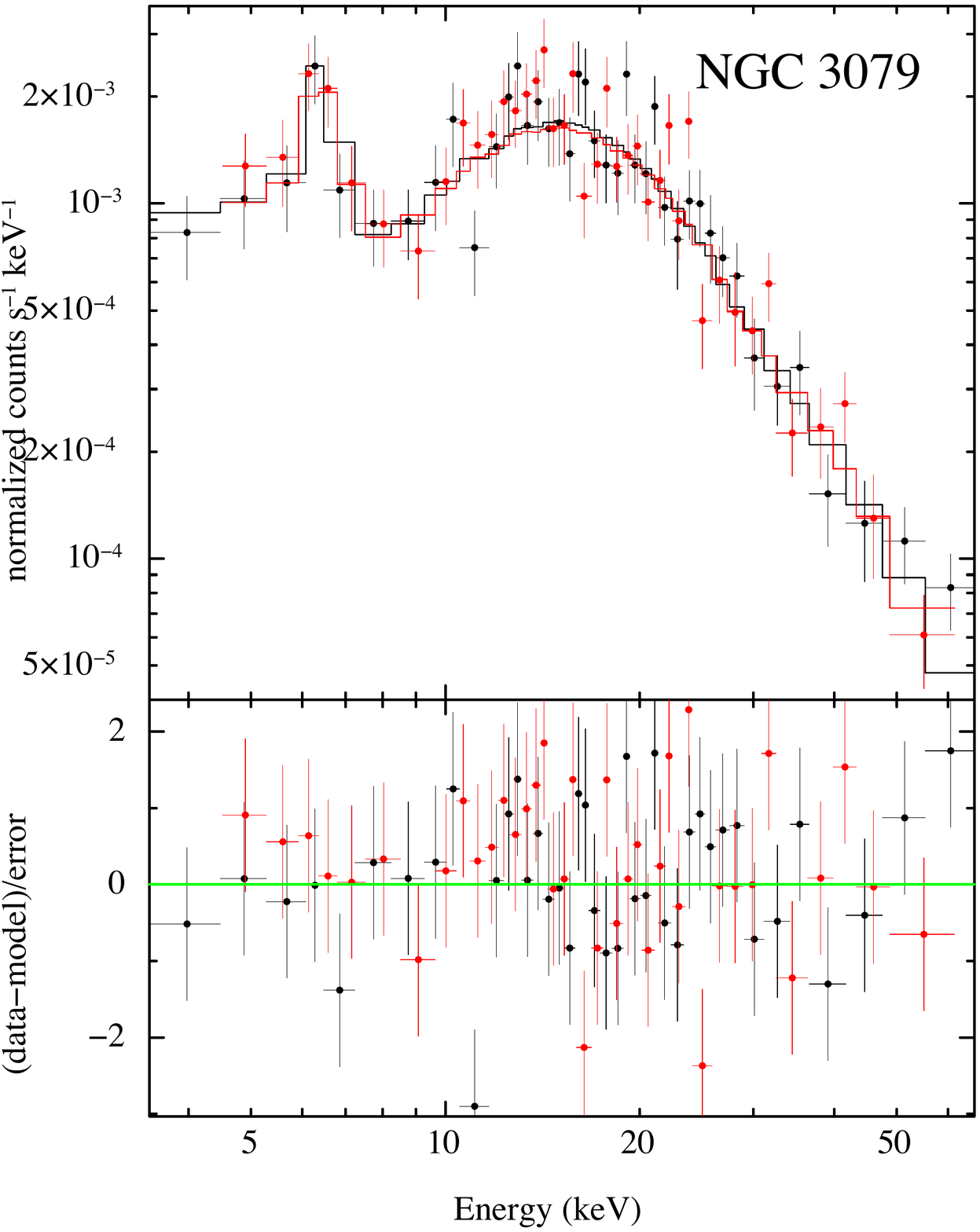}
	\includegraphics[width=\mywidth]{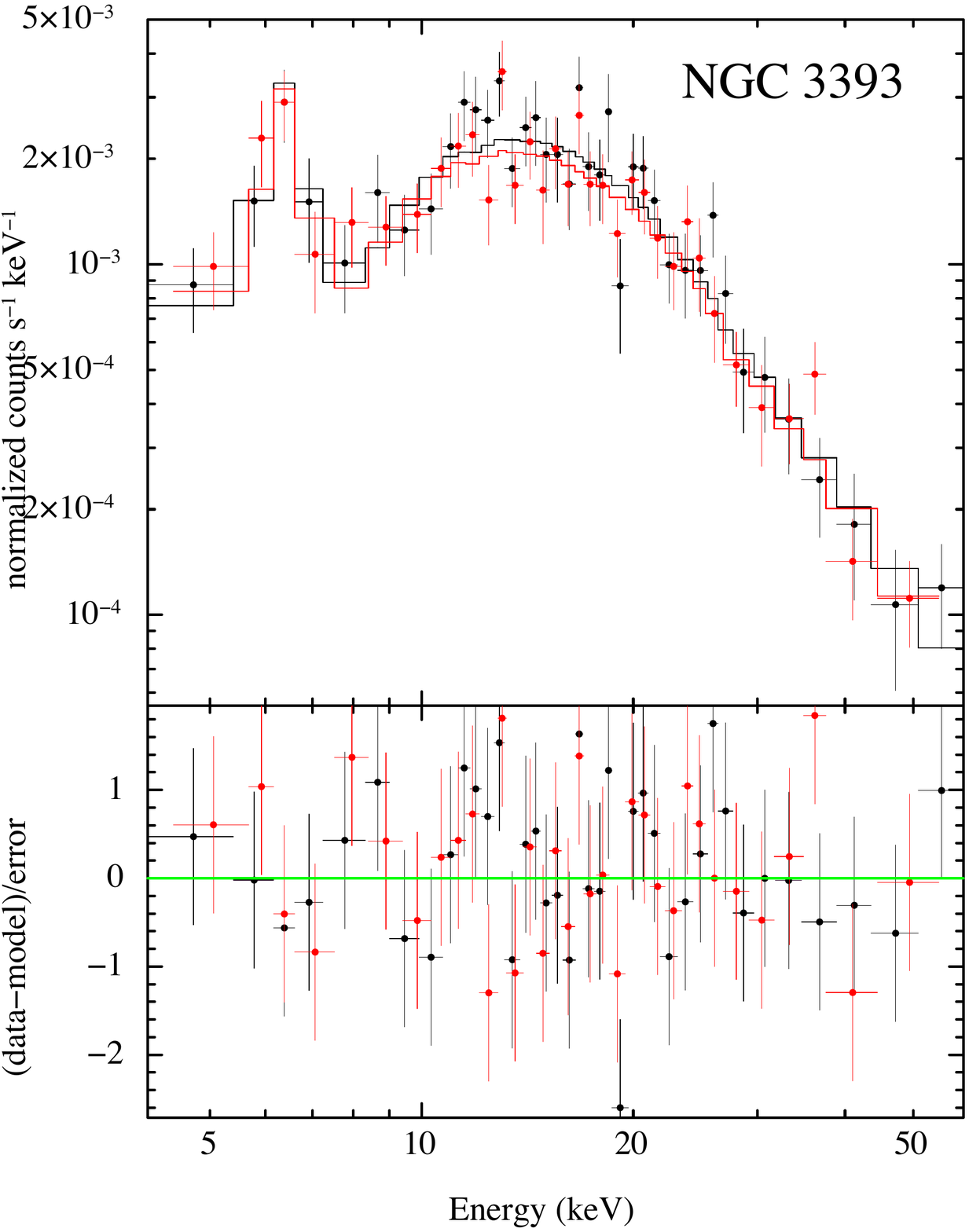}
	\includegraphics[width=\mywidth]{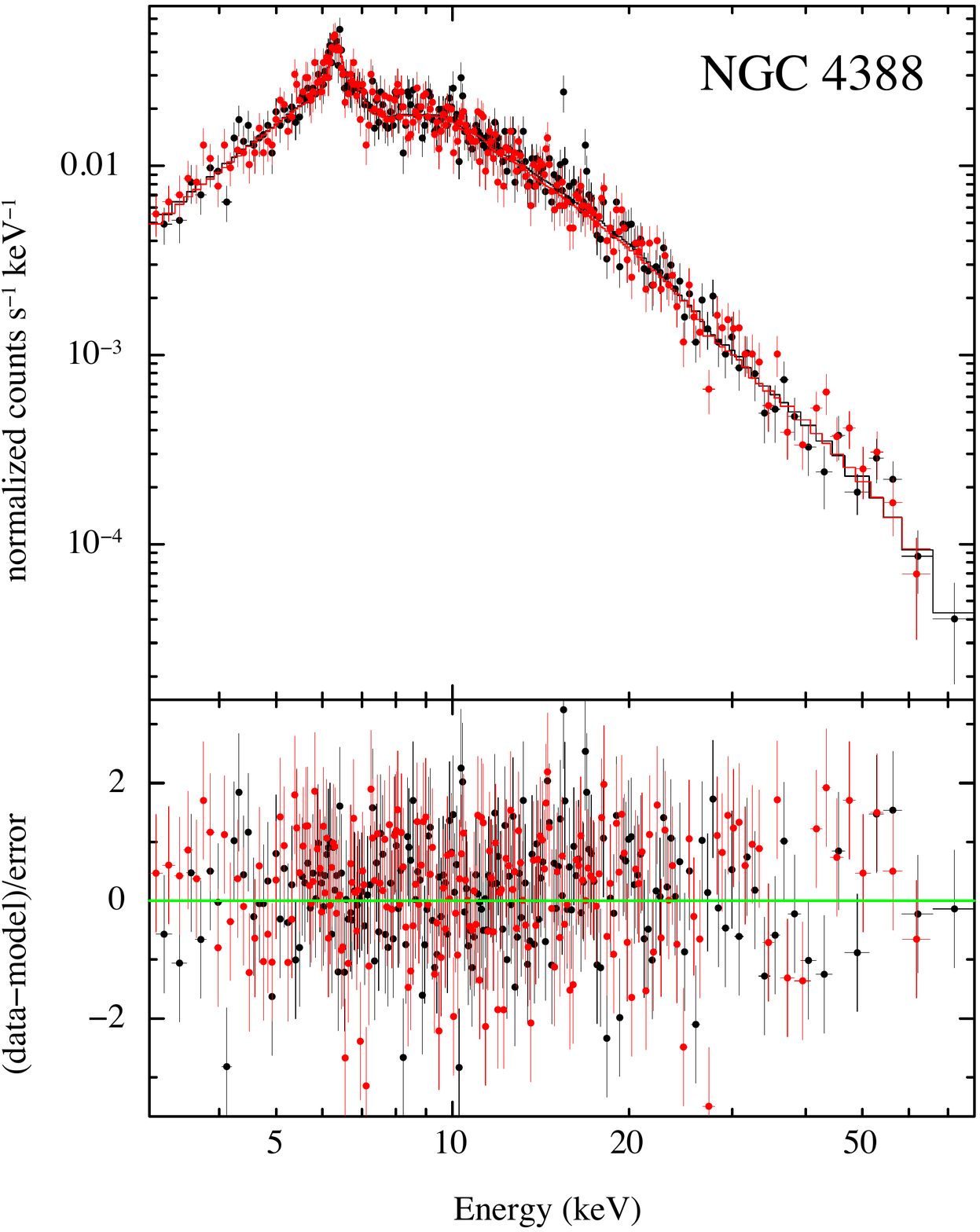}\\
	\includegraphics[width=\mywidth]{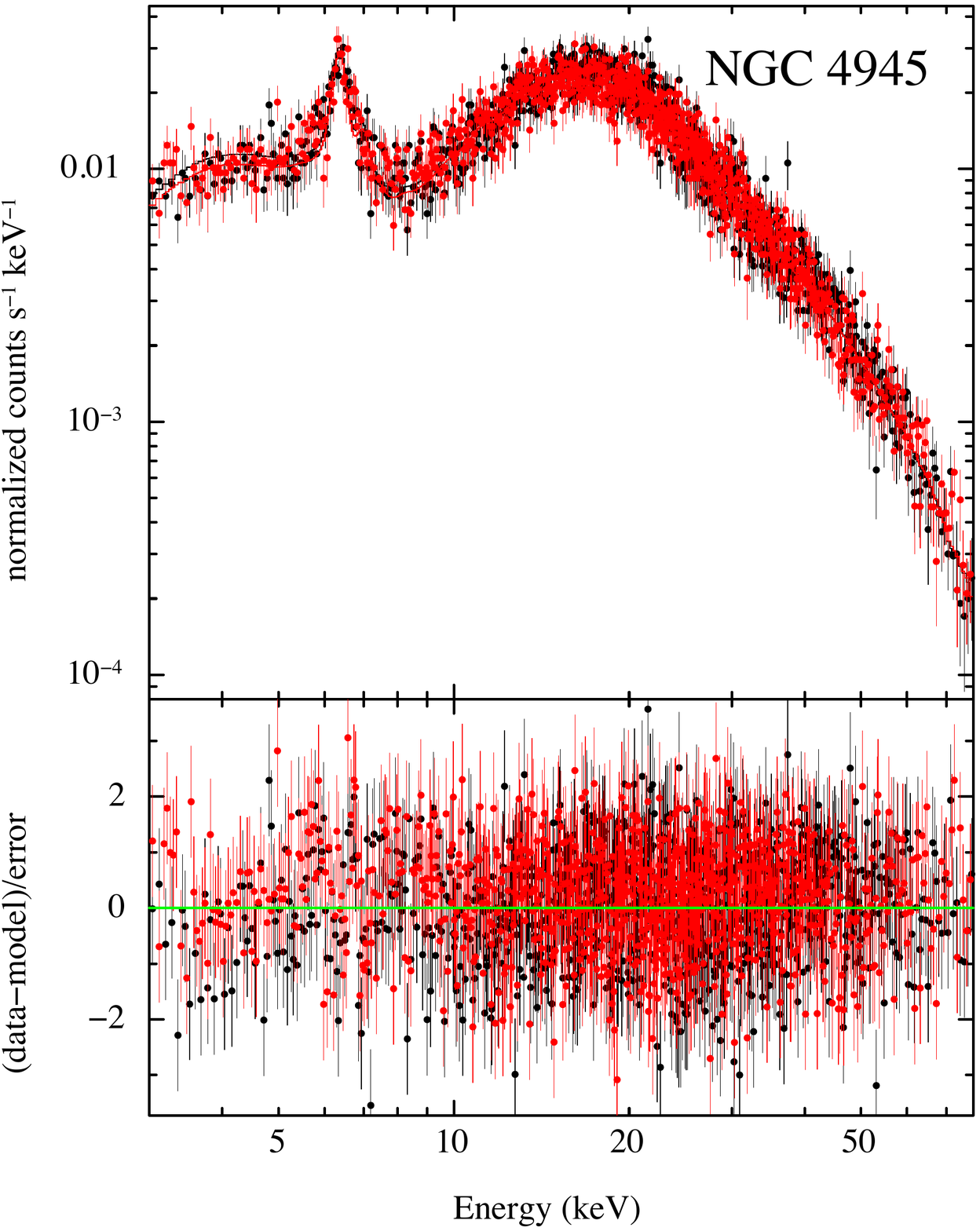}
	\includegraphics[width=\mywidth]{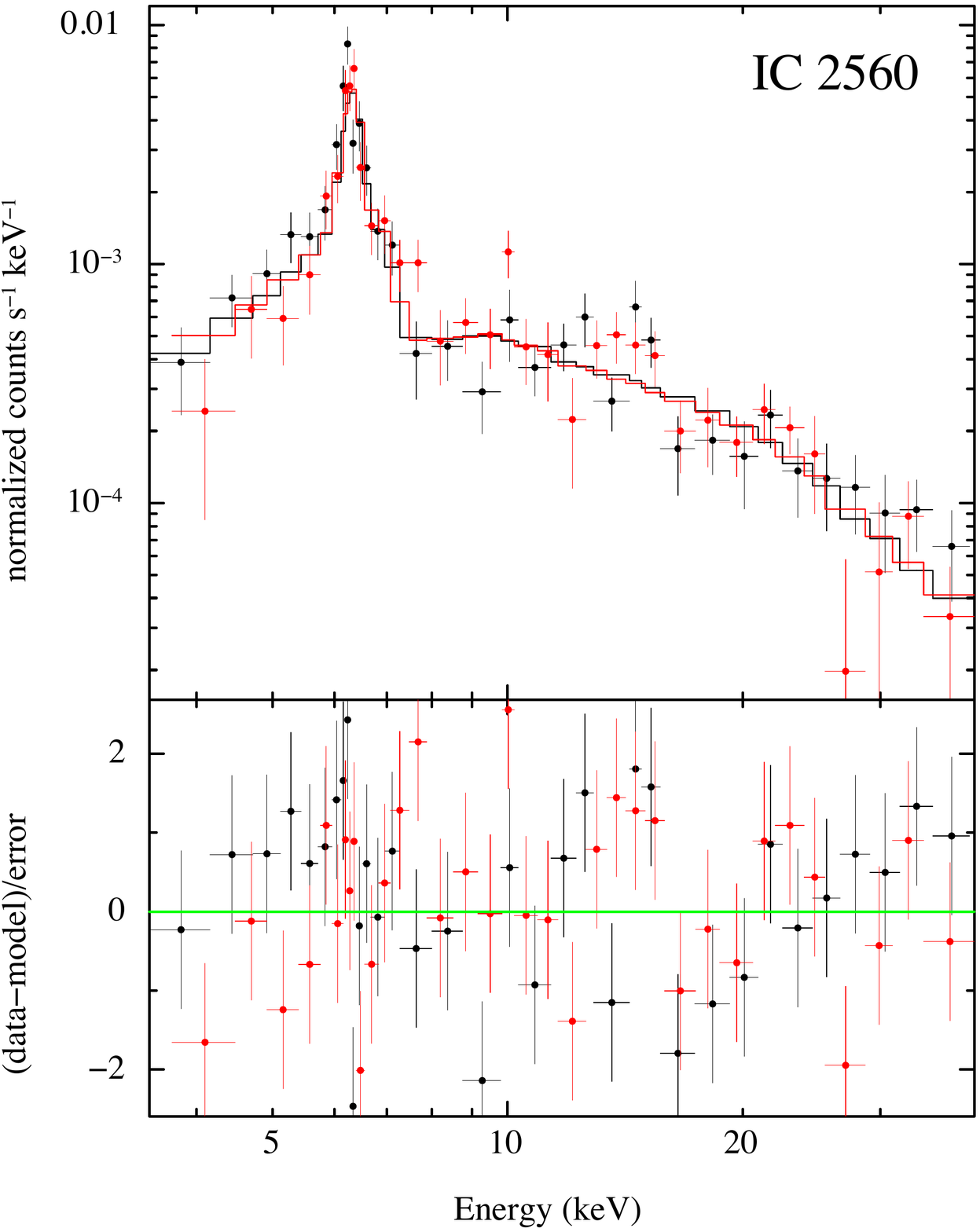}
	\includegraphics[width=\mywidth]{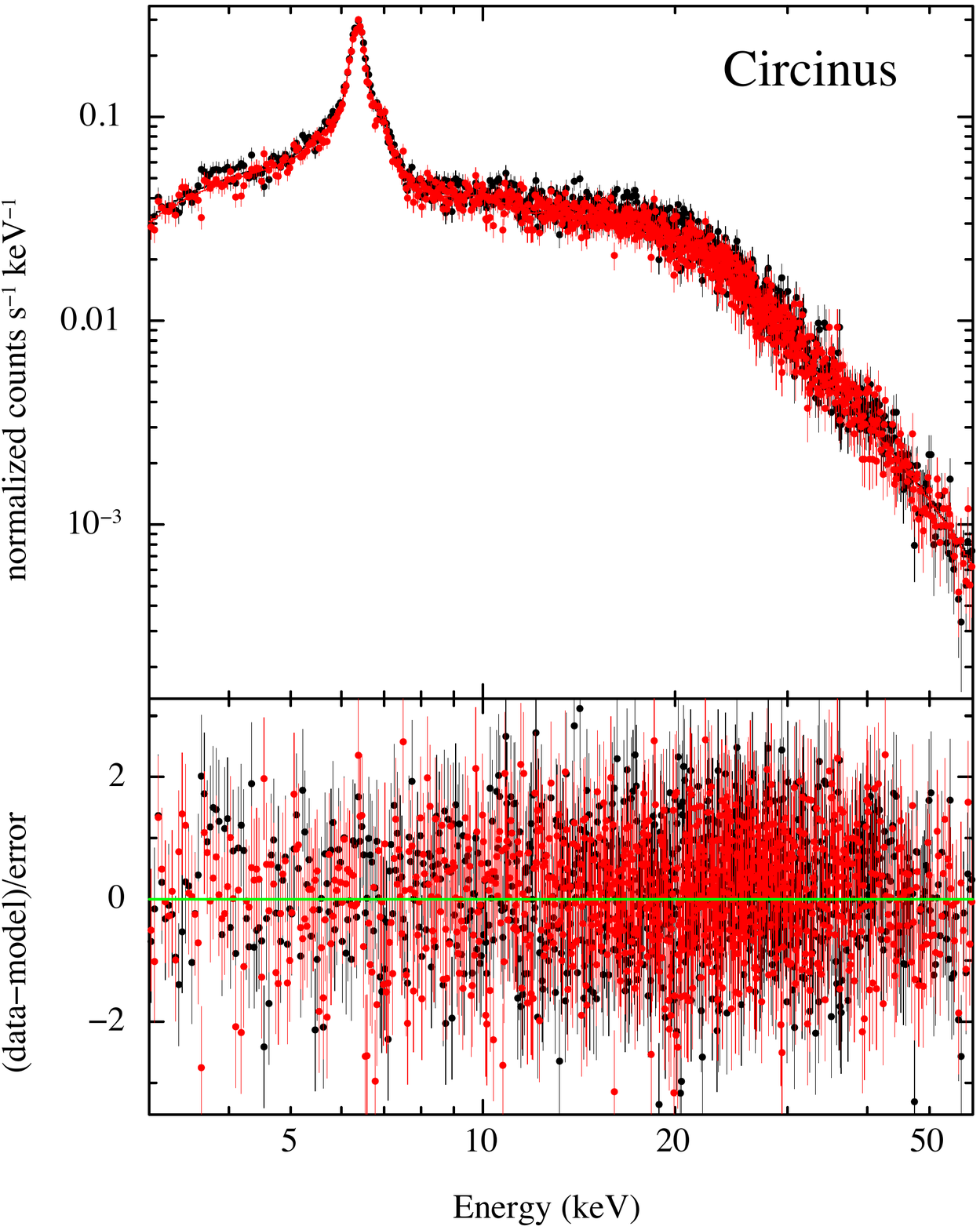}\\
\caption{The top panels show the \nustar\ spectra (black data points indicate FPMA data whereas the red ones indicate FPMB data) with the best-fit models, whereas the bottom panels show the data-to-model ratios.}
\label{figure:fig1}
\end{figure*}
%%%%%%%%%%%%%%%%%%%%%%%%%%%%%%%%%%%%%%%%%%%%%%%%%%%%%%%
%%%%%%%%%%%%%%%%%%%%%%%%%%%%%%%%%%%%%%%%%%%%%%%%%%%%%%%

\section{Spectral analysis}
%%%%%%%%%%%%%%%%%%%%%%%%%%%%%%%%%%%%%%%%%%%%%%%%%%%%%
%%%%%%%% TAB2: Spectral results %%%%%%%%%%%%%%%%%%%%%%%%%%%%%%%%%
\begin{table*}
	\caption{Spectral Results}		
	\begin{center}
		\begin{tabular}{lrrrrrrrrrrr} 
			\toprule
			\toprule       
			\mcol{Source} & \mcol{$N_{{\textrm{H}}_{\textrm{Gal}}}$} &  \mcol{$\log(N_{{\textrm{H}}_{\textrm{bor}}})$} & \mcol{$\cos{\theta}$} & \mcol{CFtor} & \mcol{$A_{\textrm{Fe}}$} & 
			\mcol{$N_{{\textrm{H}}_{\textrm{mytz}}}$} & \mcol{$\Gamma$} & \mcol{\nbmc} & \mcol{$\log{A}$} & \mcol{$f_{\textrm{s}}$} & \mcol{($\chi^2/$dof)}\\
			
			\mcol{name} & \mcol{($10^{20}$ cm$^{-2}$)} & \mcol{~} & \mcol{~} & \mcol{(\%)} & \mcol{~} & 
			\mcol{($10^{24}$ cm$^{-2}$)} & \mcol{~} & \mcol{~} & \mcol{~} & \mcol{(\%)} & \mcol{~}\\
			
			\mcol{(1)} & \mcol{(2)} & \mcol{(3)} & \mcol{(4)} & \mcol{(5)} & \mcol{(6)} & \mcol{(7)} & \mcol{(8)} & \mcol{(9)} & \mcol{(10)} & \mcol{(11)} & \mcol{(12)}\\
			%			\noalign{\smallskip}
			\midrule
			NGC 1068 & $2.59$ & $23.3\pm0.1$ & $0.1$ & 15  & 1.0& $3.5\pm0.1$ & $1.98_{-0.01}^{+0.01}$ & $2.2_{-0.1}^{+0.1}\times10^{-3}$ & 0.08 & 1.6 & 754.9/713\\
			\noalign{\smallskip}
			NGC 1194 & $5.53$ & $23.9\pm0.1$ & $0.1$ & 91 & 3.2 & $0.8\pm0.1$ & $1.62_{-0.05}^{+0.05}$ & $6.4_{-0.8}^{+1.1}\times10^{-5}$ & 0.57 & 4.0 & 194.3/167\\
			\noalign{\smallskip}
			NGC 2273 & $5.80$ & $25.0\pm0.7$ & $0.1$ & 15 & 1.0 & $6.8\pm0.4$ & $1.95_{-0.05}^{+0.05}$ & $3.1_{-0.2}^{+0.2}\times10^{-3}$ & 2.0 & \ldots & 54.0/57\\
			\noalign{\smallskip}
			NGC 3079 & $0.87$ & $24.5\pm0.1$ & $0.1$ & 20 & 1.0 & $2.9\pm0.1$ & $1.91_{-0.06}^{+0.05}$ & $1.1_{-0.2}^{+0.2}\times10^{-3}$ & 0.8 & 0.3 & 80.1/75\\
			\noalign{\smallskip}
			NGC 3393 & $6.13$ & $25.2\pm0.2$ & $0.1$ & 15 & 1.0 & $2.4\pm0.1$ & $1.86_{-0.10}^{+0.10}$ & $9.6_{-1.2}^{+3.1}\times10^{-4}$ & 0.22 & \ldots & 54.3/65\\
			\noalign{\smallskip}
			NGC 4388 & $2.57$ & $23.6\pm0.1$ & $0.1$ & 91 & 1.0 & $0.4\pm0.1$ & $1.66_{-0.04}^{+0.04}$ & $3.3_{-0.4}^{+0.5}\times10^{-4}$ & -0.55 & 17.0 & 435.2/420\\
			\noalign{\smallskip}
			NGC 4945 & $14.0$ & $24.4\pm0.1$ & $0.1$ & 91 & 0.8 & $3.0\pm0.7$ & $1.74_{-0.05}^{+0.05}$ & $1.6_{-0.1}^{+0.1}\times10^{-3}$ & 2.15 & 0.5& 1699.8/1716\\
			\noalign{\smallskip}
			IC 2560 & $6.51$ & $25.1\pm0.1$ & $0.1$ & 15 & 2.3 & $6.9\pm0.1$ & $2.08_{-0.08}^{+0.08}$ & $1.8_{-0.3}^{+0.4}\times10^{-3}$ & 2.0 & \ldots &  88.3/64\\
			\noalign{\smallskip}
			Circinus & $52.5$ & $23.6\pm0.1$ & $0.1$ & 24 & 1.7 & $1.6\pm0.1$ & $2.17_{-0.01}^{+0.01}$ & $9.8_{-0.2}^{+0.1}\times10^{-3}$ & -0.43 & 3.3 & 1713.2/1714\\
			\bottomrule
		\end{tabular}
	\end{center}
	\begin{flushleft}
		Columns: 1 = megamaser AGN name. 
		2 = Galactic column density from NASA's HEASARC. 3 = column density calculated with the \borus\ model. 4 = cosine of the inclination angle. 5 = covering factor. 6 = iron abundance relative to the solar value. 7 = column density calculated with the \mytorus\ model. 8 = photon index. 9 = normalization of the BMC model. 10 = logarithm of $A$, where $A=(f+1)/f$ and $f$ is the fraction of seed photons that are scattered. 11 = fraction of the primary emission scattered along the line of sight by an extended ionized reflector. 12 = $\chi^2$ divided by degrees of freedom. 
	\end{flushleft}
	\label{table:tab2}
\end{table*}
%%%%%%%%%%%%%%%%%%%%%%%%%%%%%%%%%%%%%%%%%%%%%%%%%%%%%
%%%%%%%%%%%%%%%%%%%%%%%%%%%%%%%%%%%%%%%%%%%%%%%%%%%%%
The X-ray spectral analysis  was performed using the \xspec\ \texttt{v.12.9.0} software package \citep{arn96}, and  the errors quoted on the spectral parameters represent the 1$\sigma$ confidence level.

The \nustar\ spectra of this sample have already been reasonably well fitted with self-consistent physically motivated models such as \mytorus\ \citep{muya09} and \torus\ \citep{brina11} to account for the continuum scattering and absorption, as well as the fluorescent line emission produced by the torus, whereas the primary emission was parametrized with a phenomenological power-law model. However, in order to apply the X-ray scaling method (whose key features are described in the following section), the primary emission needs to be parametrized by the Bulk Motion Comptonization model (\texttt{BMC}), which is a generic Comptonization model that convolves thermal seed photons producing a power law \citep{tita97}. This model, which can be used to parametrize both the bulk motion and the thermal Comptonization, is described by four spectral parameters: the normalization \nbmc, the spectral index $\alpha$, the temperature of the seed photons $kT$, and $\log A$, where $A$ is related to the fraction of scattered seed photons $f$ by the relationship $A=(f+1)/f$.
Unlike the phenomenological power-law model, the \texttt{BMC} parameters are computed in a self-consistent way, and the power-law component produced by the \texttt{BMC} does not extend to arbitrarily low energies.

We carried out a homogeneous systematic reanalysis of the \nustar\ spectra of these sources.
We started from the best-fit models reported in the literature but utilized the \borus\ model \citep{balo18}, which can be considered as an evolution of the previous torus models. Specifically, \borus\ has the same geometry implemented in \torus\ but can also be used in a decoupled mode, where the column density \nh\ responsible for the continuum scattering and fluorescent line emission is allowed to be different from the \nh\ responsible for the attenuation of the primary component. Additionally, unlike \torus, this model correctly accounts for the absorption experienced by the photons backscattered from the far side of the inner torus. With respect to \mytorus, \borus\ contains additional emission lines, has a larger range for \nh, and directly yields the value of the covering fraction. However, since \borus\ only parametrizes the scattered continuum and the fluorescent line components associated with the torus, to account for the absorption and scattering experienced by the primary emission, we utilized the zeroth-order component of \mytorus\ (\texttt{MYTZ}), which 
properly includes the effects of the Klein-Nishina Compton scattering cross section that are relevant in heavily absorbed AGN at energies above 10 keV. 
In summary, our procedure can be summarized in three steps: 1) we started from the spectral best fits reported in the literature; 2) we then substituted \borus\  (more specifically, we used the \texttt{borus02\_v170323a.fits} table) for either \torus\ or \mytorus\ to account for the scattered and line components, and used the zeroth-order component of \mytorus\ for the transmitted one; 3) finally, we substituted \texttt{BMC} for the power-law model used for the primary emission. 

\begin{figure}
\includegraphics[width=\columnwidth]{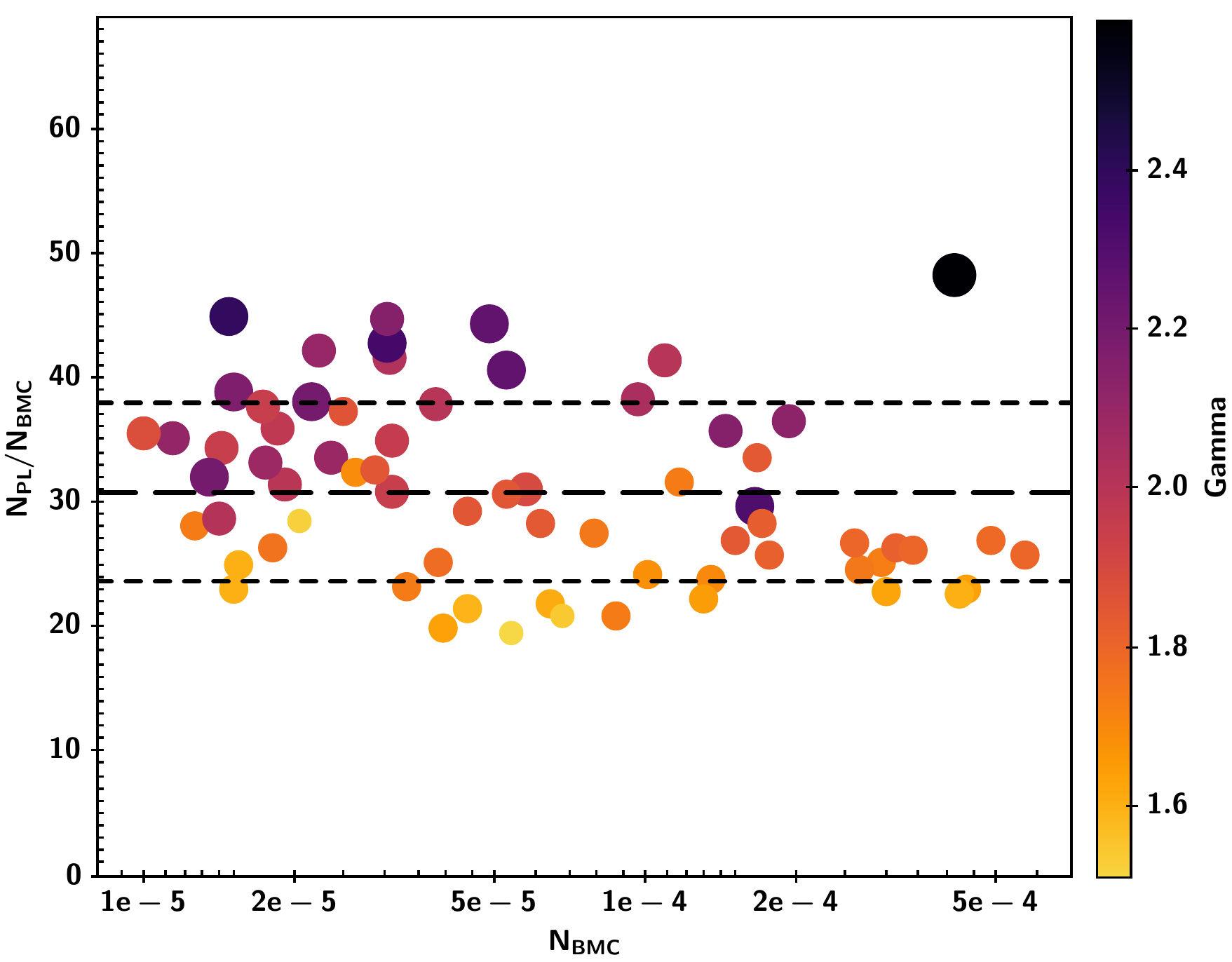}
\caption{ \nplbmc\ plotted vs. \nbmc\ for a sample of ``clean'' type 1 AGN (i.e., AGN with negligible warm or cold absorbers). The black longer-dashed line represents the average value, whereas the shorter-dashed lines indicate the one standard deviation levels from the average. Both the data point's size and color provide information on the source's photon index $\Gamma$: the larger the symbol and the darker the color, the steeper the  $\Gamma$.}
\label{figure:fig2}
\end{figure}

In the spectral fitting, in order to preserve the self-consistency of these physically motivated torus models, which are created by Monte Carlo simulations using a power law to parametrize the X-ray primary emission, one needs to link the primary emission parameters -- the photon index $\Gamma$ and the normalization \npl\ -- to the input parameters of the scattered continuum and emission-line components. In the case of the \texttt{BMC}  model, the power-law slope is described by the spectral index $\alpha$, which is related to the photon index by the relationship $\Gamma=\alpha+1$. However, there is not a known mathematical equation linking the normalizations \nbmc\ and \npl. We therefore derived this relationship empirically by using a sample of clean type 1 AGN (i.e., AGN without cold or warm absorbers), whose details are described in \citet{will18,glioz20}. We fitted the 2--10 keV \xmm\ spectra twice, first with the \texttt{BMC} model and then with a power law.  The results of this analysis are illustrated in Fig.~\ref{figure:fig2}, where \nplbmc\ is plotted versus 
\nbmc, showing that, regardless of the value of \nbmc, the normalization ratios cluster around the average value, $30.8\pm0.9$, represented by the longer-dashed line, with moderate 
scattering  of $\sigma=7.2$, represented by the shorter-dashed lines.  Fig.~\ref{figure:fig2}, where the data point's size and color provide information about the photon index,  also reveals a tendency for the AGN with steeper spectra to have larger values of \nplbmc. This trend  is formally confirmed by a least-squares best fit of  \nplbmc\  vs. $\Gamma$, which yields  \nplbmc\ =-18.9 + 26.3$\Gamma$, with a Pearson's correlation coefficient of 0.85. 
 
 These results are  in agreement with those obtained from a series of simulations carried out with the \texttt{fakeit} command in \xspec. Simulating spectra of the \texttt{BMC} model with  the parameters varying over a broad range, and then fitting  them with a power-law model, we found that \nplbmc\  shows a horizontal trend when plotted vs. \nbmc\ with an average  value consistent with 30 for $\Gamma=1.9$, whereas the horizontal trend is consistent with an average value of 24 for $\Gamma=1.6$ and 33 for $\Gamma=2.2$.
 
 Based on these findings, in our spectral fitting of the megamaser sample we forced \nbmc\ to be equal to \npl/30 by linking these parameters to reflect this relationship. For completeness,  and to take into account the weak dependence of \nplbmc\  on $\Gamma$, we have also carried out the spectral analysis assuming  \nplbmc\ = 24 (i.e., the average value minus one standard deviation) for flat spectrum sources and   \nplbmc\ = 38 (average $+\sigma$) for steep spectrum sources.
 
 We note that, compared to  $\Gamma$ and \nbmc, the remaining \texttt{BMC} parameters $kT$ and $\log A$ play a marginal role in the shape of the spectrum 
and in the determination of the \mbh, as explicitly assessed in \citet{glioz11}. Therefore, to limit the number of free parameters, we fixed $kT$ to 0.1 keV, which is consistent with the values generally obtained when the \texttt{BMC} model is fitted to X-ray AGN spectra \citep[e.g.,][]{glioz11,will18}, whereas $\log A$ was fixed to the best-fit value obtained in the first fitting iteration.

 %%%%%%%%%%%%%%%%%%%%%%%%%%%%%%%%%%%%%%%%%%%%%%%%%%%%
%%%%%%%% FIG.2 N_PL/N_BMC vs.N_BMC   %%%%%%%%%%%%%%%%%%%%%%%%%%%%%%
%%%%%%%%%%%%%%%%%%%%%%%%%%%%%%%%%%%%%%%%%%%%%%%%%%%%%
%%%%%%%%%%%%%%%%%%%%%%%%%%%%%%%%%%%%%%%%%%%%%%%%%%%%%

Our baseline model for all type 2 AGN fitted in this work is expressed in the \xspec\ syntax as follows:
\begin{verbatim}
phabs * (atable(Borus) + MYTZ * BMC + const * BMC)
\end{verbatim}
where the first absorption model \texttt{phabs} accounts for our Galaxy contribution, the \borus\ table model parametrizes the continuum scattering and fluorescent emission line components associated with the torus, and \texttt{MYTZ} models the absorption and Compton scattering acting on the transmitted primary emission, which is described by the Comptonization model \texttt{BMC}. The last additive component \texttt{const*BMC} parametrizes the fraction of primary emission directly scattered towards the observer by a putative optically thin ionized medium, which is often observed below 5 keV in spectra of heavily obscured AGN \citep[e.g.,][]{yaq12}.
 
Depending on the source and the complexity of its X-ray spectrum, additional components (such as the host galaxy contribution, individual lines, additional absorption and scattering components, or models describing off-nuclear sources contained in the \nustar\ extraction region) are included and described in the individual notes of each source reported in the Appendix.
 
The spectral parameters obtained by fitting this baseline model are reported in Table~\ref{table:tab2}, and the best fits and model-to-data ratios are shown in Fig~\ref{figure:fig1}.

\section{Black hole masses}
\subsection{\texorpdfstring{\mbh}{MBH} from the X-ray scaling method}
The X-ray scaling method was first introduced by \citet{shapo09}, who showed that the BH mass and distance $D$ of any stellar mass BH can be obtained by scaling these properties  from those of an appropriate reference source (i.e., a  BH system with \mbh\ dynamically determined and distance tightly constrained). In its original form this technique exploits the similarity of the trends displayed
by different BH systems in two plots -- the photon index $\Gamma$ vs. quasi-periodic oscillation (QPO) frequency plot and the \nbmc--$\Gamma$ diagram -- to derive their \mbh\  and $D$. 

Based on the assumption that the process leading to the ubiquitous emission of X-rays -- the Comptonization of seed photons 
produced by the accretion disk -- is the same in all BH systems regardless of their mass, this method can in principle be extended to any BH including the supermassive BHs at the cores of AGN. In the latter case, since the detection of QPOs is extremely rare but the distance is generally well constrained by redshift or Cepheid measurements, only the  \nbmc--$\Gamma$ diagram is used to determine the \mbh.  Indeed, over the years, this method has been successfully applied to stellar mass BHs  (e.g., \citealt{sei14,tita16a}) and to
ultraluminous X-ray sources (e.g., \citealt{tita16b,jang18}), as well as to a handful of AGN that showed high spectral and temporal variability during deep X-ray exposures (e.g., \citealt{glioz10,gia14,sei18}.) 

Although the vast majority of AGN do not possess long-term X-ray observations and do not show strong intrinsic spectral variability (i.e., variability described by
substantial changes of $\Gamma$ not caused by obscuration events), the X-ray scaling method can be extended to any type 1 AGN with one good-quality X-ray observation. Indeed, \citet{glioz11} demonstrated that the \mbh\ values determined with this method are fully consistent with the corresponding values obtained from the reverberation mapping technique. The reference sources, used in that study and then also in this work, are three stellar mass BHs residing in X-ray 
binaries -- GRO J1655-40, GX 339-4, and  XTE J1550-564 -- with \mbh\ dynamically determined and spectral evolution during the rising and decaying phases of their outbursts mathematically parametrized by \citet{shapo09}.  The physical properties of the stellar references and the mathematical description of their spectral trends, as well as the details of the method, are reported in \citet{glioz11}. 

In summary, all the reference trends yielded \mbh\ measurements consistent 
with the reverberation mapping values within their nominal uncertainties, with the decaying trends showing a slightly better agreement than the rising trends, which have a tendency to underestimate \mbh\ to a moderate degree. Unfortunately, the most reliable reference source -- GRO J1655-40 during the 2005 decaying phase (hereafter GROD05) -- has a fairly small range of $\Gamma$ during its spectral transition limiting its application to sources with relatively flat photon indices. Using the reverberation mapping values as calibration, it was determined that for AGN with steep spectra ($\Gamma > 2$) the best estimate of \mbh\ is obtained using the value derived  from the rising phase of the 1998 outburst of XTE J1550-564 multiplied by a factor of 3 (hereafter 3*XTER98). Below, we summarize the general principles at the base of this technique; a more detailed explanation can be found in \citet{shapo09} and \citet{glioz11}. For completeness, in the Appendix we report the basic information on the reference sources, including the mathematical expression of their spectral trends, which is necessary to derive \mbh\ using the equation reported below.

The scaling method assumes that all BH systems accreting at a moderate or high rate undergo similar spectral transitions, characterized by the ``softer when brighter'' trend (i.e., the X-ray spectrum softens when the accretion and hence the luminosity increases). These spectral transitions are routinely observed in stellar BHs \citep[e.g.,][]{remi06} and often found in samples of AGN (e.g., \citealt{shem08,risa09,bright13,bright16}), which are characterized by considerably longer dynamical timescales, making it nearly impossible to witness a genuine state transition in a supermassive BH system, although  a few long monitoring studies have observed this spectral trend in individual AGN \citep[e.g.,][]{sobo09}. The ``softer when brighter'' trend, usually illustrated by plotting the photon index versus the Eddington ratio $\lambda_ {\mathrm{Edd}}$, is seen with some scattering in numerous type 1 AGN samples and also in the heavily absorbed type 2 AGN, which are the focus of our work \citep{bright16}. This lends support to the hypothesis that the photon index $\Gamma$ is a reliable indicator of the accretion state of any BH.

\begin{figure}
\includegraphics[width=\columnwidth]{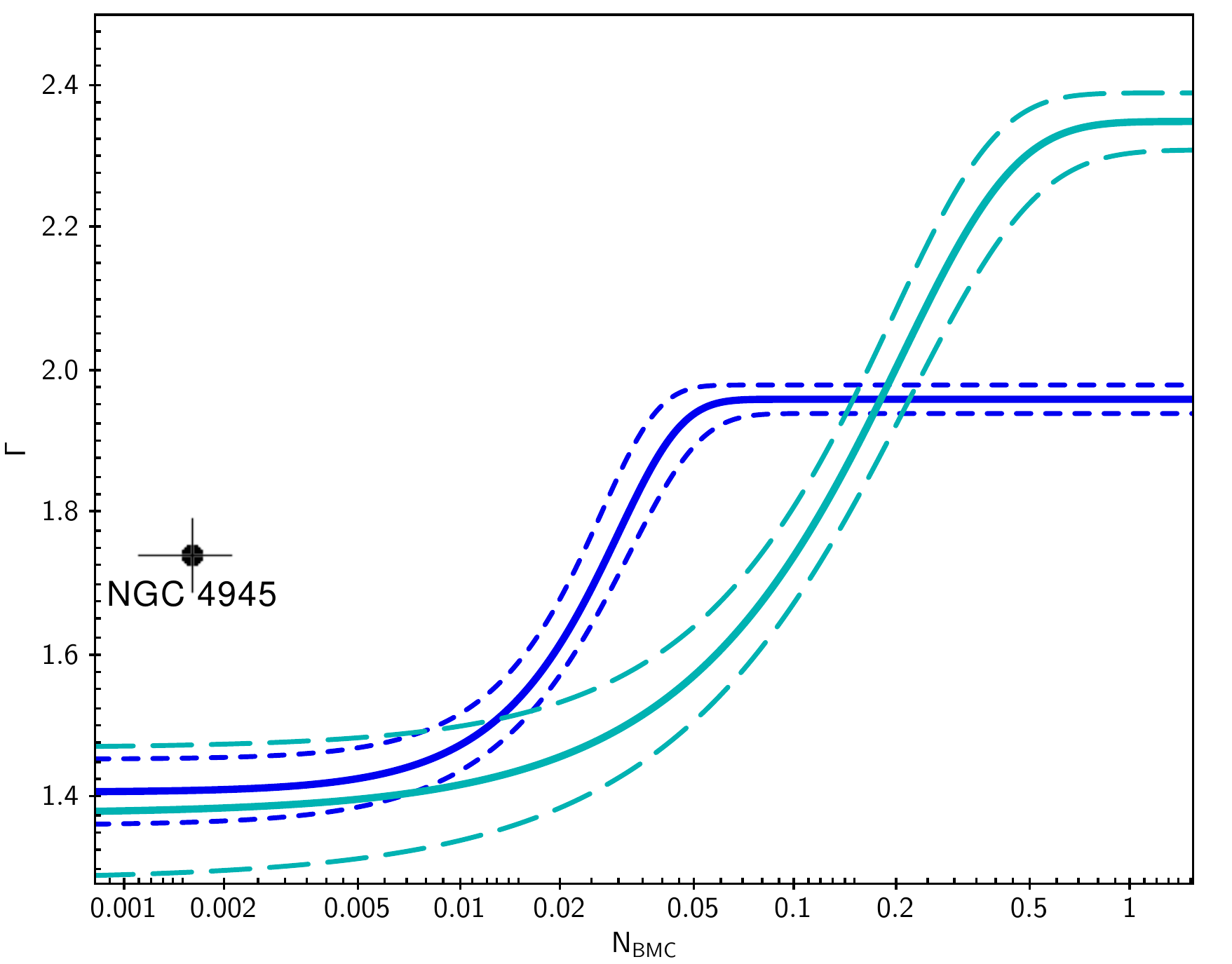}
\caption{\nbmc--$\Gamma$ plot, showing the data point corresponding to NGC 4945 and two reference patterns; the darker trend refers to GROD05, the spectral evolution of GRO 1655-40 during the decay of an outburst that occurred in 2005, and the lighter color trend indicates GROR05, the spectral evolution shown by the same source during the outburst rise. The dashed lines indicate the uncertainties in the reference spectral trends, whereas the error bars represent the uncertainties of the AGN spectral parameters.}
\label{figure:fig3}
\end{figure}

Indeed, this is the fundamental assumption of the X-ray scaling method: $\Gamma$ is indicative of the accretion state of the source, and BH systems in the same accretion state are characterized by the same accretion rate (in Eddington units) and the same radiative efficiency $\eta$. As a consequence, when we compare the accretion luminosity ($L \propto \eta M_{\mathrm{BH}} \dot m$) in BH systems that are in the same accretion state (i.e., with the same $\Gamma$), we are directly comparing their \mbh. This explains why the comparison of the values of the normalization of the \texttt{BMC} model, \nbmc\ (which is defined as the accretion luminosity in units of $10^{39}$ erg s$^{-1}$ divided by the distance squared in units of 10 kpc), computed at the same value of $\Gamma$ between the AGN and a known stellar BH reference source, yields the \mbh. This is illustrated in Fig.~\ref{figure:fig3} and mathematically described by 
\[
M_{\mathrm{BH,AGN}}=M_{\mathrm{BH,ref}}  \times \left(\frac{N_{\mathrm{BMC,AGN}}}{N_{\mathrm{BMC,ref}}}\right)  \times \left(\frac{d_{\mathrm{AGN}}^2}{d_{\mathrm{ref}}^2}\right)
\]
where $N_{\mathrm{BMC,ref}}$ and $d_{\mathrm{ref}}$ are the \texttt{BMC} model normalization and distance of the stellar mass BH system used as a reference.

%%%%%%%%%%%%%%%%%%%%%%%%%%%%%%%%%%%%%%%%%%%%%%%%%%%%
%%%%%%%% FIG.3 Gamma-N_BMC plot  %%%%%%%%%%%%%%%%%%%%%%%%%%%%%%
%%%%%%%%%%%%%%%%%%%%%%%%%%%%%%%%%%%%%%%%%%%%%%%%%%%%%
%%%%%%%%%%%%%%%%%%%%%%%%%%%%%%%%%%%%%%%%%%%%%%%%%%%%%

Fig.~\ref{figure:fig3} illustrates the X-ray scaling method and its inherent uncertainties that are related to the statistical errors on the spectral parameters $\Gamma$ and $N_{\mathrm{BMC}}$ and on the uncertainty of the reference source spectral trend (shown by the dashed lines), as well as on the specific reference source trend utilized.  Although similar in shape, the reference spectral trends show some differences (e.g., in their plateau levels and slopes), leading to slightly different \mbh\ values. From Fig.~\ref{figure:fig3}, it is clear that these differences exist also between the rise and decay phases of the same reference source.

It is important to note that at very low accretion rates both stellar mass and supermassive BHs show an anti-correlation between $\Gamma$ and  $\lambda_ {\mathrm{Edd}}$ (e.g., \citealt{const09,gu09,gulte12}). Since the X-ray scaling method is based on the positive correlation between these two quantities, it cannot be applied to determine the \mbh\ of
objects in the very low-accretion regime. This was explicitly demonstrated by the work of  \citet{jang14}, who analyzed a sample of low-luminosity low-accreting AGN.

In the following, we systematically estimate the \mbh\ using all the reference sources available (depending on the  AGN's $\Gamma$, not all reference sources can be used since their photon index ranges vary from reference source to reference source) and then compute the \mbh\ average value and its uncertainty $\sigma/\sqrt{n}$ (where $\sigma$ is the standard deviation and $n$ is the number of reference trends utilized).  As already explained above, for AGN with steep spectra, the most reliable estimate of \mbh\ is obtained using the 3*XTER98 reference trend; therefore, we also include this value in Table~\ref{table:tab3}.  All \mbh\ values listed in this table were computed assuming \nplbmc\ = 30; however, for completeness, we also report the \mbh\ obtained assuming \nplbmc\ = 24 and 38 for flat- and steep-spectrum sources, respectively.  We note that such changes in \nplbmc\ lead to \mbh\ values that are consistent with the  values obtained  with the original assumption \nplbmc\ = 30, within the respective 
\mbh\ uncertainties  that are of the order of 10\%--40\%.

\begin{figure*}
	\includegraphics[width=\mywidth]{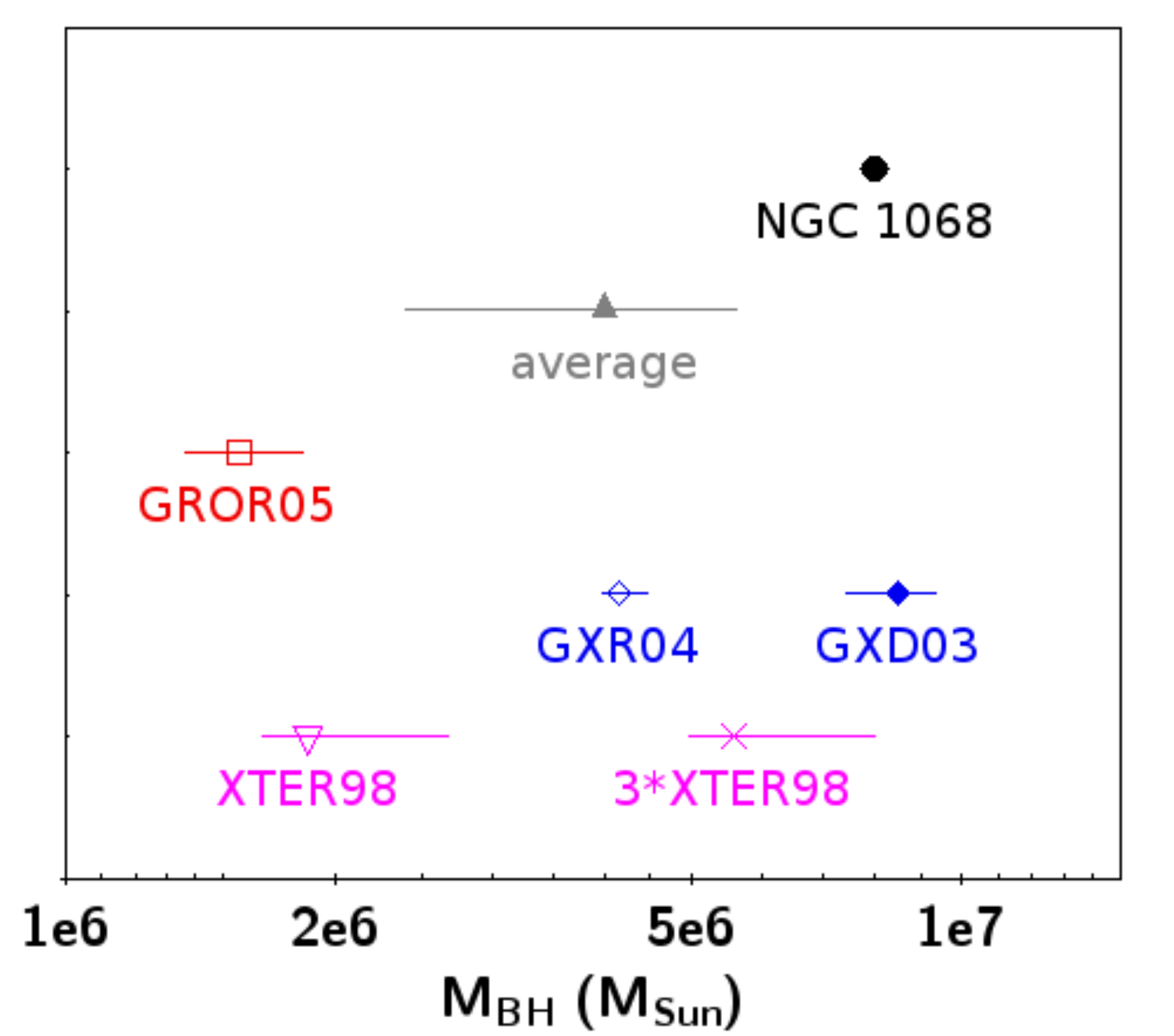}
	\includegraphics[width=\mywidth]{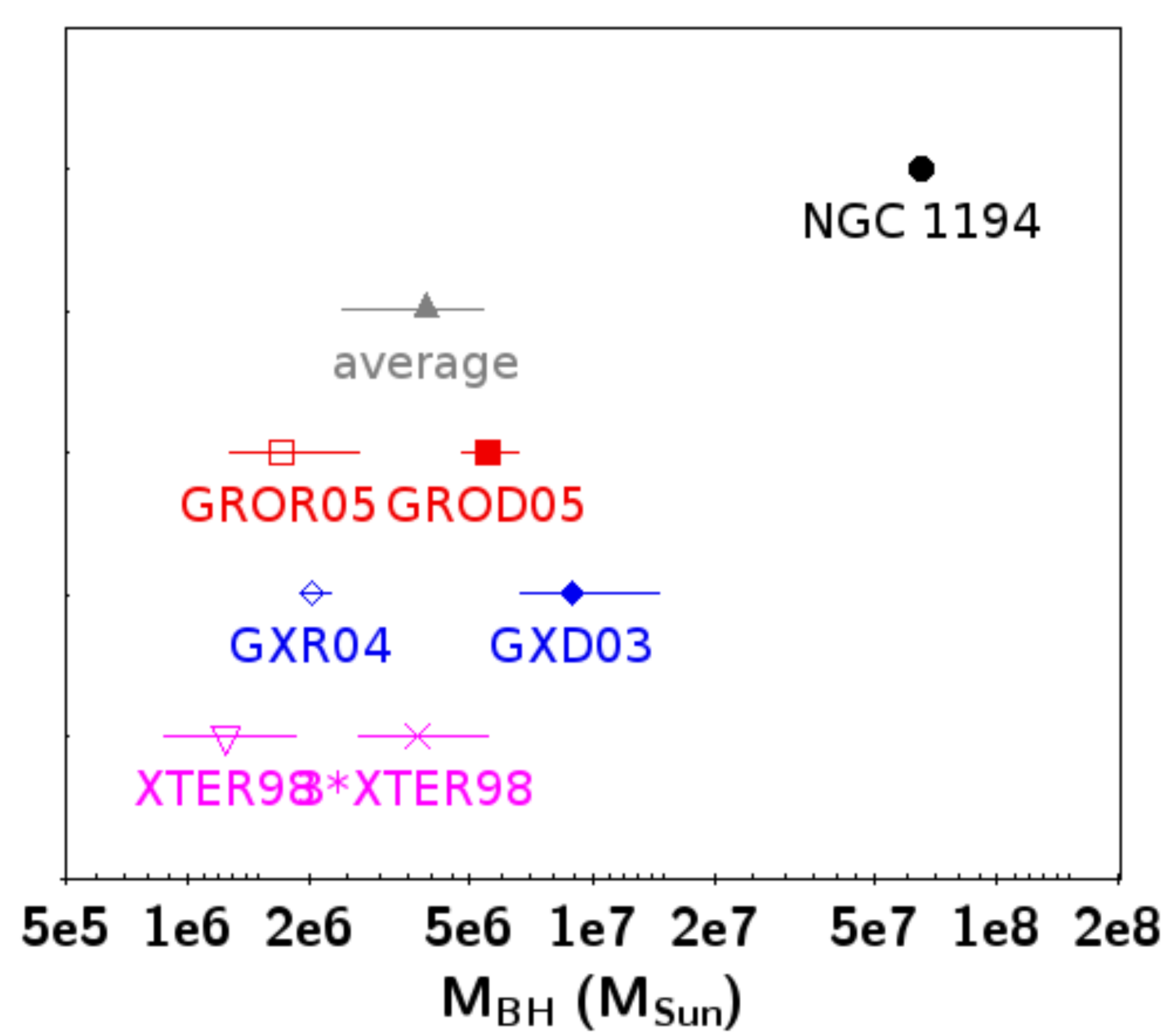}
	\includegraphics[width=\mywidth]{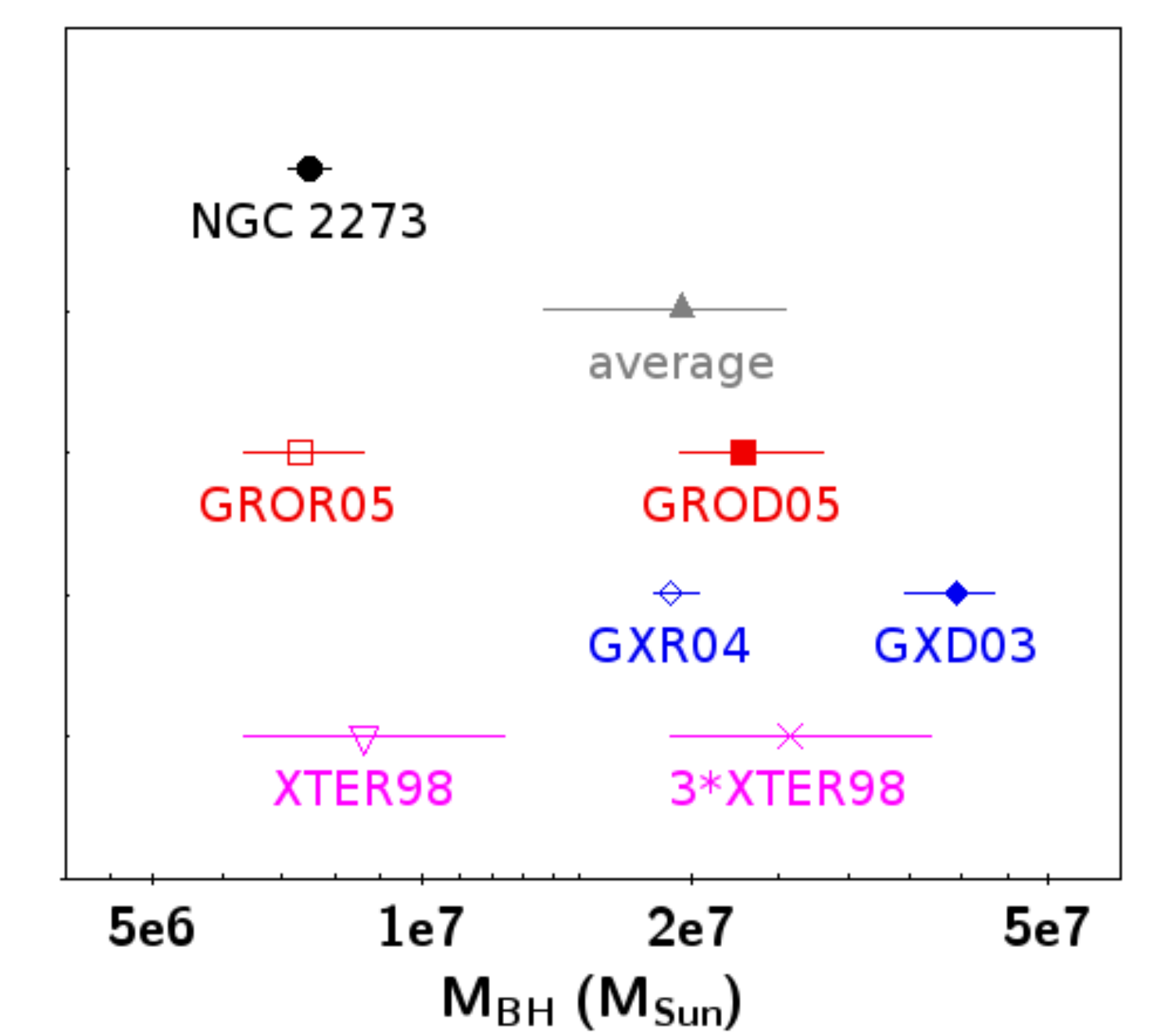}\\
	\includegraphics[width=\mywidth]{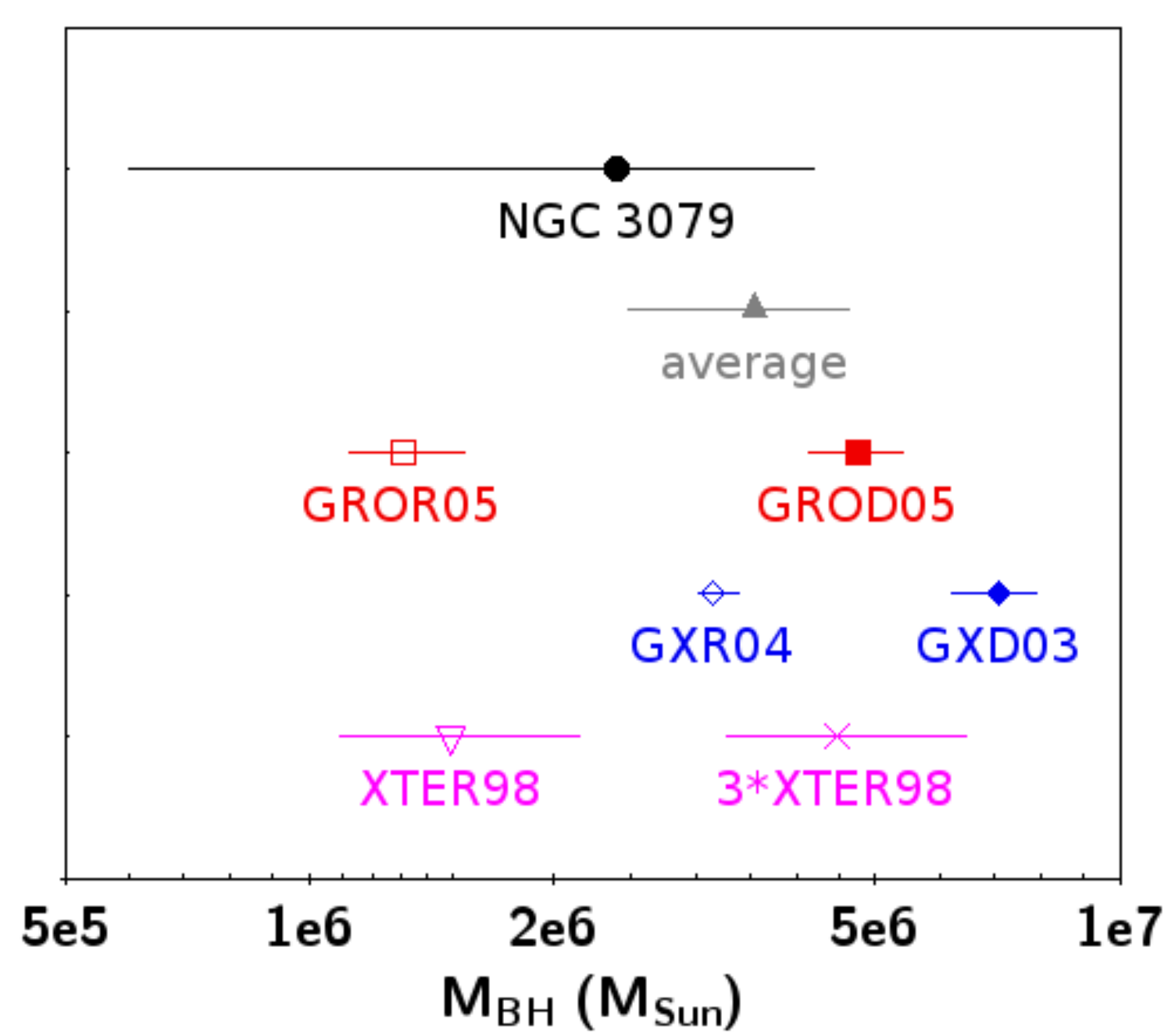}
	\includegraphics[width=\mywidth]{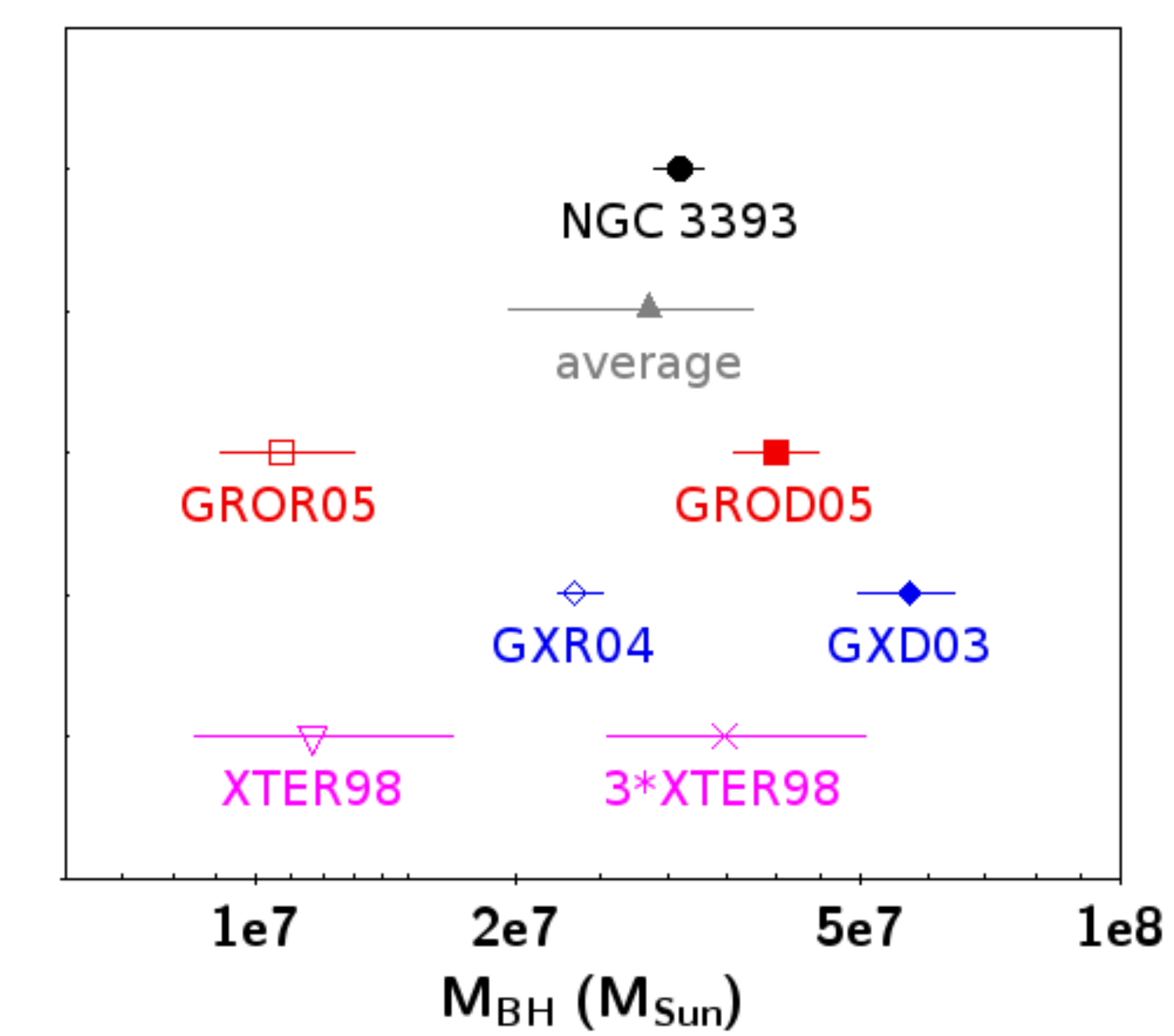}
	\includegraphics[width=\mywidth]{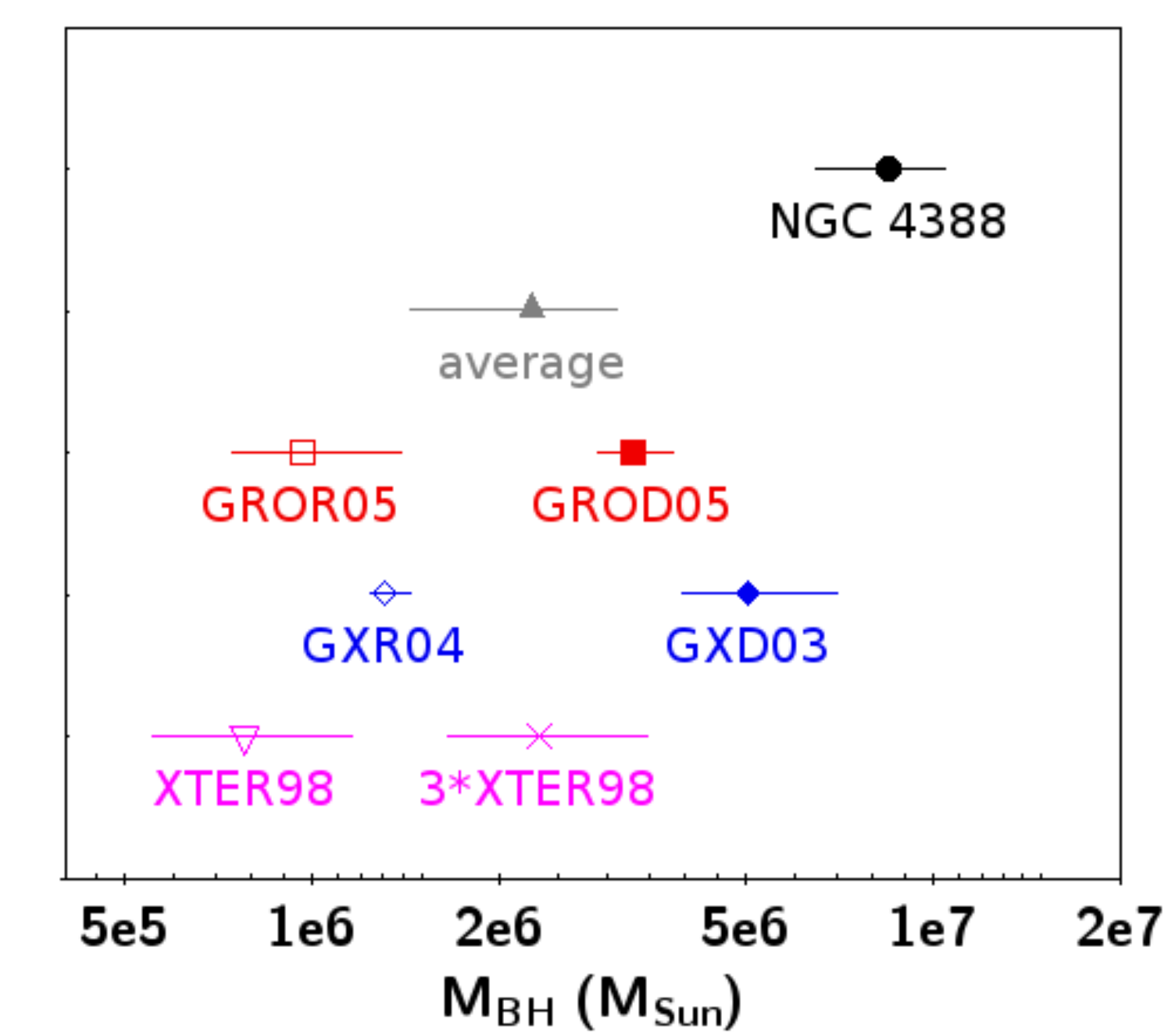}\\
	\includegraphics[width=\mywidth]{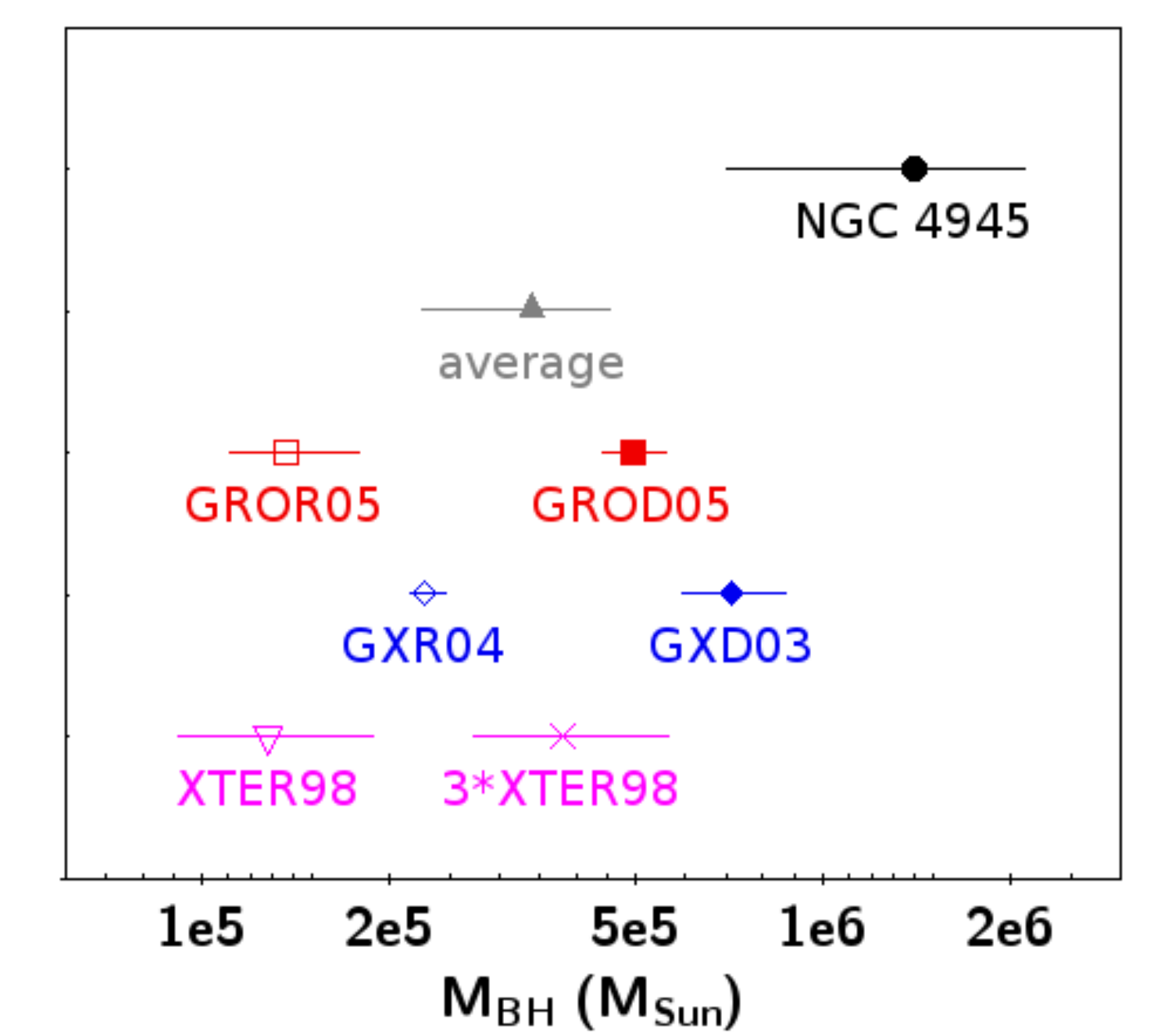}
	\includegraphics[width=\mywidth]{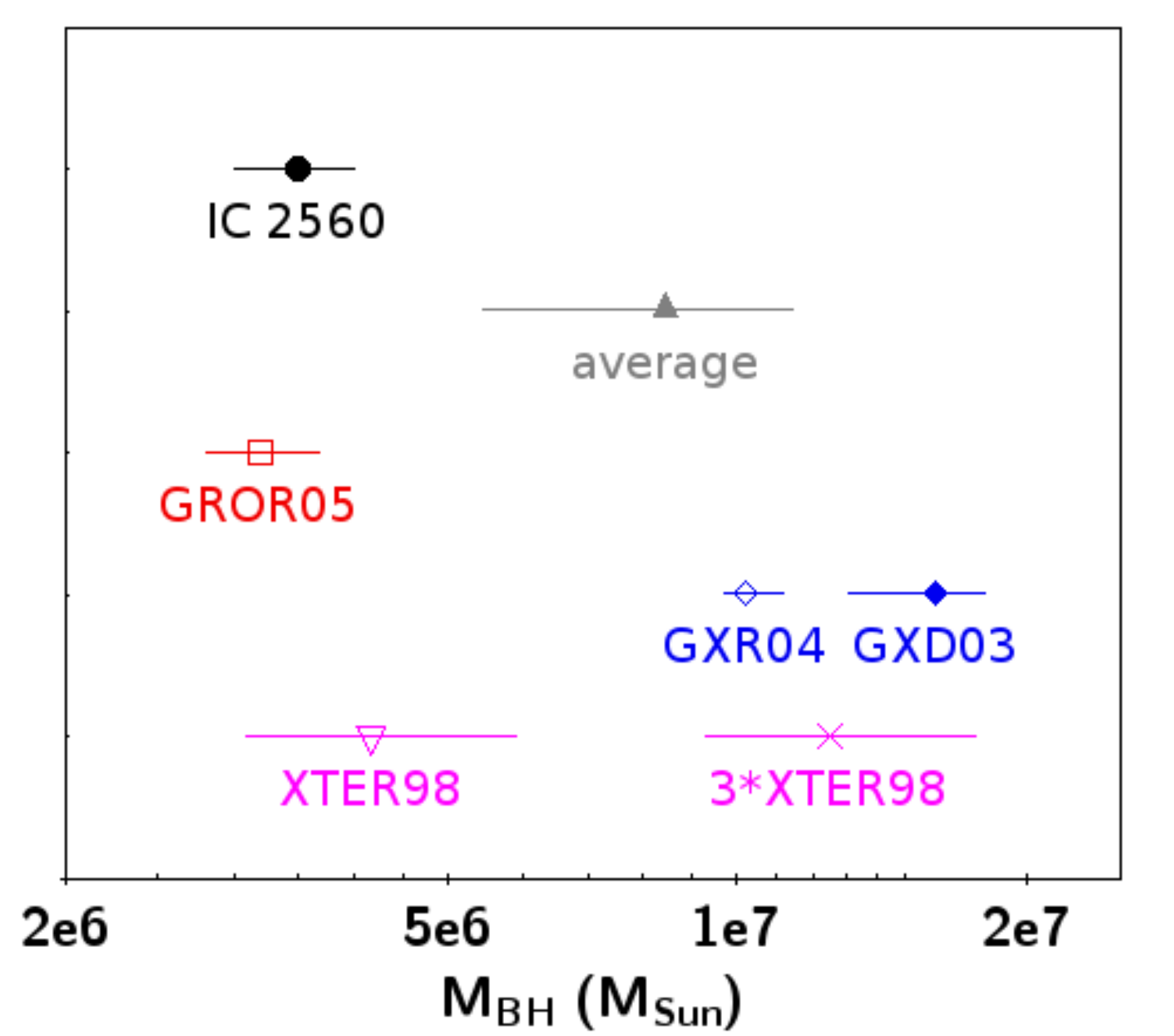}
	\includegraphics[width=\mywidth]{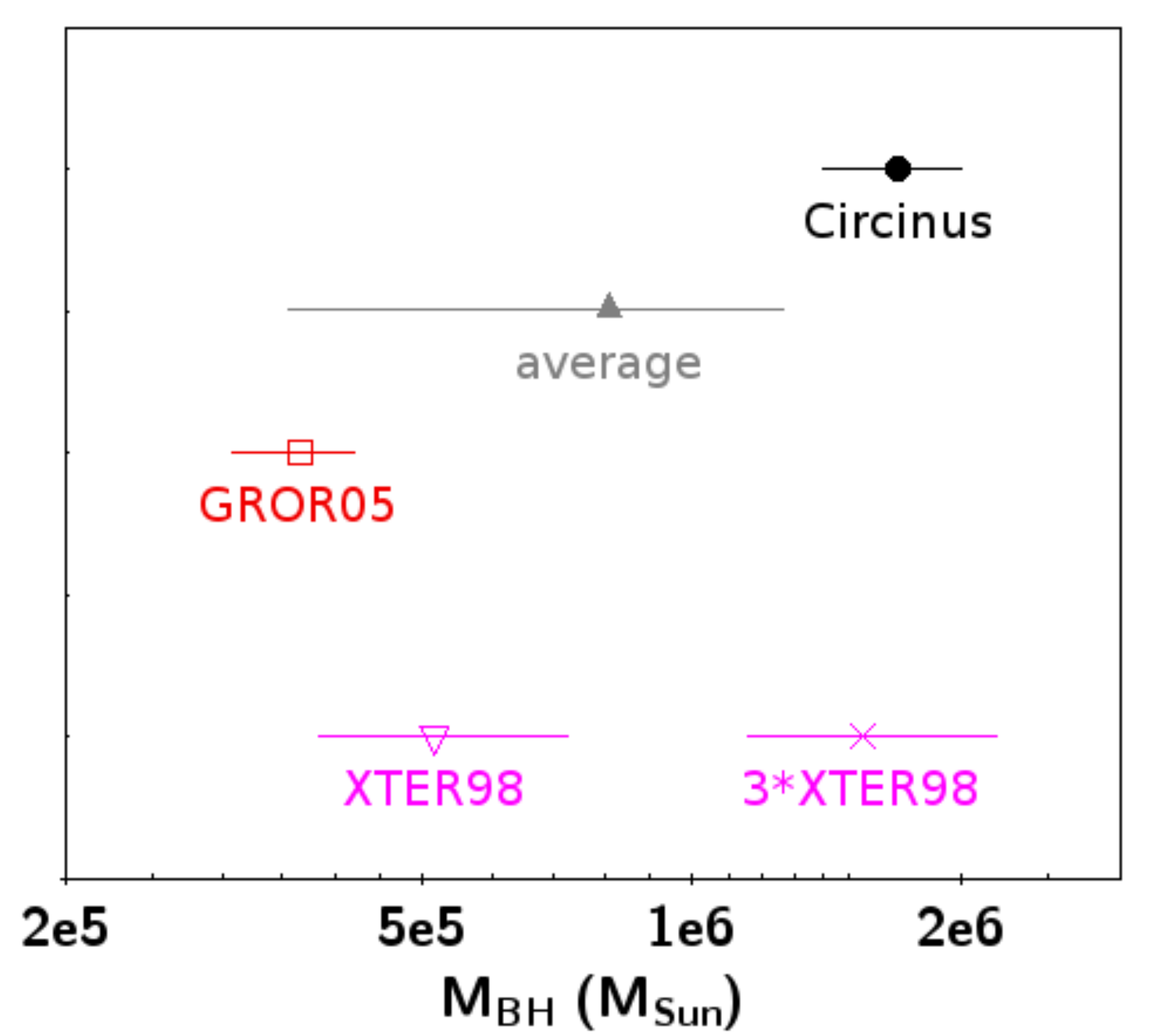}\\
\caption{$M_{\rm BH}$ values obtained with the X-ray scaling method using different reference sources, compared with $M_{\rm BH}$ obtained from megamaser measurements, which are represented by the black symbols at the top of each panel.}
\label{figure:fig4}
\end{figure*}

The \mbh\ values obtained with the different reference sources and their average are illustrated in Fig.~\ref{figure:fig4}. As already found in \citet{glioz11} for the reverberation mapping AGN sample, the reference trends of decaying outbursts yield systematically larger \mbh\ values compared to those obtained from the rising trends. For each obscured AGN, several \mbh\ values obtained from different reference sources and their average appear to be broadly consistent with the value obtained from megamaser measurements (a quantitative comparison is carried out in the next subsection). The only noticeable exception is NGC 1194, for which the X-ray scaling method yields values significantly lower than the maser one. This discrepancy however is not surprising, since this source has a fairly low accretion rate and in that regime the X-ray scaling method cannot be safely applied.
%, as shown by \citet{jang14}, based on the systematic analysis of a sample of low-accreting AGN.

%%%%%%%%%%%%%%%%%%%%%%%%%%%%%%%%%%%%%%%%%%%%%%%%%%
%%%%%%%% FIG4: MBH  %%%%%%%%%%%%%%%%%%%%%%%%%%%%%%%%%%%%
%%%%%%%%%%%%%%%%%%%%%%%%%%%%%%%%%%%%%%%%%%%%%%%%%%%%%
%%%%%%%%%%%%%%%%%%%%%%%%%%%%%%%%%%%%%%%%%%%%%%%%%%%%%

%%%%%%%%%%%%%%%%%%%%%%%%%%%%%%%%%%%%%%%%%%%%%%%%%%%%%
%%%%%%%% TAB3: MBH %%%%%%%%%%%%%%%%%%%%%%%%%%%%%%%%%%%%%%%
\begin{table*}
	\caption{Black hole masses with the X-ray scaling method}		
	\begin{center}
		\begin{tabular}{lrrrrrrr} 
			\toprule
			\toprule       
			\mcol{Source} & \mcol{$M_{{\textrm{BH}}_{\textrm{GROD05}}}$} & \mcol{$M_{{\textrm{BH}}_{\textrm{GROR05}}}$} & \mcol{$M_{{\textrm{BH}}_{\textrm{GXD03}}}$} & 
			\mcol{$M_{{\textrm{BH}}_{\textrm{GXR04}}}$} & 
			\mcol{$M_{{\textrm{BH}}_{\textrm{XTER98}}}$} &
			\mcol{$M_{{\textrm{BH}}_{\textrm{aver}}}$} &
			\mcol{$M_{{\textrm{BH}}_{\textrm{3*XTER98}}}$} \\
			\noalign{\smallskip}
			\mcol{name} & \mcol{($10^6$\ \msun)} & \mcol{($10^6$\ \msun)} & \mcol{($10^6$\ \msun)} & \mcol{($10^6$\ \msun)} & \mcol{($10^6$\ \msun)} & \mcol{($10^6$\ \msun)} &
			\mcol{($10^6$\ \msun)} \\
			\mcol{(1)} & \mcol{(2)} & \mcol{(3)} & \mcol{(4)} & \mcol{(5)} & \mcol{(6)} & \mcol{(7)} & \mcol{(8)} \\
			\midrule
			NGC 1068 & \ldots & $1.6_{-0.2}^{+0.3}$ & $8.5_{-1.0}^{+0.8}$ & $4.1_{-0.2}^{+0.3}$ & $1.9_{-0.5}^{+0.8}$ & $4.0_{-1.6}^{+1.6}$ & $5.6_{-1.5}^{+2.4}$ \\		
			\noalign{\smallskip}
			\midrule	
			\noalign{\smallskip}		
			NGC 1194 & $5.6_{-0.7}^{+1.0}$ & $1.7_{-0.4}^{+1.0}$ & $9.0_{-2.3}^{+5.7}$ & $2.0_{-0.1}^{+0.2}$ & $1.2_{-0.4}^{+0.6}$ & $3.9_{-1.5}^{+1.5}$ & $3.7_{-1.1}^{+1.9}$ \\
			\noalign{\smallskip}	
			\nplbmc\ =24 & $6.9_{-1.0}^{+1.5}$ & $2.2_{-0.6}^{+1.6}$ & $12_{-3.5}^{+13}$ & $2.4_{-0.1}^{+0.4}$ & $1.5_{-0.4}^{+0.7}$ & $5.0_{-2.0}^{+2.0}$ & $4.4_{-1.3}^{+2.2}$ \\
			\noalign{\smallskip}
			\midrule	
			\noalign{\smallskip}	
			NGC 2273 & $22.8_{-3.4}^{+5.2}$ & $7.3_{-1.0}^{+1.3}$ & $39.6_{-4.8}^{+4.0}$ & $19.0_{-0.8}^{+1.4}$ & $8.6_{-2.3}^{+3.8}$ & $19.5_{-5.8}^{+5.8}$ & $25.8_{\ -6.9}^{+11.3}$ \\
			\noalign{\smallskip}
			\midrule	
			\noalign{\smallskip}	
			NGC 3079 & $4.8_{-0.6}^{+0.6}$ & $1.3_{-0.2}^{+0.3}$ & $7.1_{-0.9}^{+0.8}$ & $3.2_{-0.1}^{+0.2}$ & $1.5_{-0.4}^{+0.7}$ & $3.6_{-1.1}^{+1.1}$ & $4.5_{-1.2}^{+2.0}$ \\
			\noalign{\smallskip}
			\midrule	
			\noalign{\smallskip}	
			NGC 3393 & $40.1_{-4.2}^{+4.8}$ & $10.7_{-1.6}^{+2.3}$ & $57.1_{-7.5}^{+7.3}$ & $23.3_{-1.0}^{+1.8}$ & $11.6_{-3.1}^{+5.2}$ & $28.5_{-8.9}^{+8.9}$ & $34.7_{\ -9.4}^{+15.7}$ \\
			\noalign{\smallskip}
			\midrule	
			\noalign{\smallskip}	
			NGC 4388 & $3.3_{-0.4}^{+0.5}$ & $1.0_{-0.2}^{+0.4}$ & $5.0_{-1.1}^{+2.0}$ & $1.3_{-0.1}^{+0.1}$ & $0.8_{-0.2}^{+0.4}$ & $2.3_{-0.8}^{+0.8}$ & $2.3_{-0.7}^{+1.1}$ \\
			\noalign{\smallskip}
	              \nplbmc\ =24 & $4.1_{-0.5}^{+0.7}$ & $1.2_{-0.3}^{+0.5}$ & $6.3_{-1.4}^{+2.5}$ & $1.6_{-0.1}^{+0.2}$ & $1.0_{-0.3}^{+0.5}$ & $2.9_{-1.0}^{+1.0}$ & $2.9_{-0.8}^{+1.4}$ \\
                         \noalign{\smallskip}
			\midrule	
			\noalign{\smallskip}	
	       NGC 4945 & $0.5_{-0.05}^{+0.06}$ & $0.14_{-0.03}^{+0.04}$ & $0.7_{-0.1}^{+0.2}$ & $0.2_{-0.01}^{+0.01}$ & $0.13_{-0.04}^{+0.06}$ & $0.3_{-0.1}^{+0.1}$ & $0.4_{-0.1}^{+0.2}$ \\
			\noalign{\smallskip}
      \nplbmc\ =24 & $0.6_{-0.1}^{+0.1}$ & $0.17_{-0.03}^{+0.05}$ & $0.9_{-0.1}^{+0.2}$ & $0.3_{-0.01}^{+0.02}$ & $0.16_{-0.05}^{+0.08}$ & $0.4_{-0.1}^{+0.1}$ & $0.5_{-0.1}^{+0.2}$ \\			\noalign{\smallskip}
			\midrule	
			\noalign{\smallskip}	
			IC 2560 & \ldots & $3.2_{-0.4}^{+0.5}$ & $16.1_{-3.0}^{+2.0}$ & $10.2_{-0.5}^{+1.0}$ & $4.2_{-1.1}^{+1.7}$ & $8.4_{-3.0}^{+3.0}$ & $12.5_{-3.2}^{+5.2}$ \\
			\noalign{\smallskip}
			\nplbmc\ =38 & \ldots & $2.8_{-0.3}^{+0.4}$ & $14.6_{-2.0}^{+1.6}$ & $8.8_{-0.6}^{+0.7}$ & $3.6_{-0.9}^{+1.5}$ & $7.4_{-2.7}^{+2.7}$ & $10.8_{-2.8}^{+4.6}$ \\
			\noalign{\smallskip}
			\midrule	
			\noalign{\smallskip}	
			Circinus & \ldots & $0.4_{-0.1}^{+0.1}$ & \ldots & \ldots & $0.5_{-0.1}^{+0.2}$ & $0.4_{-0.1}^{+0.1}$ & $1.6_{-0.4}^{+0.6}$ \\
			\noalign{\smallskip}
			 \nplbmc\ =38 & \ldots & $0.3_{-0.1}^{+0.1}$ & \ldots & \ldots & $0.4_{-0.1}^{+0.2}$ & $0.6_{-0.4}^{+0.4}$ & $1.2_{-0.3}^{+0.5}$ \\
			\bottomrule
		\end{tabular}
	\end{center}
	\begin{flushleft}
		Columns: 1 = AGN name. 2--8 = black hole masses determined with the X-ray scaling method. Subscripts denote GROD05 = reference source GRO J1655-40 in the decreasing phase; GROR05 = reference source GRO J1655-40 in the rising phase; GXD03 = reference source GX 339-4 in the decreasing phase; GXR03 = reference source GX 339-4 in the rising phase; XTER98 = reference source XTE J1550-564 in the rising phase; 3*XTER98 = reference source XTE J1550-564 in the rising phase with a multiplicative correction of a factor 3 applied. Note, the average value (in column 7) is obtained averaging all the \mbh\ obtained from all the reference sources but excluding 3*XTER98. Note: For each source
		the first line reports the \mbh\ values obtained using \nplbmc\ = 30 in the spectral fitting; the second line (present only for sources with relatively flat or steep spectra) explicitly states the different value of  \nplbmc\ used.		
	\end{flushleft}
	\label{table:tab3}
\end{table*}  
%%%%%%%%%%%%%%%%%%%%%%%%%%%%%%%%%%%%%%%%%%%%%%%%%%%%%
%%%%%%%%%%%%%%%%%%%%%%%%%%%%%%%%%%%%%%%%%%%%%%%%%%%%%

%%%%%%%%%%%%%%%%%%%%%%%%%%%%%%%%%%%%%%%%%%%%
%%%%%%% FIG5: difference of MBH divided by uncertainty %%%%%%%%%%%%%
\begin{figure*}
	\includegraphics[width=\mywidthtwo]{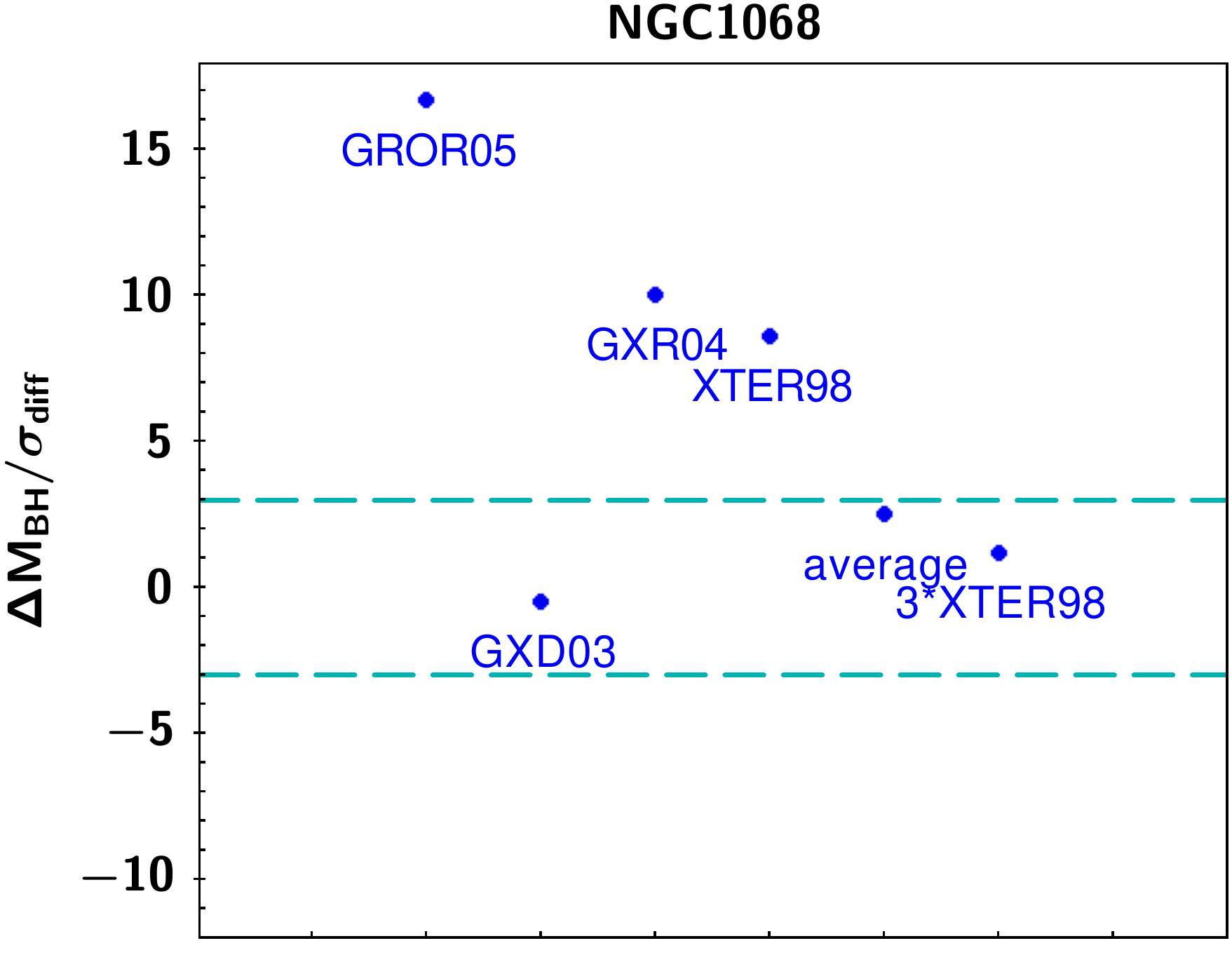}
	\hspace{1em}
	\includegraphics[width=\mywidthtwo]{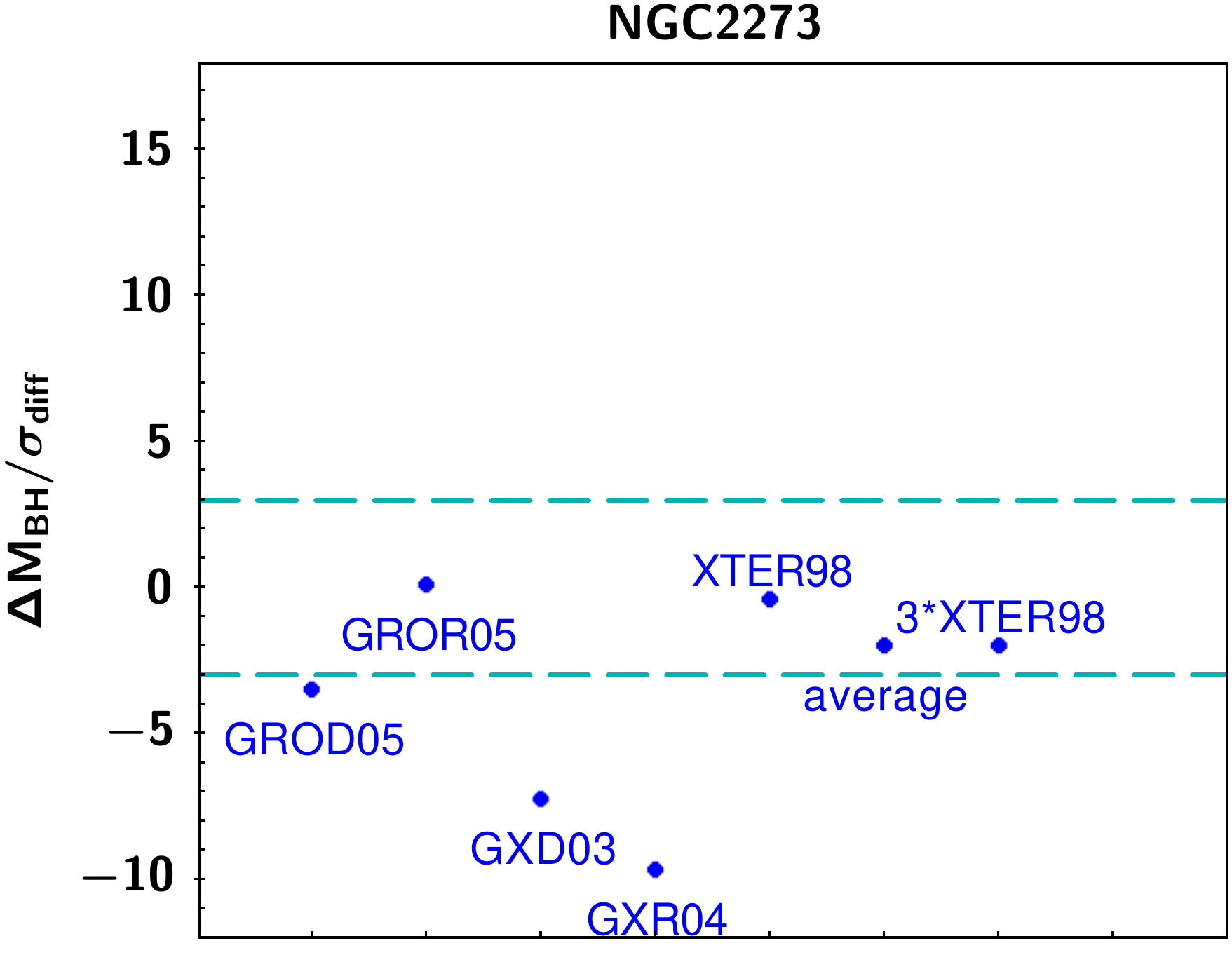}\\
	\includegraphics[width=\mywidthtwo]{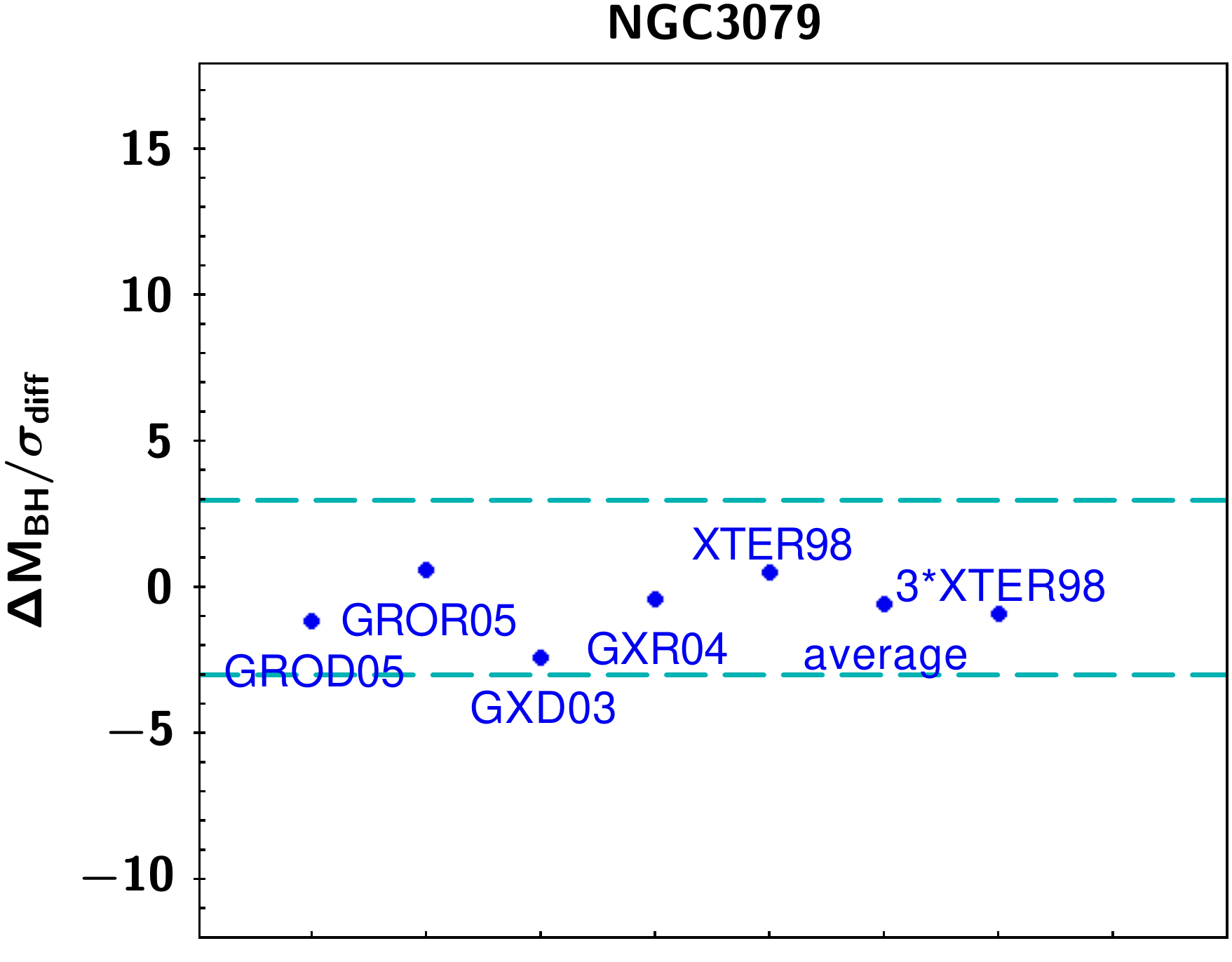} \hspace{1em}
	\includegraphics[width=\mywidthtwo]{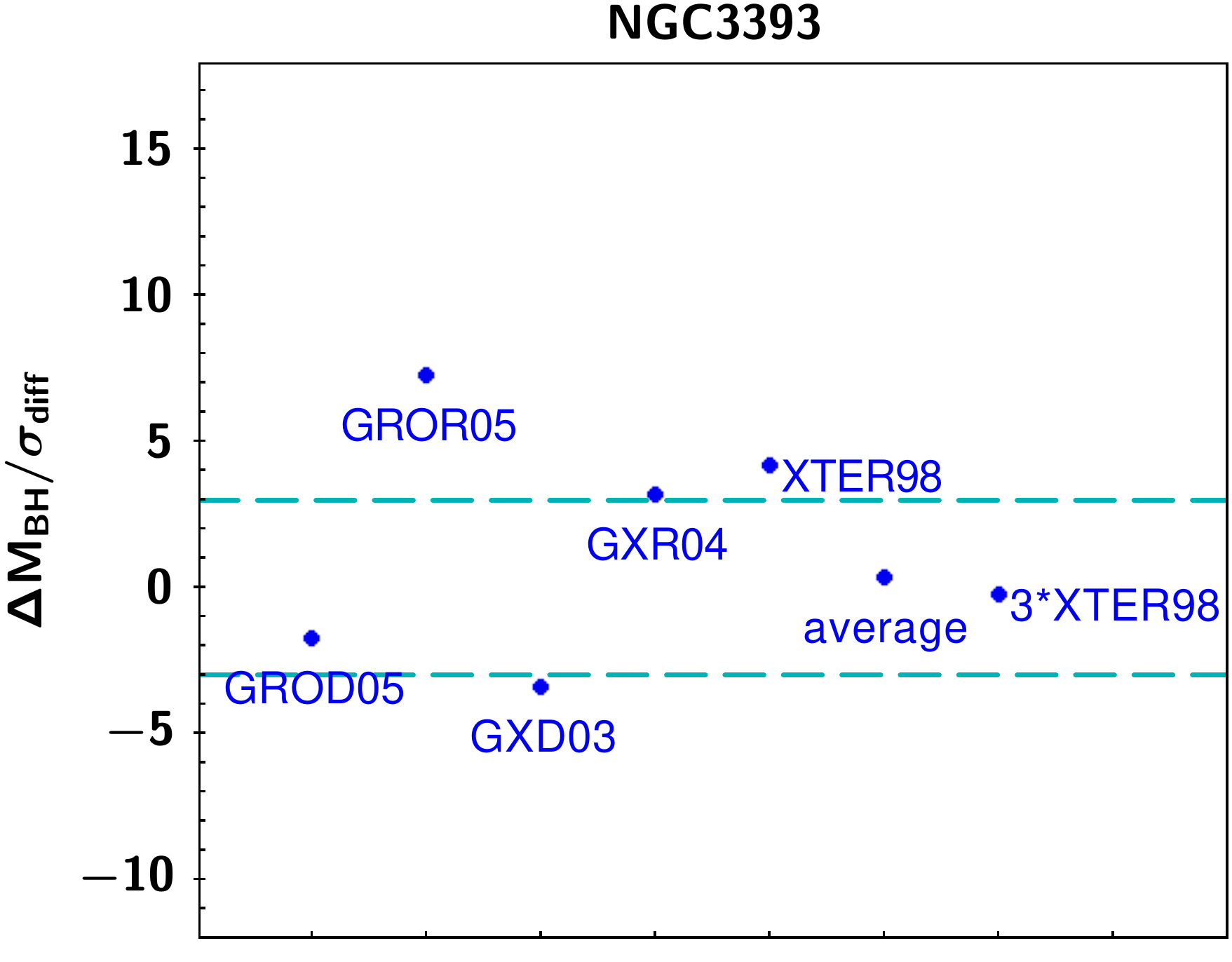}\\
	\includegraphics[width=\mywidthtwo]{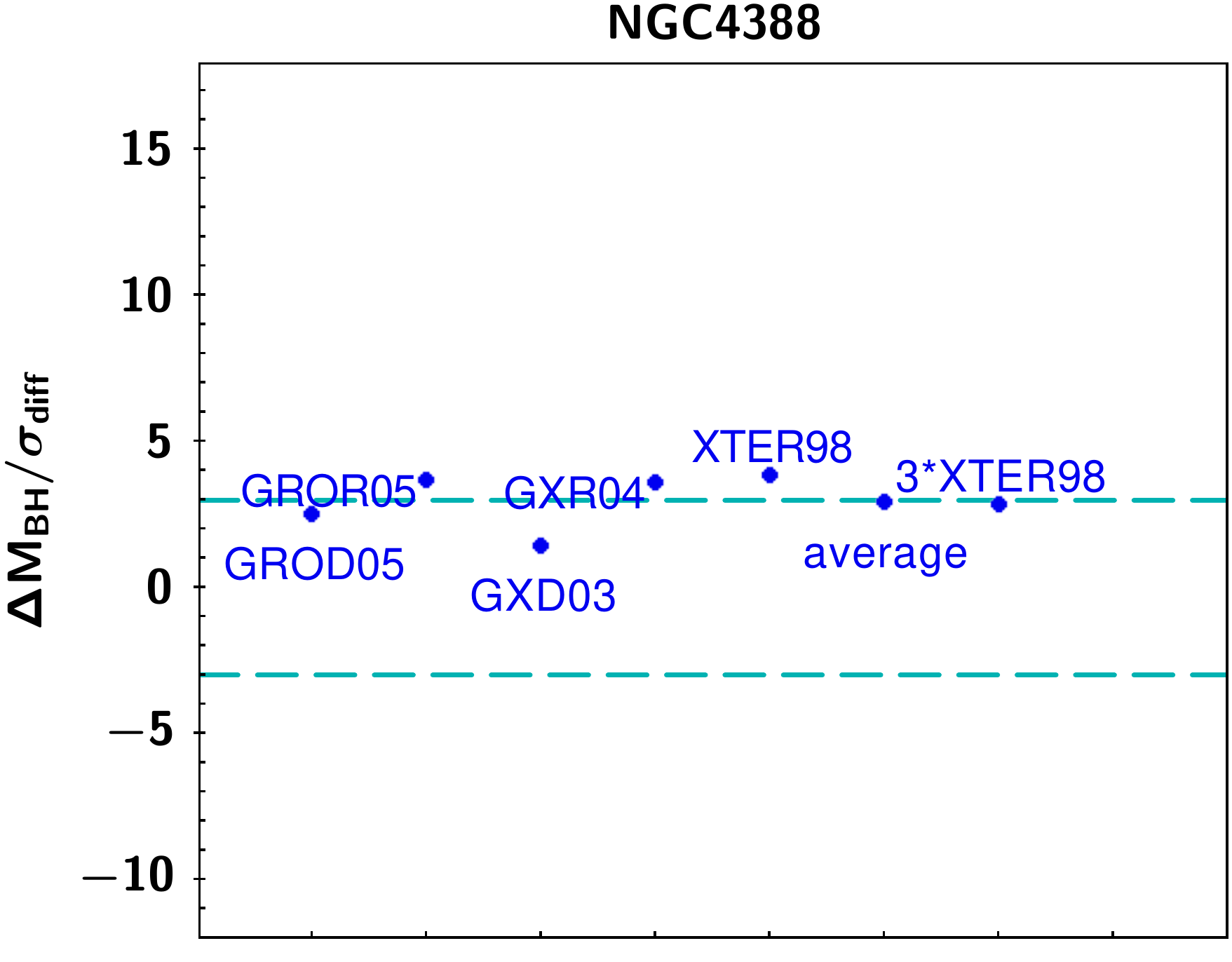} \hspace{1em}
	\includegraphics[width=\mywidthtwo]{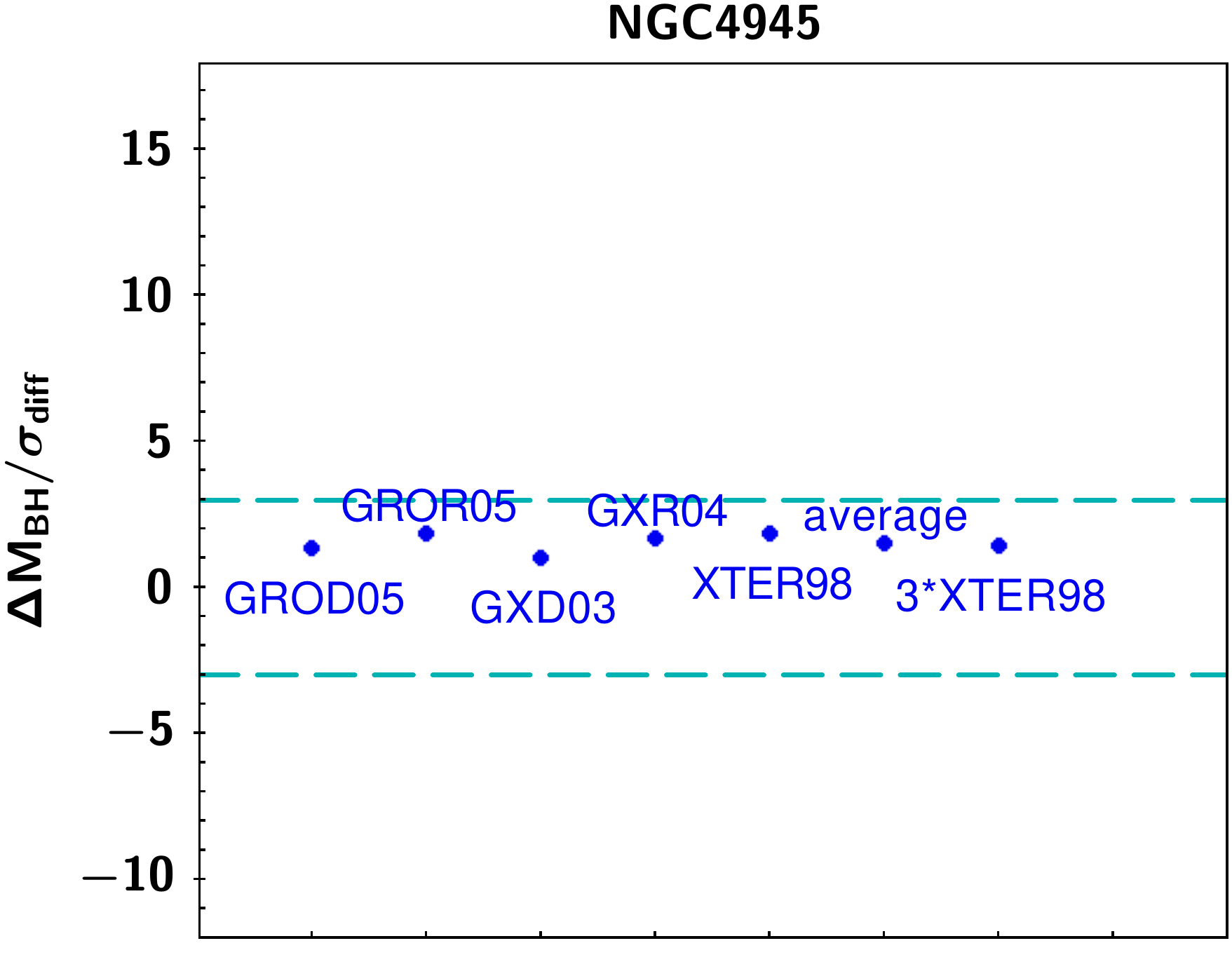}\\
	\includegraphics[width=\mywidthtwo]{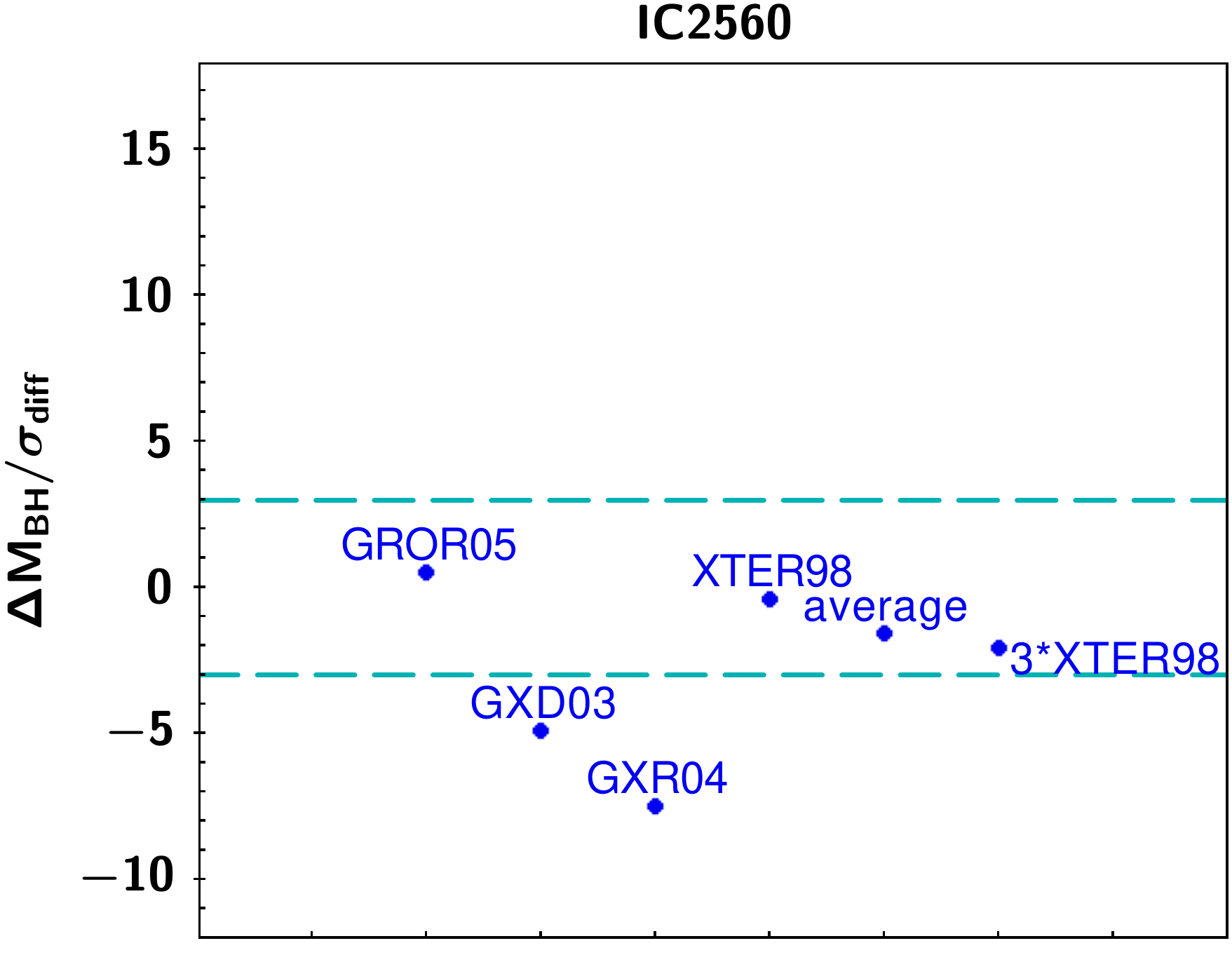} \hspace{1em}
	\includegraphics[width=\mywidthtwo]{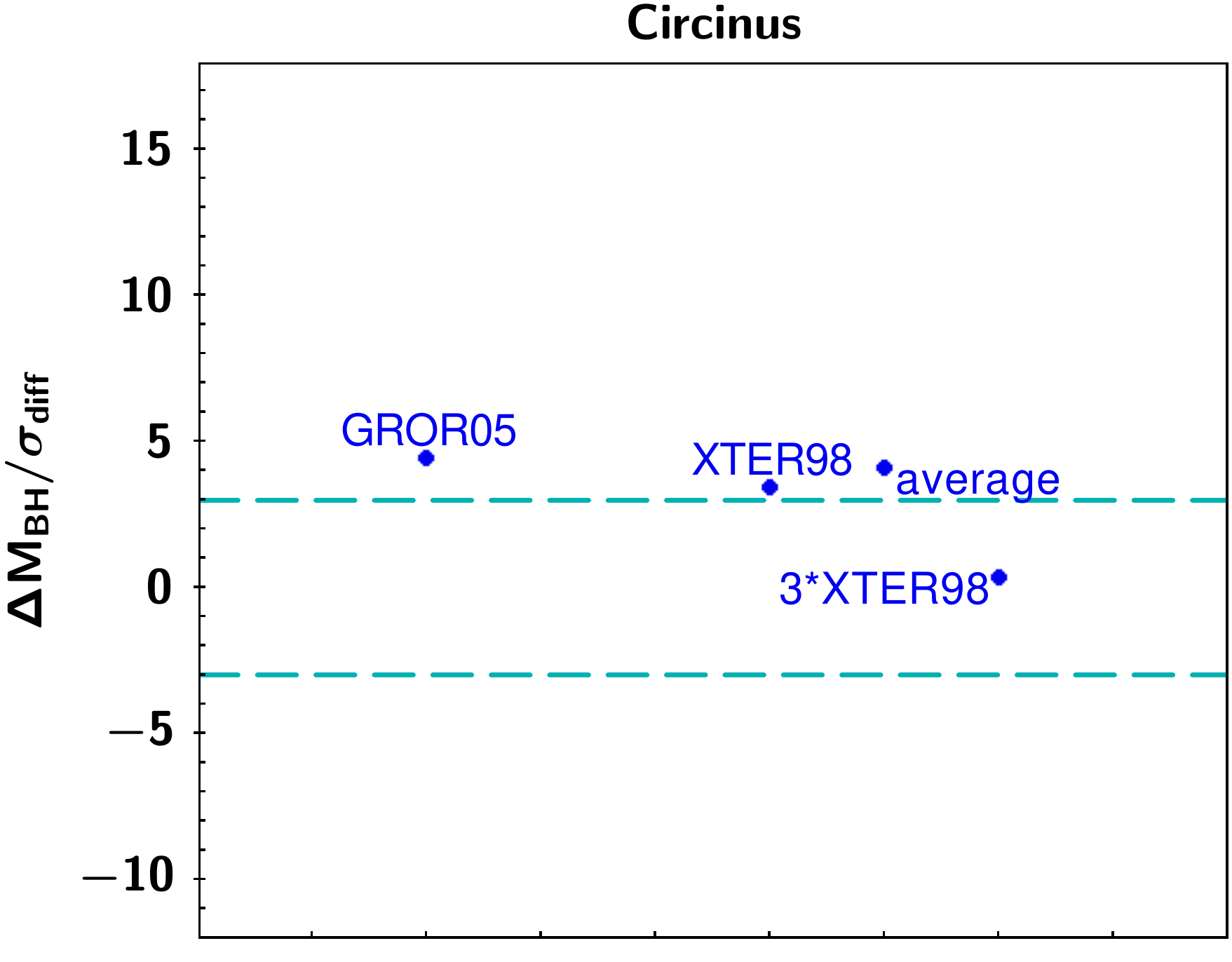}\\
\caption{Plots showing the difference between the BH mass determined from megamaser measurements and the values obtained with the X-ray scaling method for the different reference sources, divided by the uncertainty of the difference, $\Delta M_{\rm BH}/\sigma_{\rm diff}$. The horizontal dashed lines enclose the region where the difference between the BH masses is within 3$\sigma$.}
\label{figure:fig5}	
\end{figure*}
%%%%%%%%%%%%%%%%%%%%%%%%%%%%%%%%%%%%%%%%%%%%%%%%%%%%%%%%%%%%%%%%%%%%%%%%
%%%%%%%%%%%%%%%%%%%%%%%%%%%%%%%%%%%%%%%%%%%%%%%%%%%%%%%%%%%%%%%%%%%%%%%%

\subsection{Black hole mass comparison}
To compare the \mbh\ values obtained from the X-ray scaling method with the maser ones in a quantitative way, we computed, using all the available reference trends, the difference $\Delta M_{\mathrm{BH}} =  M_{\mathrm{BH,maser}}-M_{\mathrm{BH,scaling}}$ and its uncertainty $\sigma_{\mathrm{diff}}$, obtained by adding the respective errors in quadrature. As explained before, the error on the \mbh\ inferred from the scaling method includes the uncertainties on the spectral AGN parameters and on the reference trends in the \nbmc--$\Gamma$ diagram. Depending on the reference trend utilized, the percentage uncertainties range from 10\%--15\% for GROD05 and GXD03 to 30\%--40\% 
for XTER98, which is also the percentage uncertainty of the average \mbh.

The error on the \mbh\ obtained with megamaser measurements accounts for the uncertainties associated with the source position and with the fitting of the Keplerian rotation curve \citep{kuo11}.  For the uncertainties on the \mbh\ determined via megamaser measurements we used the errors quoted in the literature with the exception of NGC 4388, for which we multiplied the quoted uncertainty by a factor of 10; this yields a percentage error of $\sim$24\%, which better reflects the actual uncertainty on the \mbh\ in this source, where there is no systemic maser detected and the five maser spots detected are not sufficient to demonstrate that the rotation is Keplerian \citep{kuo11}. Note that both methods explicitly depend on the sources' distances and hence, in principle, their total uncertainties should account also for the distance uncertainties (indeed some of the sources of this maser sample are fairly close and thus their distances cannot be obtained from the redshift and Hubble's law). However, since our goal is to compare the two methods, we can avoid the uncertainty associated with the distance by assuming the exact same distance used in the maser papers.
%%%%%%%%%%%%%%%%%%%%%%%%%%%%%%%%%%%%%%%%%%%%%%%%%%%%%%
%%%%%%%% FIG.6 MBH_X vs. MBH_maser  %%%%%%%%%%%%%%%%%%%%%%%%%%%%%%
\begin{figure*}
	\includegraphics[width=\mywidthtwo]{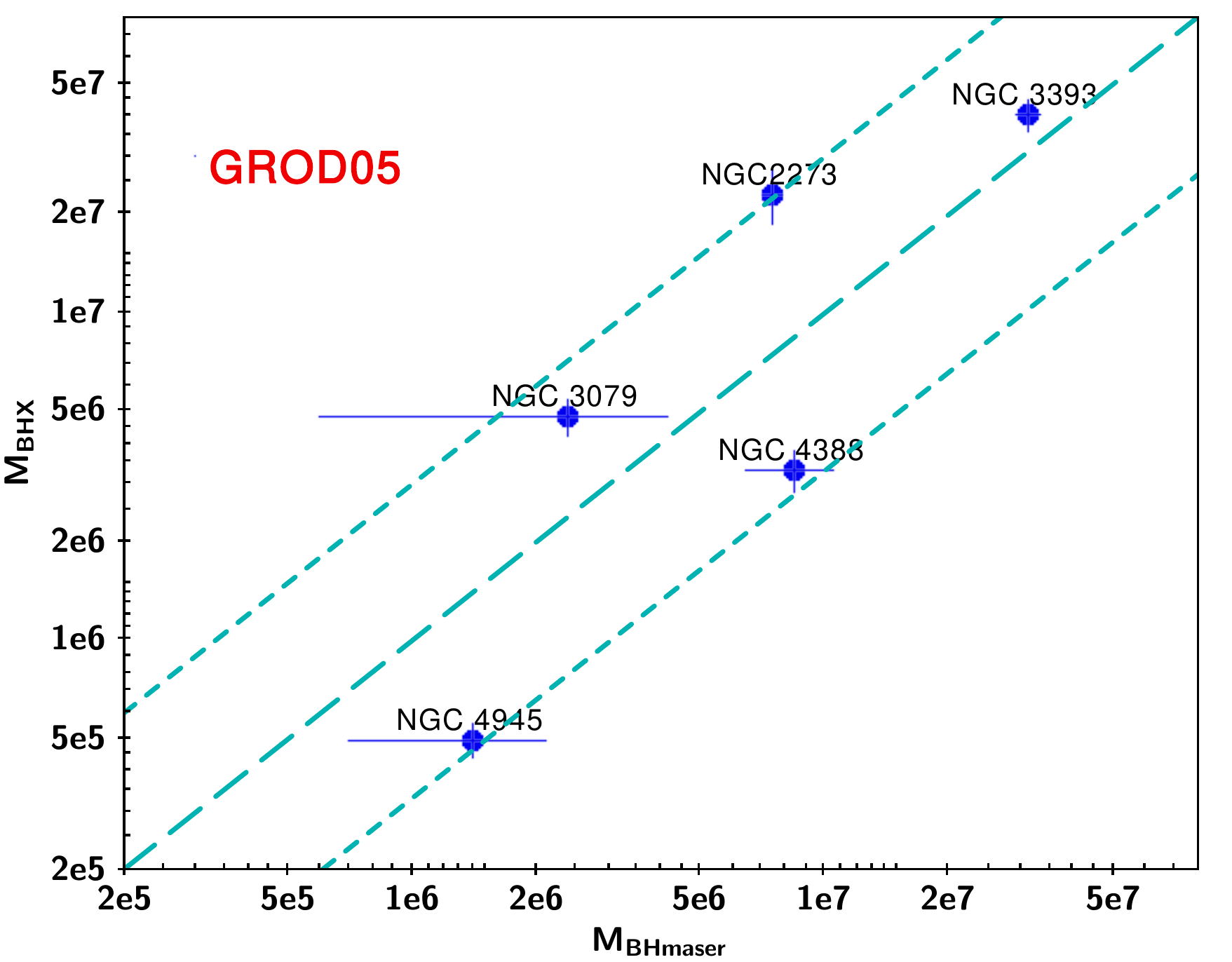} \hspace{1em}
	\includegraphics[width=\mywidthtwo]{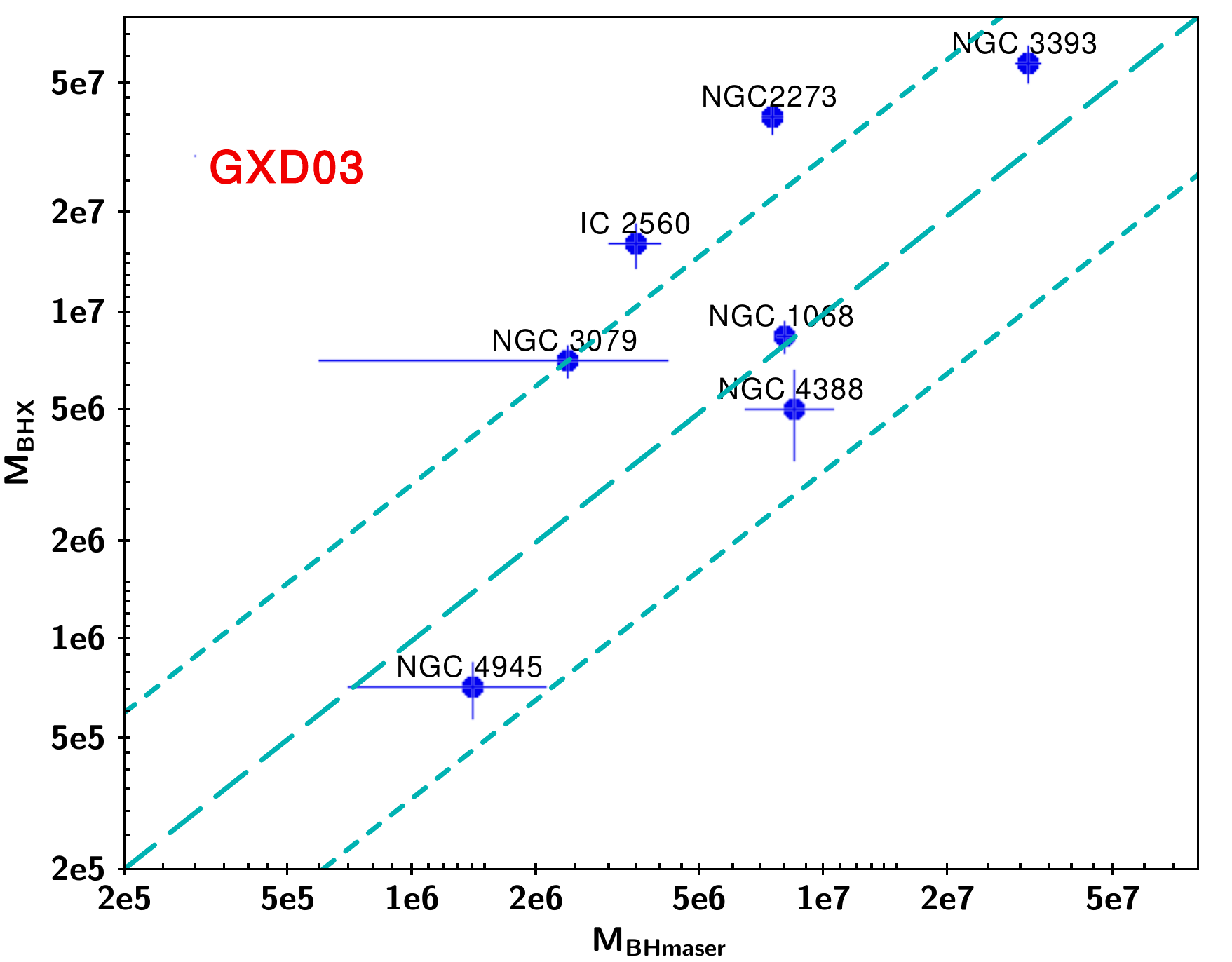}\\
	\includegraphics[width=\mywidthtwo]{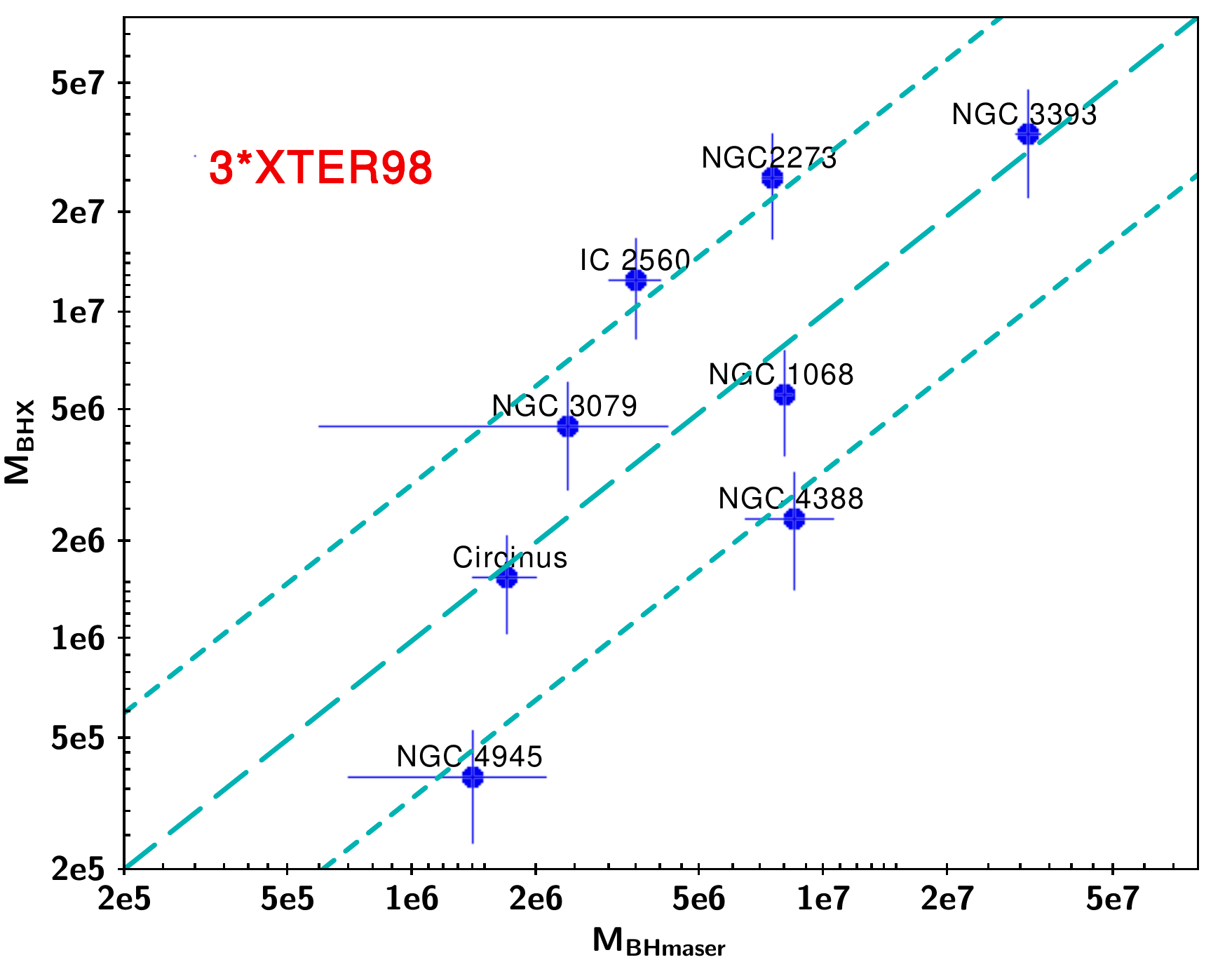} \hspace{1em}
	\includegraphics[width=\mywidthtwo]{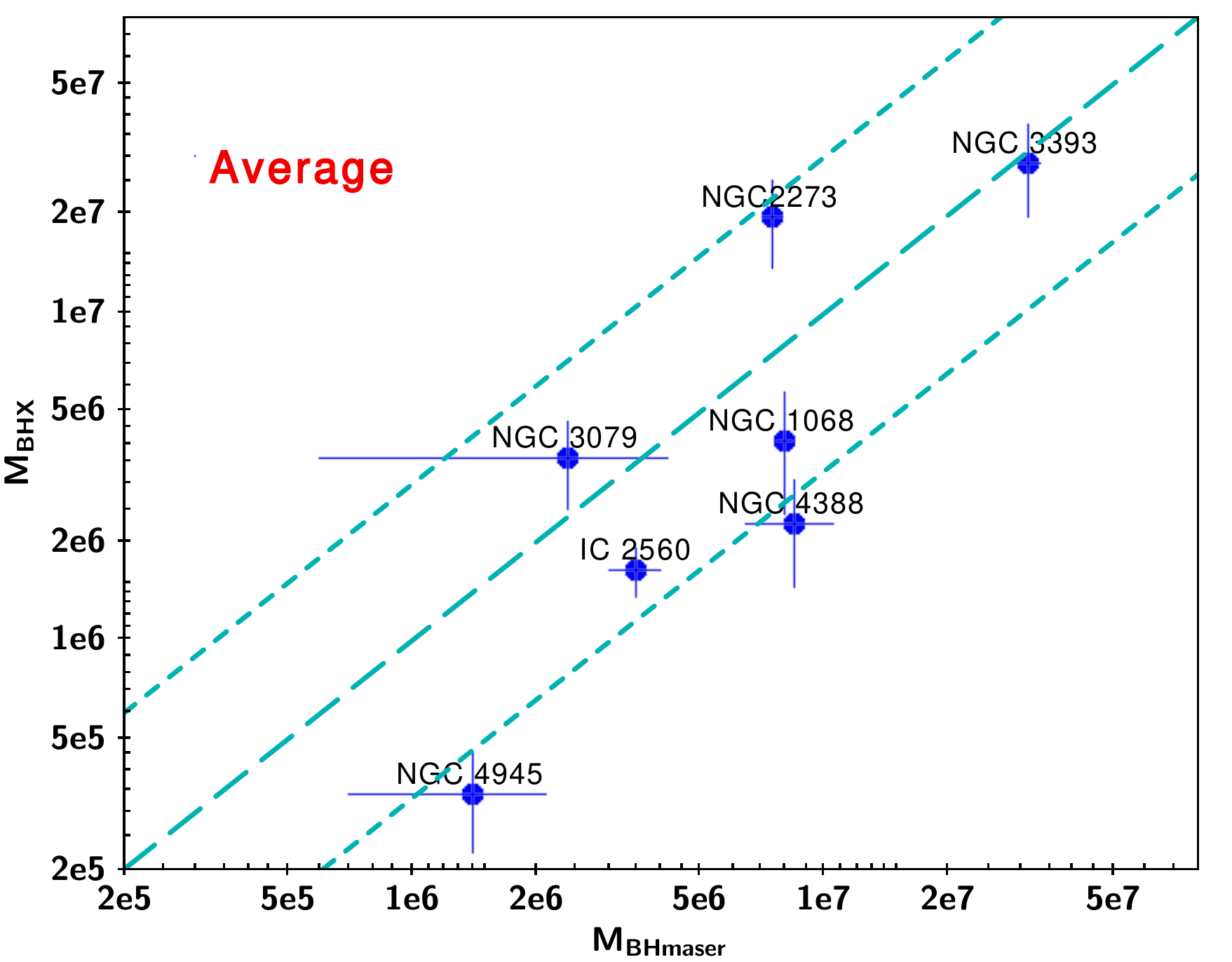}\\
\caption{$M_{\rm BH,X}$, the BH mass obtained with the scaling method plotted versus $M_{\rm BH,maser}$ obtained from the megamaser. The top left panel shows the X-ray scaling values derived from the GROD05 reference, the top right panel those from GXD03, the bottom left the values from 3*XTER98, and the bottom right panel the \mbh\ values obtained from the average of all the available reference sources. The longer-dashed line represents the perfect one-to-one correspondence between the two methods, i.e., a ratio $M_{\rm BH,maser}/M_{\rm BH,X}=1$, whereas the shorter-dashed lines indicate the ratios of 3 and 1/3, respectively.}
\label{figure:fig6}
\end{figure*}
%%%%%%%%%%%%%%%%%%%%%%%%%%%%%%%%%%%%%%%%%%%%%%%%%%%%%%%%%%%%%%%%%%%%%%%%
%%%%%%%%%%%%%%%%%%%%%%%%%%%%%%%%%%%%%%%%%%%%%%%%%%%%%%%%%%%%%%%%%%%%%%%%

We used the criterion $\Delta M_{\mathrm{BH}}/\sigma_{\mathrm{diff}}< 3$ to assess whether the \mbh\ values derived with these two methods are statistically consistent.  In other
words, the X-ray scaling measurements of the \mbh\ are considered formally consistent with the corresponding megamaser values if their difference is less than three times the 
uncertainty $\sigma_{\mathrm{diff}}$.

The results of these comparisons are summarized in Table~\ref{table:tab4} and illustrated in Fig.~\ref{figure:fig5}, where the dashed lines represent the 3$\sigma$ levels. From this figure it is evident that every source has at least one \mbh\ scaling value that is consistent with the maser one, with GROD05 and GXD03 being the most reliable ones, along with the average \mbh\ and the value obtained with 3*XTER98. The latter ones are always within 3$\sigma$ from the megamaser value, also by virtue of their slightly larger uncertainties.

An alternative way to compare the two methods is offered by the ratio $M_{\mathrm{BH,maser}}/M_{\mathrm{BH,scaling}}$. The ratios, obtained by dividing the megamaser \mbh\ by each of the available reference sources, as well as by the \mbh\ average and by 3*XTER98, are reported in Table~\ref{table:tab5} and illustrated in Fig.~\ref{figure:fig6}, where the \mbh\ values obtained with the scaling method for the most reliable references (GROD05, GXD03, 3*XTER98) and the average values are plotted versus their respective megamaser values.  From this figure, one can see that, for GROD05 (top left panel), 3*XTER98 (bottom left panel), and the average (bottom right panel), all values are consistent with the ratio of 1 within a factor of 3, and a good agreement is found also with GXD03 (top right panel) with two sources (IC 2560 and NGC 2273) that have slightly larger values. 

Based on the values reported in Table~\ref{table:tab5}, all ratios obtained from these reference trends are consistent with unity at the 3$\sigma$ limit (i.e., their ratio $\pm3\sigma$ is consistent with 1) confirming the statistical agreement between the two methods. Finally, we note that using \nplbmc\ = 24 (for flat spectrum sources) and 38 (for steep spectrum sources) confirms and reinforces the conclusions derived from the original assumption \nplbmc\ = 30.

%%%%%%%%%%%%%%%%%%%%%%%%%%%%%%%%%%%%%%%%%%%%%%%%%%%%%
%%%%%%%% TAB4: Delta MBH/sigma %%%%%%%%%%%%%%%%%%%%%%%%%%%%%%%%
\begin{table*}
	\caption{$\Delta{M_{\mathrm{BH}}}/\sigma_{\mathrm{diff}}$: Comparison between $M_{\textrm{BH}}$ from maser and X-ray scaling}		
	\begin{center}
		\begin{tabular}{lrrrrrrr} %{lccccccc}
			\toprule
			\toprule       
			\mcol{Source} & \multicolumn{7}{c}{$\Delta{M_{\mathrm{BH}}}/\sigma$}\\
			\cmidrule{2-8}
			
			\mcol{name} & \mcol{GROD05} & \mcol{GROR05} &  \mcol{GXD03} &  \mcol{GXR04} &  \mcol{XTER98}  &  \mcol{average}& \mcol{3$*$XTER98}\\
			
			\mcol{(1)} & \mcol{(2)} & \mcol{(3)} & \mcol{(4)} & \mcol{(5)} & \mcol{(6)} & \mcol{(7)} & \mcol{(8)} \\
			%			\noalign{\smallskip}
			\midrule
			NGC 1068 & \ldots & 16.7 & $-0.5$ & 10 & 8.6 & 2.5& 1.2 \\
			\noalign{\smallskip}
			\midrule
			\noalign{\smallskip}
			NGC 1194 & $19.0$ & 20.5 & $11.3$ & $21.0$ &  $21.0$ &  $18.3$ & $20.4$\\
			\noalign{\smallskip}
			 \nplbmc\ =24& $17.9$ & 19.7 & $6.0$ & $20.8$ &  $20.8$ &  $16.7$ & $20.0$\\
			\noalign{\smallskip}
			\midrule
			\noalign{\smallskip}
			NGC 2273 & $-3.5$ & 0.1 & $-7.3$ & $-9.7$ &  $-0.4$  &  $-2.0$ & $-2.0$\\
			\noalign{\smallskip}
			\midrule
			\noalign{\smallskip}
			NGC 3079 & $-1.2$ & 0.6 & $-2.4$ & $-0.4$ &  0.5 & $-0.6$ & $-0.9$ \\
			\noalign{\smallskip}
			\midrule
			\noalign{\smallskip}
			NGC 3393 & $-1.8$ & 7.3 & $-3.4$ & 3.2 & 4.2 & 0.3 & $-0.3$ \\
			\noalign{\smallskip}
			\midrule
			\noalign{\smallskip}
			NGC 4388 & $2.5$ & 3.7 & $1.4$ & 3.6 & 3.8 & 2.9 & 2.8 \\
			\noalign{\smallskip}
			\nplbmc\ =24 & $2.1$ & 3.6 & $0.8$ & 3.4 & 3.7 & 2.5 & 2.4 \\
			\noalign{\smallskip}
			\midrule
			NGC 4945 & $1.3$ & 1.8 & $1.0$ & 1.7 & 1.8 & 1.5 & 1.4 \\
			\noalign{\smallskip}
			 \nplbmc\ =24 & $1.1$ & 1.8 & $0.6$ & 1.6 & 1.8 & 1.4 & 1.3 \\
			\noalign{\smallskip}
			\midrule
			IC 2560 & \ldots & 0.5 & $-4.9$ & $-7.5$ & $-0.4$ & $-1.6$ & $-2.1$ \\
			\noalign{\smallskip}			
		       \nplbmc\ =38 & \ldots & 1.1 & $-5.7$ & $-6.3$ & $-0.1$ & $-1.4$ & $-2.0$ \\
			\noalign{\smallskip}
			\midrule
			\noalign{\smallskip}
			Circinus & \ldots & 4.4 & \ldots & \ldots & 3.4 & 4.1 & 0.3 \\
			\noalign{\smallskip}
			 \nplbmc\ =38 & \ldots & 4.7 & \ldots & \ldots & 4.0 & 2.3 & 1.0 \\
			\bottomrule
		\end{tabular}
	\end{center}
	\begin{flushleft}
		Columns: 1 = AGN name. 2--8 = Change in black hole mass over error for each reference source. Reference sources: GROD05 = reference source GRO J1655-40 in the decreasing phase; GROR05 = reference source GRO J1655-40 in the rising phase; GXD03 = reference source GX 339-4 in the decreasing phase; GXR03 = reference source GX 339-4 in the rising phase; XTER98 = reference source XTE J1550-564 in the rising phase; 3*XTER98 = reference source XTE J1550-564 in the rising phase with a multiplicative correction of a factor 3 applied. Note, the average value (in column 7) is obtained averaging all the \mbh\ obtained from all the reference sources but excluding 3*XTER98. Note: For each source
		the first line reports the values obtained using \nplbmc\ = 30 in the spectral fitting; the second line (present only for sources with relatively flat or steep spectra) explicitly states the different value of  \nplbmc\ used.		
\\
	\end{flushleft}
	\label{table:tab4}
\end{table*}
%%%%%%%%%%%%%%%%%%%%%%%%%%%%%%%%%%%%%%%%%%%%%%%%%%%%%
%%%%%%%%%%%%%%%%%%%%%%%%%%%%%%%%%%%%%%%%%%%%%%%%%%%%%
%%%%%%%%%%%%%%%%%%%%%%%%%%%%%%%%%%%%%%%%%%%%%%%%%%%%%
%%%%%%%% TAB5: MBH ratio %%%%%%%%%%%%%%%%%%%%%%%%%%%%%%%%%%%%
\begin{table*}
	\caption{Ratio between \mbh\ values obtained from maser measurements and the X-ray scaling method: $M_{\mathrm{BH,maser}}/M_{\mathrm{BH,scaling}}$}		
	\begin{center}
		\begin{tabular}{lrrrrrrr} %{lccccccc} 
			\toprule
			\toprule       
			\mcol{Source} & \multicolumn{7}{c}{Ratio}\\
			\cmidrule{2-8}
			
			\mcol{name} & \mcol{GROD05} & \mcol{GROR05} &  \mcol{GXD03} &  \mcol{GXR04} &  \mcol{XTER98}&  \mcol{average} & \mcol{3$*$XTER98} \\
			
			\mcol{(1)} & \mcol{(2)} & \mcol{(3)} & \mcol{(4)} & \mcol{(5)} & \mcol{(6)} & \mcol{(7)} & \mcol{(8)}\\
			%			\noalign{\smallskip}
			\midrule
			NGC 1068 & \ldots & $5.1\pm0.8$ & $0.9\pm0.1$ & $1.9\pm0.1$ & $4.3\pm1.5$ & $2.0\pm0.8$& $1.4\pm0.5$ \\
			\noalign{\smallskip}
			\midrule
			\noalign{\smallskip}
			NGC 1194 &  $12\pm2$ & $38\pm16$ & $7\pm3$ & $32\pm3$ & $52\pm21$ & $17\pm7$ & $17\pm7$\\
			\noalign{\smallskip}
			 \nplbmc\ =24 &  $10\pm2$ & $30\pm15$ & $6\pm4$ & $27\pm3$ & $44\pm18$ & $13\pm6$ & $15\pm6$\\
			\noalign{\smallskip}
			\midrule
			\noalign{\smallskip}
			NGC 2273 & $0.3\pm0.1$ & $1.0\pm0.2$ & $0.20\pm0.02$ & $0.40\pm0.03$ & $0.9\pm0.3$ & $0.4\pm0.1$& $0.3\pm0.1$ \\
			\noalign{\smallskip}
			\midrule
			\noalign{\smallskip}
			NGC 3079 & $0.5\pm0.4$ & $1.8\pm1.4$ & $0.3\pm0.3$ & $0.8\pm0.6$ & $1.6\pm1.3$ & $0.7\pm0.5$& $0.5\pm0.4$ \\
			\noalign{\smallskip}
			\midrule
			\noalign{\smallskip}
			NGC 3393 & $0.8\pm0.1$ & $2.9\pm0.6$ & $0.5\pm0.1$ & $1.3\pm0.1$ & $2.7\pm1.0$ & $1.1\pm0.3$& $0.9\pm0.3$ \\
			\noalign{\smallskip}
			\midrule
			\noalign{\smallskip}
			NGC 4388 & $2.6\pm0.7$ & $8.8\pm3.6$ & $1.7\pm0.7$ & $6.5\pm1.6$ & $10.9\pm5.0$  & $3.7\pm1.6$& $3.6\pm1.7$\\
			\noalign{\smallskip}
			 \nplbmc\ =24 & $2.1\pm0.6$ & $7.0\pm2.8$ & $1.3\pm0.5$ & $5.2\pm1.3$ & $8.7\pm4.0$  & $3.0\pm1.3$& $2.9\pm1.3$\\
			\noalign{\smallskip}
			\midrule
			\noalign{\smallskip}
			NGC 4945 & $2.8\pm1.5$ & $10.3\pm5.7$ & $2.0\pm1.1$ & $6.2\pm3.1$ & $11\pm6.9$  & $4.1\pm2.5$& $3.7\pm2.3$\\
			\noalign{\smallskip}
			 \nplbmc\ =24 & $2.3\pm1.2$ & $8.2\pm4.6$ & $1.6\pm1.2$ & $4.8\pm2.4$ & $8.7\pm5.4$  & $3.3\pm2.0$& $2.9\pm1.8$\\
			\noalign{\smallskip}
			\midrule
			\noalign{\smallskip}
			IC 2560 & \ldots & $1.1\pm0.2$ & $0.2\pm0.1$ & $0.3\pm0.1$ &  $0.8\pm0.3$ & $0.4\pm0.2$& $0.3\pm0.1$ \\
			\noalign{\smallskip}
			 \nplbmc\ =38 & \ldots & $1.3\pm0.2$ & $0.2\pm0.1$ & $0.4\pm0.1$ &  $1.0\pm0.4$ & $0.5\pm0.2$& $0.3\pm0.1$ \\
			\noalign{\smallskip}
			\midrule
			\noalign{\smallskip}
			Circinus & \ldots & $4.6\pm0.8$ & \ldots & \ldots & $3.3\pm1.2$ & $3.9\pm0.9$& $1.1\pm0.4$ \\
			\noalign{\smallskip}
			\nplbmc\ =38  & \ldots & $5.9\pm1.3$ & \ldots & \ldots & $4.3\pm1.6$ & $2.7\pm1.6$& $1.4\pm0.5$ \\
			\bottomrule
		\end{tabular}
	\end{center}
	\begin{flushleft}
		Columns: 1 = AGN name. 2--8 = Ratio of maser to X-ray scaling for each reference source. Reference sources: GROD05 = reference source GRO J1655-40 in the decreasing phase; GROR05 = reference source GRO J1655-40 in the rising phase; GXD03 = reference source GX 339-4 in the decreasing phase; GXR03 = reference source GX 339-4 in the rising phase; XTER98 = reference source XTE J1550-564 in the rising phase; 3*XTER98 = reference source XTE J1550-564 in the rising phase with a multiplicative correction of a factor 3 applied. Note, the average value (in column 7) is obtained averaging all the \mbh\ obtained from all the reference sources but excluding 3*XTER98. Note: For each source
		the first line reports the values obtained using \nplbmc\ = 30 in the spectral fitting; the second line (present only for sources with relatively flat or steep spectra) explicitly states the different value of  \nplbmc\ used.		
\\
	\end{flushleft}
	\label{table:tab5}
\end{table*}
%%%%%%%%%%%%%%%%%%%%%%%%%%%%%%%%%%%%%%%%%%%%%%%%%%%%%
%%%%%%%%%%%%%%%%%%%%%%%%%%%%%%%%%%%%%%%%%%%%%%%%%%%%%

\section{Discussion and Conclusions}
Constraining the \mbh\ in AGN is of crucial importance, since it determines the space and temporal scales of BHs, constrains their accretion rate via the Eddington ratio, and plays an essential role in our understanding of the BH growth  and co-evolution with the host galaxy. The most reliable ways to determine the \mbh\ are direct dynamical methods, which measure the orbital parameters of ``test particles'', whose motion is dominated by the gravitational force of the supermassive BH. For example, the mass of the supermassive BH at the center of our Galaxy has been tightly constrained by detailed studies of the orbits of a few innermost stars observed over several years (e.g., \citealt{ghez08,gill09}). In nearby weakly active galaxies, the \mbh\ is determined by the gas dynamics within the sphere of influence of the BH (e.g., \citealt{geb03}). On the other hand, in bright type 1 AGN, the \mbh\ measurement is obtained from the dynamics of the BLR via the reverberation mapping technique  (e.g., \citealt{pete04}). Finally, in heavily absorbed type 2 AGN, where the BLR is completely obscured, the only possible direct measurement of the \mbh\ is based on megamaser measurements  (e.g., \citealt{kuo11} and references therein).

The main problem with direct dynamical methods is that they are fairly limited in their application. For instance, direct measurements of \mbh\ via gas dynamics are limited to nearby weakly active galaxies, where the sphere of influence is not outshined by the AGN and are sufficiently close to be resolvable at the angular resolution of ground-based observatories. Similarly, the reverberation mapping technique, which is heavily time and instrument consuming, is limited to type 1 AGN with small or moderate masses. Finally, the megamaser emission in type 2 AGN is relatively rare, and only when the megamaser originates in the accretion disk (as opposed to the jet and outflows) can this technique be used to constrain the \mbh\ (e.g., \citealt{pane20} and references therein).

Fortunately, there are a few robust indirect methods that make it possible to constrain the \mbh\ beyond the range of applicability of the direct dynamical ones. For example, the tight correlation between \mbh\ and the stellar velocity dispersion in the bulge $\sigma_{\ast}$, observed in nearby nearly quiescent galaxies (e.g., \citealt{treme02}), can be extrapolated to constrain the \mbh\ in many distant and more active galaxies. Similarly, the empirical relationship between the BLR radius and optical luminosity makes it possible to determine the mass of numerous type 1 AGN with only one spectral measurement without the need of long monitoring campaigns (e.g., \citealt{kaspi00}). 

Although indirect methods have proven to be very useful to derive general results for large samples of AGN, caution must be applied when these methods are extrapolated well beyond the original range of applicability of the direct methods. To  check for consistency and avoid potential biases associated with the various assumptions inherent in these indirect methods, it is important to develop and utilize alternative techniques to constrain the \mbh. In this perspective, X-ray-based methods may offer a useful complementary way to the more commonly used optically based ones, since X-rays that are produced very close to the BH are less affected by absorption and by star and galaxy contamination. Indeed, model-independent methods based on X-ray variability yielded \mbh\ values broadly consistent with those obtained with dynamical methods (e.g. \citealt{papa04,niko06,mcha06,pon12}). In a previous work focused on a sample of AGN with reverberation mapping measurements and good quality \xmm\ data, we demonstrated that the X-ray scaling method also provides results in agreement with 
reverberation mapping within the respective uncertainties \citep{glioz11}.

It is important to bear in mind that the X-ray scaling method is not equivalent to making some general assumptions on the accretion rate and the bolometric correction and deriving the BH mass from the X-ray luminosity using the formula $M_\mathrm{BH}=\kappa_\mathrm{bol}L_\mathrm{X}/(1.3\times 10^{38} \lambda_\mathrm{Edd})$, where $\kappa_\mathrm{bol}$ is the bolometric correction that may range from 15 to 150 depending on the accretion rate of the source \citep{vasu09}, and $L_\mathrm{X}$ the X-ray luminosity in erg/s. With this simple equation, without an a priori knowledge of the accretion rate of the source, one could at best obtain the order of magnitude of the \mbh. Since $\lambda_\mathrm{Edd}$ can vary over a broad range (for example, for this small sample of obscured AGN, the Eddington ratio varies from 0.01 to 0.3), it is not possible to obtain a specific value of \mbh\ that can be quantitatively compared with the value obtained from the dynamical method and find a good agreement, as we did with the scaling method.

One may then argue that the only important parameter in the scaling method is \nbmc\ (because of its direct dependence on the accretion luminosity and distance) and that it is still possible to obtain a good agreement with the dynamically estimated \mbh\ with any value of the photon index. To test this hypothesis, we have selected the two sources with the flattest spectra of our sample (NGC 4388 and NGC 4945) and the two sources with the steepest spectra (IC 2560 and Circinus), and recalculated their \mbh\ with the scaling method assuming $\Gamma=2.17$ for the flattest sources and $\Gamma=1.66$ for the steepest sources. This led to changes of \mbh\ by a factor slightly larger than 2 (note that considerably larger changes of \mbh\ would have resulted if we had used a larger difference in the photon indices instead of the minimum and maximum values of this small sample). If the photon index did not play any role, then these \mbh\ changes should have not made a difference in the agreement with the values obtained via the dynamical method, with some objects showing a slightly better agreement and others a slightly worse agreement. Instead, all four sources, which were originally consistent with their maser respective estimates based on the mass ratio criterion described above (see Table~\ref{table:tab5} and Figure~\ref{figure:fig6}), showed a clear departure from the dynamical \mbh\ values with three sources (NGC 4388, NGC 4945, and IC 2560) that were not formally consistent with the maser values anymore (their new mass ratios were 8.0, 7.1, and 6.7, respectively) and only Circinus (ratio of 0.5) still consistent, but only by virtue of the fact that the original ratio was basically 1. 
We therefore conclude that the scaling method works because the photon index accurately characterizes the accretion state of accreting black holes and allows the correct selection of the reference source's \nbmc\ value to be compared with the AGN's value.

In this study, we have extended the X-ray scaling method to a sample of heavily obscured type 2 AGN with \mbh\ already constrained by megamaser measurements.
This dynamical method is rightly considered one of the most reliable; however, the accuracy of the \mbh\ derived with this technique depends on the quality of the radio data, on the assumption that the megamaser emission is produced in an edge-on disk, and that its rotation curve is strictly Keplerian. Additionally, one should bear in mind that this technique measures the mass enclosed within the megamaser emission. As a consequence, the actual \mbh\ may be slightly smaller if the measured enclosed mass encompasses a nuclear cluster or the inner part of a massive disk, or alternatively slightly larger if radiation pressure (not included in the \mbh\ derivation) plays an important role \citep{kuo11}. 

Specifically, for the sources of our sample, the rotation curve traced by the megamaser in 
NGC 1068 is non-Keplerian; the \mbh\  was derived assuming a self-gravitating accretion disk model \citep{loda03}. NGC 1194 displays one of the largest maser disks (with inner and outer radii of 0.54 and 1.33 parsecs) which appears to be
slightly bent and is consistent with Keplerian rotation \citep{kuo11}.  NGC 2273 also shows indications of a warped but much smaller disk (with inner
and outer disk radii of 0.028 and 0.084 pc) with Keplerian rotation \citep{kuo11}. In NGC 3079 the disk appears to be thick and flared \citep{kondra05}, whereas in  NGC 3393 the maser seems to describe a flat disk perpendicular to the kpc radio jet, and the positions of the
maser points have substantial uncertainties \citep{kondra08}. NGC 4388, located in the Virgo cluster, has only five megamaser spots, which make it
impossible to demonstrate that they lie on a disk or that the rotation is Keplerian \citep{kuo11}. For this reason, to reflect the actual uncertainty on the \mbh\ derived by megamaser measurements, we have increased the statistical error by a factor of 10, leading to an uncertainty of $\sim$24\%. The megamaser in NGC 4945 has been 
modeled as an edge-on thin disk, although this is not the only possible interpretation of the data; the non-Keplerian rotation of the blue-shifted emission and the substantial position errors lead to a relatively large uncertainty in the \mbh\ of $\sim$50\% \citep{green97}. In  IC 2560 the 
megamaser emission has been attributed to an edge-on thin disk with Keplerian rotation with some additional contribution from a jet \citep{yama12}.
Finally, the megamaser emission in Circinus appears to be associated with a warped accretion disk and a wide-angle outflow \citep{green03}. In
summary, because of the presence of outflows, jets, disk warps, or non-Keplerian rotation curves, we should consider the \mbh\ values determined from megamaser measurements as robust estimates but not as extremely accurate values, and the errors reported in Table~\ref{table:tab1} are likely lower
 limits on their actual uncertainties.

With respect to type 1 AGN, the main difficulty of applying the X-ray scaling method to heavily obscured AGN is the need to properly constrain the parameters of the primary emission in sources whose X-ray spectra are dominated by absorption and reflection. However, the \nustar\ spectra
of these specific sources, often complemented with \chandra\ and \xmm\ data, were the object of very detailed analyses, which led to the
disentanglement and a careful characterization of the different contributions of the AGN direct and reprocessed emission, of the host galaxy, and of the 
off-nuclear sources located in the spectral extraction region (e.g., \citealt{yaq12,puc14,are14,bau15}). Guided by these findings, we were able to parametrize the 
torus contribution using the physically motivated self-consistent model \borus\ \citep{balo18} instead of the \mytorus\ or \torus\ models used in the previous analyses. To characterize the primary emission,  instead of the phenomenological power-law model, we utilized the \texttt{BMC} Comptonization model, since the scaling method directly scales the normalization of this model \nbmc\ between AGN and an appropriate stellar reference to determine \mbh. 

With our baseline spectral model, where we assumed \nplbmc\ = 30,  as described in detail in Section 3 (see also the Appendix for details on the
spectral fittings of individual sources), and applying the scaling technique summarized in Section 4.1, we obtained the following results:
\begin{itemize}
\item Many of the \mbh\ values, obtained with different reference trends, are broadly in agreement with the corresponding megamaser ones. In particular, the estimates derived
using GROD05, 3*XTER98, and the ones obtained by averaging the values inferred from all the available reference sources, are consistent 
 at the 3$\sigma$ level, based on measurements of $\Delta M_{\mathrm{BH}}/\sigma_{\mathrm{diff}}  =  (M_{\mathrm{BH,maser}}-M_{\mathrm{BH,scaling}})/\sigma_{\mathrm{diff}}$, which are reported in Table~\ref{table:tab4} and shown in  Fig.~\ref{figure:fig5}.

\item The agreement between the two methods is confirmed by the $M_{\mathrm{BH,maser}}/M_{\mathrm{BH,scaling}}$ ratio: for all type 2 AGN of our sample
$(M_{\mathrm{BH,maser}}/M_{\mathrm{BH,scaling}})\pm3\sigma \leq 1$, when using the best reference sources or the average \mbh, as summarized in Table~\ref{table:tab5}. Fig.~\ref{figure:fig6} 
illustrates the good agreement between the two methods, showing that GROD05, GXD04 (partially), 3*XTER98, and the average obtained from all reference patterns are all consistent with the one-to-one ratio within a factor of three.

\item  The only object of our sample for which the \mbh\ inferred from the X-ray scaling method is statistically inconsistent with the megamaser value is NGC 1194, which is the AGN with the lowest accretion rate  ($\lambda_\mathrm{{Edd}}\simeq 7\times10^{-3}$). However, this discrepancy is expected, since the X-ray scaling method cannot be applied in this 
regime, where $\Gamma$ generally shows an anti-correlation  with $\lambda_{\mathrm{Edd}}$.
\end{itemize}

{\rm  In conclusion, our work demonstrates that the same X-ray scaling method works equally well for type 1 AGN (given the formal agreement with the reverberation mapping sample)
and type 2 AGN (based on the agreement with the megamaser sample). We thus conclude that this method
can be safely applied to any type of AGN regardless of their level of obscuration, provided that these sources accrete above a minimum threshold and that their primary
X-ray emission can be robustly characterized via spectral analysis.}
 This also proves that this method is robust and can be used to complement 
the various indirect methods, especially when they are applied well beyond the range of validity of the direct methods,
from which they were calibrated. Finally, the X-ray scaling method offers the possibility to investigate in a systematic and homogeneous way the existence of any intrinsic 
 difference in the fundamental properties of the central engines in type 1 and type 2 AGN. We plan to carry out this type of investigation in our future work.

\section*{Acknowledgements}
We thank the anonymous referee for constructive comments and suggestions that improved the clarity of the paper and helped strengthen our conclusions.
This research has made use of data, software, and/or web tools obtained from the High Energy Astrophysics Science Archive Research Center (HEASARC), a service of the Astrophysics Science Division at NASA/GSFC and of the Smithsonian Astrophysical Observatory's High Energy Astrophysics Division, and of the NuSTAR Data Analysis Software (NuSTARDAS) jointly developed by the ASI Science Data Center (ASDC, Italy) and the California Institute of Technology (Caltech, USA).
%The Acknowledgements section is not numbered. Here you can thank helpful
%colleagues, acknowledge funding agencies, telescopes and facilities used etc.

\section*{Data Availability}
The data underlying this article are available in  the High Energy Astrophysics Science Archive Research Center (HEASARC) Archive at https://heasarc.gsfc.nasa.gov/docs/archive.html.



%\bibliographystyle{mnras}

\begin{thebibliography}{99}
	
\bibitem[\protect\citeauthoryear{Antonucci}{1983}]{ant83} Antonucci R.~R.~J., 1983, Nature, 303, 158

\bibitem[\protect\citeauthoryear{Antonucci}{1993}]{ant93} Antonucci R.~R.~J., 1993, ARA\&A, 31, 473

\bibitem[\protect\citeauthoryear{Ar{\'e}valo et al.}{2014}]{are14} Ar{\'e}valo P. et al., 2014, ApJ, 791, 81

\bibitem[\protect\citeauthoryear{Arnaud}{1996}]{arn96} Arnaud K.~A., 1996, ASPC, 101, 17

\bibitem[\protect\citeauthoryear{Balokovi{\'c} et al.}{2018}]{balo18} Balokovi{\'c} M. et al., 2018, ApJ, 854, 42

\bibitem[\protect\citeauthoryear{Bauer et al.}{2015}]{bau15} Bauer F.~E. et al., 2015, ApJ, 812, 116
	
\bibitem[\protect\citeauthoryear{Blandford \& McKee}{1982}]{bland82} Blandford R. D., McKee C. F., 1982, ApJ, 255, 419

\bibitem[\protect\citeauthoryear{Brightman et al.}{2013}]{bright13} Brightman M. et al., 2013, MNRAS, 433, 2485

\bibitem[\protect\citeauthoryear{Brightman et al.}{2016}]{bright16} Brightman M. et al., 2016, ApJ, 826, 93

\bibitem[\protect\citeauthoryear{Brightman \& Nandra}{2011}]{brina11} Brightman M., Nandra K., 2011, MNRAS, 413, 1206

\bibitem[\protect\citeauthoryear{Constantin et al.}{2009}]{const09} Constantin A., Green P., Aldcroft T., Kim D.-W., Haggard D., Barkhouse W., Anderson S.~F., 2009, ApJ, 705, 1336

\bibitem[\protect\citeauthoryear{Gebhardt et al.}{2003}]{geb03} Gebhardt K. et al., 2003, ApJ, 583, 92

\bibitem[\protect\citeauthoryear{Ghez et al.}{2008}]{ghez08} Ghez A.~M. et al., 2008, ApJ, 689, 1044

\bibitem[\protect\citeauthoryear{Giacch\'e, Gilli, \& Titarchuk}{2014}]{gia14} Giacch\'e S., Gilli R., Titarchuk L., 2014, A\&A, 562A, 44

\bibitem[\protect\citeauthoryear{Gillessen et al.}{2009}]{gill09} Gillessen S., Eisenhauer F., Trippe S., Alexander T., Genzel R., Martins F., Ott T., 2009, ApJ, 692, 1075

\bibitem[\protect\citeauthoryear{Gliozzi et al.}{2010}]{glioz10} Gliozzi M., Papadakis I. E., Grupe D., Raeth C.,  Kedziora-Chudczer L., 2010, ApJ, 717, 1243

\bibitem[\protect\citeauthoryear{Gliozzi et al.}{2011}]{glioz11} Gliozzi M., Titarchuk L., Satyapal S., Price D., Jang I., 2011, ApJ, 735, 16

\bibitem[\protect\citeauthoryear{Gliozzi \& Williams}{2020}]{glioz20} Gliozzi M., Williams J.~K., 2020, MNRAS, 491, 532

\bibitem[\protect\citeauthoryear{Greenhill et al.}{1997}]{green97} Greenhill L.~J., Ellingsen S.~P., Norris R.~P., Gough R.~G., Sinclair M.~W., Moran J.~M., Mushotzky R., 1997, ApJL, 474, L103

\bibitem[\protect\citeauthoryear{Greenhill et al.}{2003}]{green03} Greenhill L.~J. et al., 2003, ApJ, 590, 162

\bibitem[\protect\citeauthoryear{Gu \& Cao}{2009}]{gu09} Gu M., Cao X., 2009, MNRAS, 399, 349

\bibitem[\protect\citeauthoryear{G{\"u}ltekin et al.}{2012}]{gulte12} G{\"u}ltekin K., Cackett E.~M., Miller J.~M., Di Matteo T., Markoff S., Richstone D.~O., 2012, ApJ, 749, 129

\bibitem[\protect\citeauthoryear{Hickox \& Alexander}{2018}]{hick18} Hickox R.~C., Alexander D.~M., 2018, ARA\&A, 56, 1

%\bibitem[\protect\citeauthoryear{Hood et al.}{2017}]{hood17} Hood C.~E., Barth A.~J., Ho L.~C., Greene J.~E., 2017, ApJ, 838, 26

\bibitem[\protect\citeauthoryear{Jang et al.}{2014}]{jang14} Jang I., Gliozzi M., Hughes C., Titarchuk L., 2014, MNRAS, 443, 72

\bibitem[\protect\citeauthoryear{Jang et al.}{2018}]{jang18} Jang I., Gliozzi M., Satyapal S., Titarchuk L., 2018, MNRAS, 473, 136

\bibitem[\protect\citeauthoryear{Kaspi et al.}{2000}]{kaspi00} Kaspi S., Smith P.~S., Netzer H., Maoz D., Jannuzi B.~T., Giveon U., 2000, ApJ, 533, 631

\bibitem[\protect\citeauthoryear{Khachikian \& Weedman}{1974}]{khach74} Khachikian E.~Y., Weedman D.~W., 1974, ApJ, 192, 581

\bibitem[\protect\citeauthoryear{Kondratko, Greenhill, \& Moran}{2005}]{kondra05} Kondratko P.~T., Greenhill L.~J., Moran J.~M., 2005, ApJ, 618, 618

\bibitem[\protect\citeauthoryear{Kondratko, Greenhill, \& Moran}{2008}]{kondra08} Kondratko P.~T., Greenhill L.~J., Moran J.~M., 2008, ApJ, 678, 87

\bibitem[\protect\citeauthoryear{Koss et al.}{2015}]{koss15} Koss M.~J. et al., 2015, ApJ, 807, 149

\bibitem[\protect\citeauthoryear{Kuo et al.}{2011}]{kuo11} Kuo C.~Y. et al., 2011, ApJ, 727, 20

%\bibitem[\protect\citeauthoryear{Liu et al.}{2018}]{liu18} Liu T. et al., 2018, MNRAS, 479, 5022

\bibitem[\protect\citeauthoryear{Lodato \& Bertin}{2003}]{loda03} Lodato G., Bertin G., 2003, A\&A, 398, 517

\bibitem[\protect\citeauthoryear{Masini et al.}{2016}]{mas16} Masini A. et al., 2016, A\&A, 589, A59

\bibitem[\protect\citeauthoryear{McHardy et al.}{2006}]{mcha06} McHardy I.~M., Koerding E., Knigge C., Uttley P., Fender R.~P., 2006, Nature, 444, 730

\bibitem[\protect\citeauthoryear{Murphy \& Yaqoob}{2009}]{muya09} Murphy K.~D., Yaqoob T., 2009, MNRAS, 397, 1549

\bibitem[\protect\citeauthoryear{Netzer}{2015}]{netz15} Netzer H., 2015, ARA\&A, 53, 365

\bibitem[\protect\citeauthoryear{Niko{\l}ajuk et al.}{2006}]{niko06} Niko{\l}ajuk M., Czerny B., Zi{\'o}{\l}kowski J., Gierli{\'n}ski M., 2006, MNRAS, 370, 1534

%\bibitem[\protect\citeauthoryear{Onori et al.}{2017}]{ono17} Onori F. et al., 2017, MNRAS, 468, L97

\bibitem[\protect\citeauthoryear{Osterbrock}{1978}]{oster78} Osterbrock D.~E., 1978, PNAS, 75, 540

\bibitem[\protect\citeauthoryear{Panessa et al.}{2020}]{pane20} Panessa F., Castangia P., Malizia A., Bassani L., Tarchi A., Bazzano A., Ubertini P., 2020, A\&A, in press, preprint (arXiv:2006.08280)

\bibitem[\protect\citeauthoryear{Papadakis}{2004}]{papa04} Papadakis I.~E., 2004, MNRAS, 348, 207

\bibitem[\protect\citeauthoryear{Peterson et al.}{2004}]{pete04} Peterson B.~M. et al., 2004, ApJ, 613, 682

\bibitem[\protect\citeauthoryear{Ponti et al.}{2012}]{pon12} Ponti G., Papadakis I., Bianchi S., Guainazzi M., Matt G., Uttley P., Bonilla N.~F., 2012, A\&A, 542, A83

\bibitem[\protect\citeauthoryear{Puccetti et al.}{2014}]{puc14} Puccetti S. et al., 2014, ApJ, 793, 26

\bibitem[\protect\citeauthoryear{Ramos Almeida \& Ricci}{2017}]{ramo17} Ramos Almeida C., Ricci C., 2017, Nature Astronomy, 1, 679

\bibitem[\protect\citeauthoryear{Remillard \& McClintock}{2006}]{remi06} Remillard R.~A., McClintock J.~E., 2006, ARA\&A, 44, 49

\bibitem[\protect\citeauthoryear{Risaliti, Young, \& Elvis}{2009}]{risa09} Risaliti G., Young M., Elvis M., 2009, ApJL, 700, L6

\bibitem[\protect\citeauthoryear{Seifina, Titarchuk, \& Shaposhnikov}{2014}]{sei14}  Seifina E., Titarchuk L., Shaposhnikov N., 2014, ApJ, 789, 57

\bibitem[\protect\citeauthoryear{Seifina, Chekhtman, \& Titarchuk}{2018}]{sei18}  Seifina E.,  Chekhtman A., Titarchuk L., 2018, A\&A, 613A, 48

\bibitem[\protect\citeauthoryear{Shaposhnikov \& Titarchuk}{2009}]{shapo09} Shaposhnikov N., Titarchuk L., 2009, ApJ, 699, 453

\bibitem[\protect\citeauthoryear{Shemmer et al.}{2008}]{shem08} Shemmer O., Brandt W.~N., Netzer H., Maiolino R., Kaspi S., 2008, ApJ, 682, 81

\bibitem[\protect\citeauthoryear{Sobolewska \& Papadakis}{2009}]{sobo09} Sobolewska M.~A., Papadakis I.~E., 2009, MNRAS, 399, 1597

\bibitem[\protect\citeauthoryear{Tadhunter}{2008}]{tad08} Tadhunter C., 2008, NewAR, 52, 227

\bibitem[\protect\citeauthoryear{Titarchuk, Mastichiadis, \& Kylafis}{1997}]{tita97} Titarchuk L., Mastichiadis A., Kylafis N.~D., 1997, ApJ, 487, 834

\bibitem[\protect\citeauthoryear{Titarchuk \& Seifina}{2016}]{tita16a} Titarchuk L., Seifina E., 2016a, A\&A, 585A, 94

\bibitem[\protect\citeauthoryear{Titarchuk \& Seifina}{2016}]{tita16b} Titarchuk L., Seifina E., 2016b, A\&A, 595A, 101

\bibitem[\protect\citeauthoryear{Tremaine et al.}{2002}]{treme02} Tremaine S. et al., 2002, ApJ, 574, 740

\bibitem[\protect\citeauthoryear{Urry \& Padovani}{1995}]{urry95} Urry C.~M., Padovani P., 1995, PASP, 107, 803

%\bibitem[\protect\citeauthoryear{Villarroel et al.}{2017}]{villa17} Villarroel B., Nyholm A., Karlsson T., Comer{\'o}n S., Korn A.~J., Sollerman J., Zackrisson E., 2017, ApJ, 837, 110
\bibitem[\protect\citeauthoryear{Vasudevan \& Fabian}{2009}]{vasu09} Vasudevan R. ~V., Fabian A.~C., 2009, MNRAS, 392, 1124

\bibitem[\protect\citeauthoryear{Williams, Gliozzi, \& Rudzinsky}{2018}]{will18} Williams J.~K., Gliozzi M., Rudzinsky R.~V., 2018, MNRAS, 480, 96

\bibitem[\protect\citeauthoryear{Yamauchi et al.}{2012}]{yama12} Yamauchi A., Nakai N., Ishihara Y., Diamond P., Sato N., 2012, PASJ, 64, 103

\bibitem[\protect\citeauthoryear{Yaqoob}{2012}]{yaq12} Yaqoob T., 2012, MNRAS, 423, 3360

\end{thebibliography}

%\begin{thebibliography}{99}
%\end{thebibliography}

%%%%%%%%%%%%%%%%%%%%%%%%%%%%%%%%%%%%%%%%%%%%%%%%%%

%%%%%%%%%%%%%%%%% APPENDICES %%%%%%%%%%%%%%%%%%%%%

\appendix

\section{Additional Spectral Results}

\noindent{\bf NGC 1068:} A detailed analysis of the \nustar, \xmm, and \chandra\ spectra of this source was carried out by \citet{bau15}. Thanks to the excellent sensitivities of \xmm\ and
\nustar\ over broad complementary energy ranges, and to the sub-arcsecond spatial resolution of \chandra, the authors were able to disentangle the contributions of the host galaxy and off-nuclear sources from the AGN emission within the \nustar\ extraction region. The overall  best-fit model is fairly complex and comprises several Fe and Ni emission lines, a Bremsstrahlung component to account for the radiative recombination continuum and lines, a cutoff power-law model to account for the off-nuclear X-ray sources, in addition to the AGN-related emission, which is parametrized by two different \mytorus\ scattered and line components, in addition to the transmitted one described by  the zeroth-order component of that model. In our fitting, in addition to our baseline model we added the Bremsstrahlung and cutoff power law with all parameters fixed at the values provided by  \citet{bau15}, and a Gaussian line to roughly model the excess around 6.5 keV. To account for the multiple absorption components, we also added a second \borus\ model, whose best-fit parameters are $\log(N_{{\textrm{H}}_{\textrm{bor}}})=24.9\pm0.1$, CFtor = 83\%, and $A_{\textrm{Fe}}=1$. Our best-fit parameters are broadly consistent with the results presented by \citet{bau15}.
The observed flux in the 2--10 keV energy band is $5.4\times10^{-12}~ \mathrm{erg~cm^{-2}s^{-1}}$, and the intrinsic one (i.e., corrected for absorption)  $1.3\times10^{-10}~ \mathrm{erg~cm^{-2}s^{-1}}$.

\noindent{\bf NGC 1194:}  The starting model for the spectral fit of this source is provided by the work of \citet{mas16}, who fitted the \nustar\ spectrum with the \mytorus\ model in the decoupled mode, with the addition of a Gaussian line at 6.8 keV, and a scattering fraction of the primary continuum of $f_{\mathrm{s}} \sim 3 \%$.
In our fitting, we used our baseline model and found the main parameters ($\Gamma$,  $N_{\mathrm{H}}$, and $f_{\mathrm{s}}$) to be fully consistent with their best-fit results.
The 2--10 keV observed flux is $1.2\times10^{-12}~ \mathrm{erg~cm^{-2}s^{-1}}$, and the intrinsic one $1.0\times10^{-11}~ \mathrm{erg~cm^{-2}s^{-1}}$.

\noindent{\bf NGC 2273:}  The starting spectral model for this source is again provided by the work of \citet{mas16}, who fitted the \nustar\ spectrum with the \torus\ model that favored a heavily absorbed scenario with $N_{\mathrm{H}} > 7\times 10^{24} \,{\mathrm{cm^{-2}}} $.
In our fitting, we used our baseline model, which yielded a best fit broadly consistent with their results. The 2--10 keV observed flux is $9.2\times10^{-13}~ \mathrm{erg~cm^{-2}s^{-1}}$, and the intrinsic one $3.6\times10^{-10}~ \mathrm{erg~cm^{-2}s^{-1}}$.

\noindent{\bf NGC 3079:}  The starting spectral model for this source is again provided by the work of \citet{mas16}, who fitted the \nustar\ spectrum with the \mytorus\ model in a coupled mode. The results obtained with our baseline model are consistent within the respective uncertainties with their results. The 2--10 keV observed flux is $6.4\times10^{-13}~ \mathrm{erg~cm^{-2}s^{-1}}$, and the intrinsic one $1.2\times10^{-10}~ \mathrm{erg~cm^{-2}s^{-1}}$.

\noindent{\bf NGC 3393:}  The starting spectral model for this source is provided by the work of \citet{koss15} and \citet{mas16}, who fitted the \nustar\ spectrum with both \mytorus\ and \torus\ models. The results obtained with our baseline model are broadly consistent with the results presented by these authors with a slightly larger value of $N_{\mathrm{H}}$ ($10^{25}$ vs. $2.2\times 10^{24} \,{\mathrm{cm^{-2}}}$). The 2--10 keV observed flux is $4.4\times10^{-13}~ \mathrm{erg~cm^{-2}s^{-1}}$, and the intrinsic one $8.3\times10^{-11}~ \mathrm{erg~cm^{-2}s^{-1}}$.

\noindent{\bf NGC 4388:}  The starting spectral model for this source is once more provided by the work of \citet{mas16}, who fitted the \nustar\ spectrum with the \mytorus\ and \torus\ models, which favor a Compton-thin scenario with a substantial scattered primary emission that dominates below 5 keV. The results from our baseline model are fully consistent with their results. The 2--10 keV observed flux is $7.9\times10^{-12}~ \mathrm{erg~cm^{-2}s^{-1}}$, and the intrinsic one $1.4\times10^{-11}~ \mathrm{erg~cm^{-2}s^{-1}}$.

\noindent{\bf NGC 4945:} A detailed analysis of the \nustar, \suzaku, and \chandra\ spectra of this source was carried out by \citet{puc14}, who in turn, were guided by the results obtained by \citet{yaq12} based on a comprehensive analysis of all the hard X-ray spectra available at that time. The wealth of  high-quality broad-band spectra obtained with several observatories made it possible to parametrize separately the different contributions of the host galaxy, the AGN, and contaminating sources within the \nustar\ extraction region. The best-fit model is fairly complex and comprises several emission lines, the galaxy optically thin thermal continuum, which is described  by the \texttt{APEC} model,  the contamination from off-nuclear sources parametrized by a power law, and the AGN emission seen through a torus described by the \mytorus\ model in the decoupled mode. 
In our fitting procedure, in addition to our baseline model we included the \texttt{APEC} and power-law models with all parameters fixed at the values provided by Yaqoob. Our results are broadly consistent with those obtained by both Yaqoob and Puccetti. The 2--10 keV observed flux is $3.7\times10^{-12}~ \mathrm{erg~cm^{-2}s^{-1}}$, and the intrinsic one $2.7\times10^{-10}~ \mathrm{erg~cm^{-2}s^{-1}}$.

\noindent{\bf IC 2560:}  The starting spectral model for this source is again provided by the work of \citet{mas16}, who fitted the \nustar\ spectrum with the \torus\ model, which favors a heavily absorbed primary emission characterized by a steep photon index. The results from our baseline model are broadly consistent with their results. The 2--10 keV observed flux is $3.7\times10^{-13}~ \mathrm{erg~cm^{-2}s^{-1}}$, and the intrinsic one $1.7\times10^{-10}~ \mathrm{erg~cm^{-2}s^{-1}}$.

\noindent{\bf Circinus:} A detailed analysis of the \nustar, \xmm, and \chandra\ spectra of this source was carried out by \citet{are14}. Combining the complementary properties of these 
observatories (i.e., the high sensitivities of \xmm\ and \nustar\ over broad energy ranges and the sub-arcsecond spatial resolution of \chandra), the authors were able to disentangle the contributions of different contamination sources (diffuse emission from the host galaxy, supernova remnant contribution, and off-nuclear X-ray binary sources) from the AGN emission within the \nustar\ extraction region. The overall  best-fit model is complex and comprises several emission lines, an \texttt{APEC} model for the diffuse emission, three \texttt{Mekal} models to parametrize the supernova remnant, and a power-law model to account for the off-nuclear point-like sources, in addition to two different \mytorus\ models used in the decoupled mode. In our fitting, in addition to our baseline model we added all the contamination models with all the parameters fixed at the values provided by  \citet{are14} and three Gaussian lines to roughly model the line excess in the 5.5--7.5 keV range. To account for the multiple absorption components, we also added a second \borus\ model, whose best-fit parameters are $\log(N_{{\textrm{H}}_{\textrm{bor}}})=24.6\pm0.1$, CFtor = 10\%, and $A_{\textrm{Fe}}=1$. Our best-fit parameters are broadly consistent with their results. The 2--10 keV observed flux is $2.0\times10^{-11}~ \mathrm{erg~cm^{-2}s^{-1}}$, and the intrinsic one $2.1\times10^{-10}~ \mathrm{erg~cm^{-2}s^{-1}}$.

\section{The X-ray scaling method}

The X-ray scaling method for determining the mass of a black hole (\mbh) was first described by \citet{shapo09} and first applied to AGN by \citet{glioz11}, where the method is described in detail. Here, we only report the essential information on the stellar reference sources -- their \mbh\ values and distances (Table~\ref{table:tabA1}) and the mathematical expression of the spectral trend with the best fit parameters for the different sources (Table~\ref{table:tabA2}) -- that is needed to reproduce the \mbh\ values. The two steps below accomplish the scaling described in Section 4.1.

Step 1. Use the  following equation to solve for $N_\textrm{BMC,r}$, the BMC normalization the reference source would have at the same photon index as the target AGN. The reference source is a Galactic, stellar-mass black hole with known mass and distance.

\begin{equation}\label{eqjang}
	N_\textrm{BMC,r}(\Gamma) = N_\textrm{tr} \times \left\lbrace 1 - \ln \left[ \exp \left( \frac{A - \Gamma}{B} \right) - 1 \right] \right\rbrace ^{(1/\beta)}
\end{equation}
where $\Gamma$ is the photon index of the target AGN as determined by the spectral fit, and $A$, $B$, $N_\mathrm{tr}$, and $\beta$ are given in Table~\ref{table:tabA2}. Note: this equation was first presented by \citet{jang18} with an error: there should be a minus sign before the logarithm.

Step 2. Use the equation presented in Section 4.1 to solve for $M_\textrm{BH,t}$. 

\begin{equation}\label{eqgliozzi}
	M_\textrm{BH,t} = M_\textrm{BH,r} \times \left( \frac{N_\textrm{BMC,t}}{N_\textrm{BMC,r}} \right) \times \left( \frac{d_\textrm{t}}{d_\textrm{r}} \right)^2
\end{equation}
where \mbh\ is the black hole mass, \nbmc\ is the BMC normalization, and $d$ is the distance. The $t$ subscript denotes the target AGN and the $r$ subscript denotes the reference source.

\begin{table*}
	\caption{Characteristics of reference sources}		
	\begin{center}
	\begin{tabular}{lcc} 
			\toprule
			\toprule       
			Name & \mbh & $d$ \\
			& (\msun) & (kpc)  \\
			%			\noalign{\smallskip}
			\midrule
			\gro & $6.3 \pm 0.3$ & $3.2 \pm 0.2$  \\
			\noalign{\smallskip}
			\gx & $12.3 \pm 1.4$ & $5.7 \pm 0.8$  \\
			\noalign{\smallskip}
			\xte & $10.7 \pm 1.5$ & $3.3 \pm 0.5$  \\			\noalign{\smallskip}
			\bottomrule
		\end{tabular}	
	\end{center}
	\begin{flushleft}
	Source: \citet{glioz11}	
	\end{flushleft}
	\label{table:tabA1}
	%\footnotesize
\end{table*}

\begin{table*}
	\caption{Parametrization of $\Gamma$--\nbmc\ reference patterns}		
	\begin{center}
	\begin{tabular}{lcccc} 
			\toprule
			\toprule       
			\mcol{Transition} & $A$ & $B$ & $N_\mathrm{tr}$ & $\beta$  \\
			\mcol{(1)} & \mcol{(2)} & \mcol{(3)} & \mcol{(4)} & \mcol{(5)} \\
			%			\noalign{\smallskip}
			\midrule
			\gro\ D05 & $1.96\pm0.02$ & $0.42\pm0.02$ & $0.023\pm0.001$ & $1.8\pm0.2$ \\
			\gro\ R05 & $2.35\pm0.04$ & $0.74\pm0.04$ & $0.131\pm0.001$ & $1.0\pm0.1$ \\
			\gx\ D03 & $2.13\pm0.03$ & $0.50\pm0.04$ & $0.0130\pm0.0002$ & $1.5\pm0.3$ \\
			\gx\ R04 & $2.10\pm0.03$ & $0.46\pm0.01$ & $0.037\pm0.001$ & $8.0\pm1.5$ \\
			\xte\ R98 & $2.96\pm0.02$ & $2.8\pm0.2$ & $0.055\pm0.010$ & $0.4\pm0.1$ \\
			\bottomrule
		\end{tabular}	
	\end{center}
	\begin{flushleft}
		Columns: 1 = reference source spectral transition. 
		2 = parameter that determines the rigid translation of the spectral pattern along the y-axis. 3 = parameter characterizing the lower saturation level of the pattern. 4 = parameter that determines the rigid translation of the spectral pattern along the x-axis. 5 = slope of the spectral pattern. Source: \citet{glioz11}
	\end{flushleft}
	\label{table:tabA2}
	%\footnotesize
\end{table*}

%%%%%%%%%%%%%%%%%%%%%%%%%%%%%%%%%%%%%%%%%%%%%%%%%%%%%%%%%%%%%%%%
% Don't change these lines
\bsp	% typesetting comment
\label{lastpage}
\end{document}